%% file: John_Zhang_Latex_Paper_5.tex
\documentclass[12pt]{article}
\usepackage{xcolor}
\usepackage{graphicx}
\usepackage{xymtex}
\usepackage{epsfig}
\usepackage{mathtools}
\usepackage{tabulary}
\usepackage{array}
\usepackage[version=4]{mhchem}
\usepackage{float}
\usepackage[utf8x]{inputenc}
\usepackage{amsmath}
\usepackage{caption}
\usepackage{multirow}
\usepackage{geometry}
\usepackage{pdflscape}
\usepackage{booktabs}
\usepackage{amssymb}

\input econfmacros.tex
\def\Title#1{\begin{center} {\Large {\bf #1} } \end{center}}

\begin{document}

\Title{Topological Properties and Characterizations }

\bigskip\bigskip


\begin{raggedright}  

{\it John Hongguang Zhang\index{Zhang, J.}\\
Altamont Research\\
Altamont, NY12009}
\bigskip\bigskip
\end{raggedright}

{\bf Abstract:}  There are three important types of structural properties that remain unchanged under the structural transformation of condensed matter physics and chemistry. They are the properties that remain unchanged under the structural periodic transformation-periodic properties. The properties that remain unchanged under the structural multi scale transformation-fractal properties. The properties that remain unchanged under the structural continuous deformation transformation-topological properties. In this paper, we will describe topological properties and characterizations in three layers: the first layer is intuitive concept,  characterizations and applications, the second layer is logical physics understanding of topological properties, characterizations and applications, and the third layer is the nature of topological properties and its power. Duality and trinity  are viewed as intrinsically topological objects and are recognized as common knowledge shared among human society activity, mathematics, physics,chemistry,biology and many other kinds of nature science and technology. Some important methods used so far to characterize the topological properties, including topological index, topological order, topological invariant, topology class and the topology partition are discussed. The theories of molecular topology,topological quantum matter including topological insulators, topological metal and topological superconductors and topological quantum computing are reviewed. The  development of the topological duality connection between the qubit and singularity via topological space time is briefly introduced. We will see the use of iterated function systems (IFS)to simulate the connection between singularities and their qubit control codes. The novel applications of topology in integrated circuits technology are also discussed in this paper.

\section{Introduction}
Among the three important types of structural properties that remain unchanged under the structural transformation, the periodic properties,people have done quite in-depth research,the whole set of concepts, theories and research methods has been widely used. The fractal properties, it was first proposed by B. Mandelbrot ~\cite{Mandelbrot}, and he named the structures with fractal properties Fractal. Fractal dimension was introduced to quantitatively describe the geometric features of the fractal structure.  This theory has been widely used in many disciplines such as physics, chemistry, biology, and geosciences~\cite{Mandelbrot1984,Mandelbrot1990, Zhang2020A,Baumann,Zhang1994A,Zhang1993B,Zhang1992A,Kaandorp, Turcotte,Falconer1985,Jadczyk,Vicsek}. Which opened up a whole new field of research. The topological properties,  although have been studied in the mathematical world for more than a century ~\cite{Stillwell}, the application of topological properties in physics and chemistry is only a matter of recent decades. If the scientific historian Kuhn's paradigm theory on the process of scientific development is used to illustrate~\cite{Kuhn}, it can be considered that the study of periodic nature has established a formal paradigm, while the study of fractal and topological properties is just before the establishment of the paradigm, and the discipline is not yet mature. At this time, different viewpoints can coexist, and knowledge is continuously accumulated, but there is no coherent theoretical system. Our previous work have described some important methods used so far to characterize the fractal properties, including the theoretical method of calculating the fractal dimension, the renormalization group method, and the experimental method of measuring the fractal dimension ~\cite{Zhang2020A,Zhang1994A,Zhang1993B, Zhang1992A,Zhang1992B,Zhang1992C}. Multiscale fractal theory method, thermodynamic representation form and phase change of multiscale fractal, and wavelet transform of multiscale fractal. The  development of the fractal concepts is briefly introduced: negative fractal dimension, complex fractal dimension and fractal space time. New concepts such as balanced and conserved universe,the wormholes connection to the whiteholes and blackholes for universes communication, quantum fractals,  platonic quantum fractals for a qubit, new manipulating fractal space time effects such as transformation function types, probabilities of measurement,manipulating codes,and hiding transformation functions are also discussed. In this work, we will describe topological properties and characterizations in three layers: the first layer is intuitive concept and application, the second layer is physical understanding of topological properties and application, and the third layer is the nature of topological properties and its power. Some important methods used so far to characterize the topological properties, including topological index, topological order, topological invariant, topology class and the topology partition are discussed. The theories of molecular topology,topological quantum matter including topological insulators, topological metal and topological superconductors and topological quantum computing are reviewed. The  development of the topological spaces and the duality connection between the qubit and singularity via topological space time is briefly introduced. In addition, we will see the use of iterated function systems (IFS)to simulate the duality connection between the qubit and singularity. The novel applications of topology in integrated circuits technology are also discussed in this paper.

\section{Intuitive Concept of Topology and Application in Layer 1}

\subsection{Intuitive Concept}
\label{subsec:Intuitive}

\begin{figure}[H]
\begin{center}
\epsfig{file=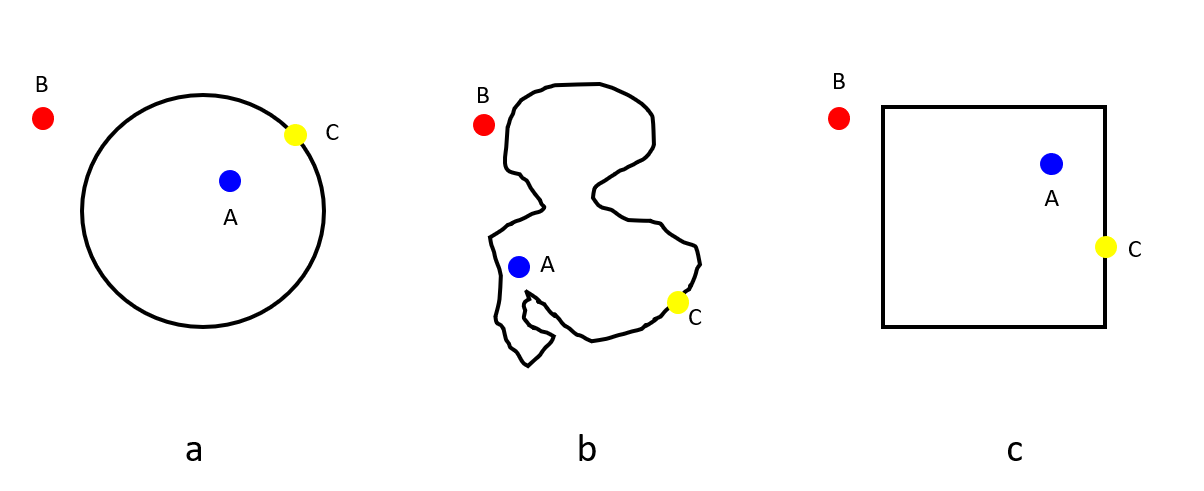,height=2.0in}
\caption{Example of intuitive topology concept}
\label{fig:tc}
\end{center}
\end{figure}

The property that the figure can maintain after continuous deformation is called topological property. At first glance, you may think there no one kind of property is topological, which means that any property of a graph can be changed by a certain deformation. This is not the case. For example, a circle in Figure ~\ref{fig:tc}a divides the points on the plane into three sets: point A inside the circle, point C on the circle and point B outside the circle. If we imagine this circle C and the two points A, B are drawn on a piece of ideal elastic rubber, and an elastic movement is performed on the figure, and the result may become a curve C and two points A and B as shown in Figure ~\ref{fig:tc}b. Point A and Point B are located inside and outside of curve C in Figure~\ref{fig:tc}a, after the elastic movement of this rubber, point A and Point B are still located inside and outside curve C in Figure ~\ref{fig:tc}b. Therefore, "A is inside curve C", "B is outside of curve C", and the property of "a circle on the plane divides the points on the plane into three sets" is the property that the figure still retains after being deformed, which is the topological property of the figure. As for the property that "A is closer to C than B" before deformation on this figure, it is not a topological property, because after elastic motion, B can be very close to C, and A is far from C. Similarly, although the shape and size of the circle and the square are different, after elastic motion, the two have the same topological properties as shown in Figure ~\ref{fig:tc}c

\subsubsection{Homeomorphic Mapping}
\label{subsubsec:Homeomorphic}

\begin{figure}[H]
\begin{center}
\epsfig{file=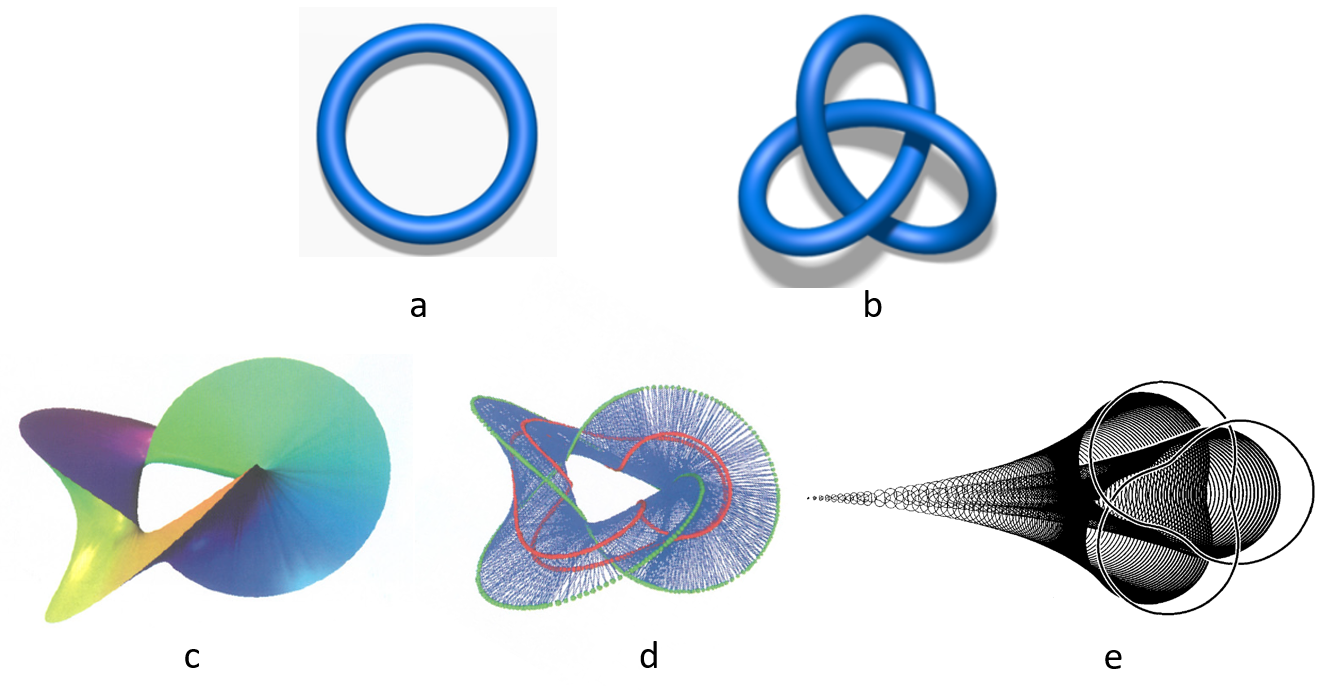,height=2.8in}
\caption{Example of intuitive concept of topologically equivalent}
\label{fig:tcm}
\end{center}
\end{figure}

The intuitive concept of topological properties establishes the equivalent meaning of the two graphic topologies. However, "elastic motion" is mixed with some intuitive by-products that we do not want. The concept of "elastic motion" contains such content: "Motion" from one place to another must follow a certain path or road. In fact, two topologically equivalent figures can not always "place" one figure to another through a kind of elastic motion in space. For example, the unknot circle shown in Figure  ~\ref{fig:tcm}a is  topologically equivalent to the knotted trefoil Figure  ~\ref{fig:tcm}b . If we imagine a rubber band into a unknot round shape, it is impossible to wear it into a knotted trefoil by pulling and stretching, but if we cut the rubber band first, tie a knot, and then connect the two ends as they were, it is easy to get a knotted trefoil. So these two kinds of knots are topologically equivalent after the cut and connection operation. In biology, careful analyses of knots and catenanes generated by enzymes have yielded important mechanistic information concerning these enzymes. Additionally, knots and catenanes have proved to be valuable synthetic tools for probing DNA–protein interactions\cite{Bates}. Now we use the concept of mapping instead of "motion" to give the topological equivalent a precise definition.

The mapping function $f$ from figure A to figure B is called a homeomorphic mapping, if the following two conditions are met: (1) This mapping is single-valued on both sides. (2) This mapping is continuous on both sides, which means that not only the mapping $f$ is continuous, but also the inverse mapping $f^{-1}$ is continuous.

From an intuitive point of view, a homeomorphic image will be neither rupture nor superimposition.  In the homeomorphic image, it does not keep the distance constant, but keeps the continuity of the distribution of the points in the graph, that is, tearing and overlapping are not allowed.

Two figures A and B, one of which is a homeomorphic image to the other, then figure A and B is homeomorphic to each other. From the perspective of topology, the two figures are equivalent.

\begin{figure}[H]
\begin{center}
\epsfig{file=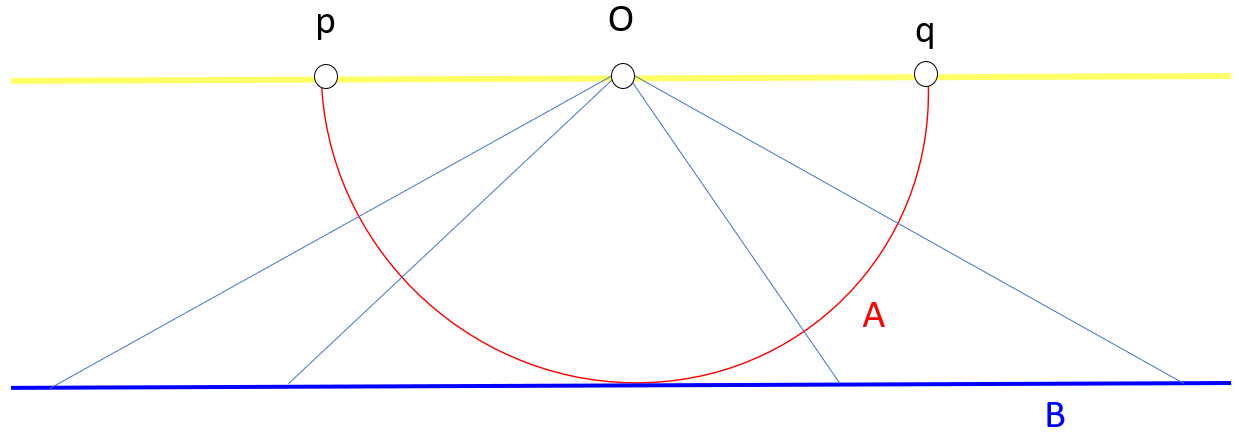,height=1.6in}
\caption{Example of intuitive concept of homeomorphic mapping}
\label{fig:ho}
\end{center}
\end{figure}

For example, in Figure ~\ref{fig:ho},  A is the half circle centered at O, and B is the straight line parallel to the diameter pq. The projection from the center O reflects the points on the semicircle into the straight line except for the end points p and q of the semicircle, we get the figure A, which is mapped homeomorphically into the straight line B. So the straight line is homeomorphic to the semicircle with no endpoints. 

Figure  ~\ref{fig:tcm}c, d and e show different topology class of trefoil obtained via the different kinds of operation on trefoil knot shown in Figure~\ref{fig:tcm}b, they are topologically equivalent to each other after the operation, this will be discussed in section  ~\ref{subsubsec:Class}.

In physics, the Yang–Baxter equation is a topological consistency equation which was first introduced in the field of statistical mechanics~\cite{Yang1967, Baxter, Jimbo}. It depends on the idea that in some scattering situations, particles may preserve their momentum while changing their quantum internal states. It states that a matrix $R$, acting on two out of three objects, satisfies. 

\begin{equation}
( R \otimes 1)( 1 \otimes R)( R \otimes 1)= ( 1 \otimes R)( R \otimes 1)( 1 \otimes R)
\label{equ:YB}
\end{equation}

\begin{figure}[H]
\begin{center}
\epsfig{file=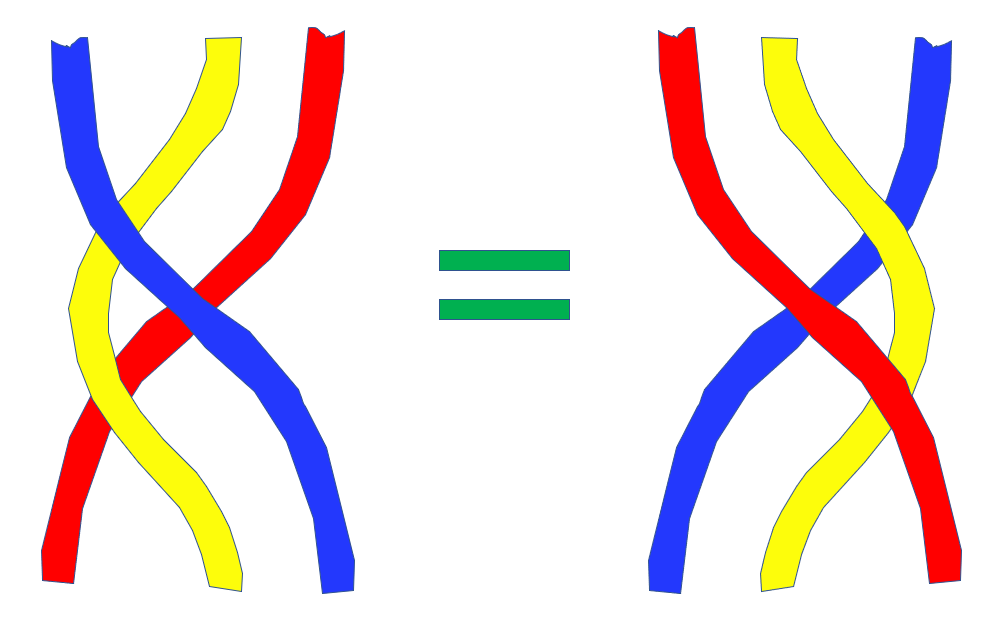,height=2.0in}
\caption{Yang–Baxter equation in knot theory and the braid groups }
\label{fig:yb}
\end{center}
\end{figure}

The Yang–Baxter equation~\ref{equ:YB} also shows up when discussing knot theory and the braid groups where $R$ corresponds to swapping two strands~\cite{Yang1989}. Since one can swap three strands two different ways, the Yang–Baxter equation~\ref{equ:YB} enforces that both paths are the same  as shown in Figure~\ref{fig:yb}.

\subsubsection{Topological Invariant}
\label{subsubsec:Invariant}
\begin{figure}[H]
\begin{center}
\epsfig{file=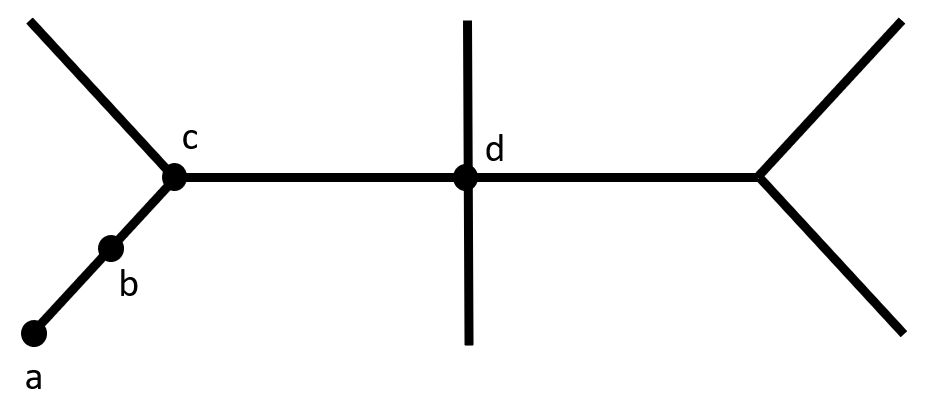,height=1.2in}
\caption{Example of intuitive concept of topological invariant}
\label{fig:tiv}
\end{center}
\end{figure}

In order to assert that two figures are homeomorphic to each other, it is enough to actually get a homeomorphic mapping from one figure to another. But how to prove that two figures are not homeomorphic from each other?  Even we can not find the homeomorphic mapping, we can not prove it not exist. If we establish a certain rule so that there is a certain number or quantity corresponding to each figure, and the numbers or quantities corresponding to two homeomorphic figures are equal, then these numbers or quantities are expressed as unchanged properties during the homeomorphic mapping of the figures, then these numbers or quantities are the topological invariant. For example, if the numbers of two figures A and B are different, then these two figures can not be homeomorphic to each other. Therefore, to prove two figures are not homeomorphic to each other, the topological invariant can be used. Generally speak, topological invariant is described in number or other algebraic expression since this is convenient for practical application. Currently the simplest graph topological invariant include: the number of connected pieces that make up the graph, called the number of connected areas; the number of "divided" points and non-"divided" points on the graph; Euler characteristic number; the degree of the point and the degree of the side of the graph etc. Here let us focus on the concept of the degree of the point. Suppose that graph A is composed of a finite number of line segments (a finite number of arcs), and x is a point on graph A. If there are several arcs concentrated at point x, then the number of arcs concentrated at point x is defined as the degree of point x . In Figure~\ref{fig:tiv}, point a has degree 1, point b has degree 2, point c has degree 3 and point d has degree 4.In the homeomorphic mapping, although the graph area and shape keep change, but the degree of the points on the graph keeps unchanged. So the degree of the point on the graph is a topological invariant. Same for the degree of the arcs, it is also a topological invariant. The concept of degree of point is a basic and important concept in molecular topological index theory~\cite{Zhang1992A, Wiener1947A, Wiener1947B, Klein, Zhang1989A, Zhang1990A,Zhang1990B}. In mathematics and physics, there are many kinds of topological invariants be defined such as Chern-Simons~\cite{Chern}, Nieh-Yan~\cite{Nieh} and Pontryagin~\cite{Pontrjagin} topological invariants and will be discussed later. Yang–Mills theory~\cite{Yang1954} is a gauge theory based on a special unitary group $SU(N)$, or more generally any compact, reductive Lie algebra. Yang–Mills theory seeks to describe the behavior of elementary particles using these non-abelian Lie groups and is at the core of the unification of the electromagnetic force and weak forces (i.e. $ U(1) × SU(2)$) as well as quantum chromodynamics, the theory of the strong force (based on $SU(3)$). Thus it forms the basis of our understanding of the Standard Model of particle physics. The equations of Yang–Mills remain unsolved at energy scales relevant for describing atomic nuclei. Using topological invariants it can be shown that potentials must exist for Yang-Mills equation has arbitrarily many independent solutions~\cite{Atiyah}. A systematic way of finding exact solutions with nontrivial topology to Yang-Mills equation has been developed~\cite{Nian}. The topological invariants also play important roles in molecular topology and quantum computing and will be discussed in sections~\ref{subsubsec:Molecular Topology Invariant} and ~\ref{subsec:Topological Quantum Computation}.

\subsubsection{Topology Class}
\label{subsubsec:Class}
The topology classes can be obtained by different kinds of operations operating on generators. Here we will consider following operations: intersection, twisting, folding and sewing, string, projection,iteration and transformation. 

(1) Intersection

\begin{figure}[H]
\begin{center}
\epsfig{file=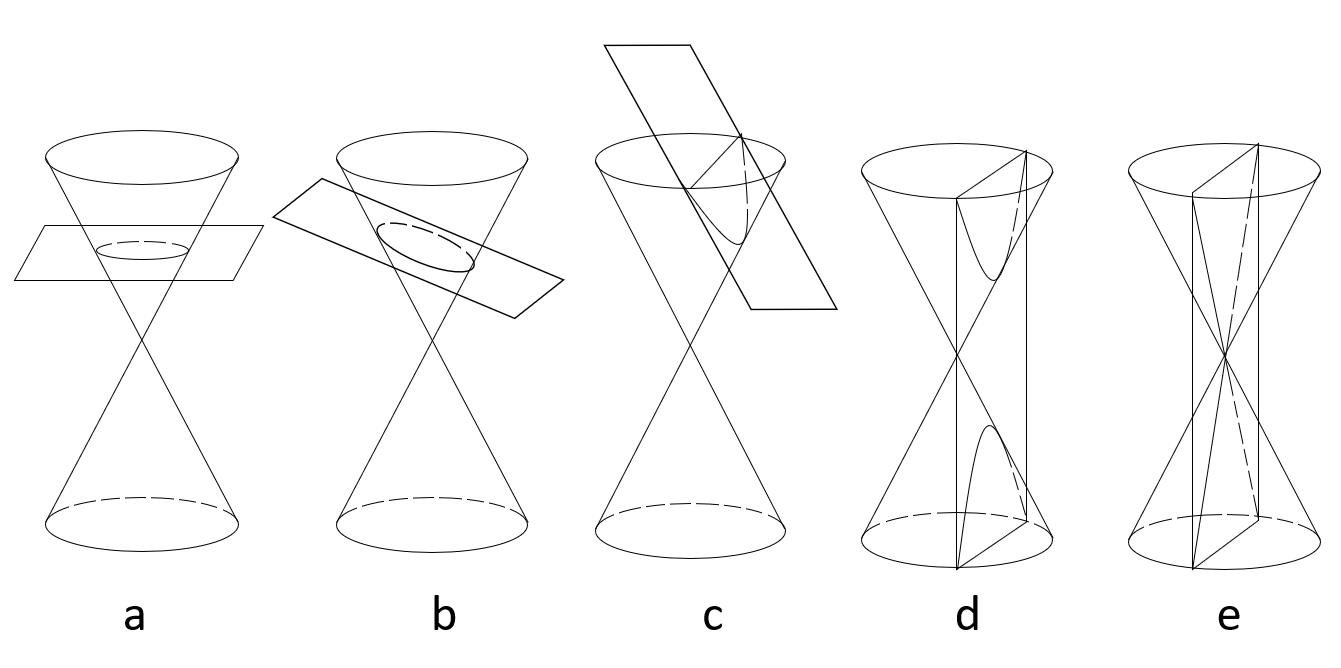,height=1.8in}
\caption{Example of topology class formed with intersection operation on light conic}
\label{fig:conic}
\end{center}
\end{figure}

In mathematics, a conic section is a curve obtained as the intersection of the surface of a light cone with a plane. The three types of topology class of conic section are the ellipse,the parabola and  the hyperbola as shown in Figure ~\ref{fig:conic}b, c and d.  Figure ~\ref{fig:conic}a the circle is a special case of the ellipse. Figure ~\ref{fig:conic}e the intersection of the light cone with the tangent plane form the Lorentzian regions~\cite{Pavlova}. Conics are everywhere. The applications of conics can be seen everyday all around us. Conics are found in architecture~\cite{Burger}, physics~\cite{Christodoulides,Basak}, astronomy~\cite{Neugebauer} and many other areas~\cite{Glaeser}. 

(2) Twisting, Folding and Sewing

\begin{figure}[H]
\begin{center}
\epsfig{file=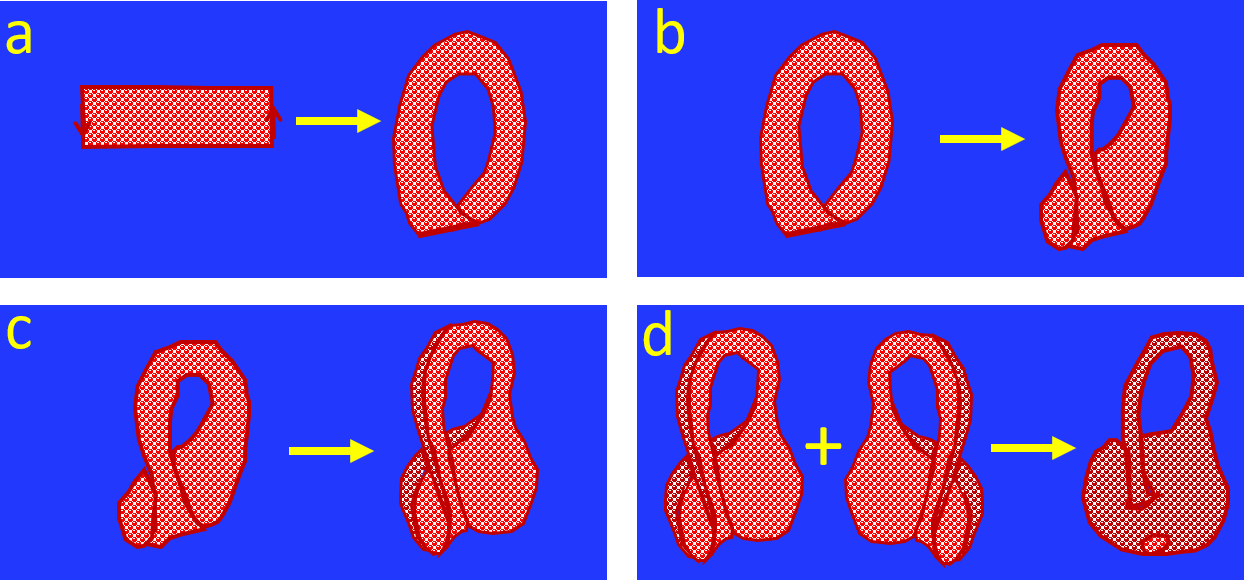,height=2.8in}
\caption{Diagram of topological transform from a 2D band to 3D Möbius band and Klein bottle via twisting and sewing operation}
\label{fig:mk}
\end{center}
\end{figure}

Figure~\ref{fig:mk} shows how a 2D band topologically transform to 3D Möbius band and Klein bottle via twisting and sewing operation. Figure~\ref{fig:mk}a first do a half twist of the two ends of the 2D band and then sewing the two ends to form the Möbius band.  Figure~\ref{fig:mk}b take the Möbius band and make its bottom wider.
Figure~\ref{fig:mk}c make the rear part of the band like a half of a bottle, also the front part of the band like a tube. Figure~\ref{fig:mk}d take another 2D band and repeat Figure~\ref{fig:mk}a, b and c  for the new one. Then sewing the new Möbius band with the old one along their boundaries. Now we get a Klein bottle. Therefore, a Klein bottle is homeomorphic to the two sewed Möbius bands.

\begin{figure}[H]
\begin{center}
\epsfig{file=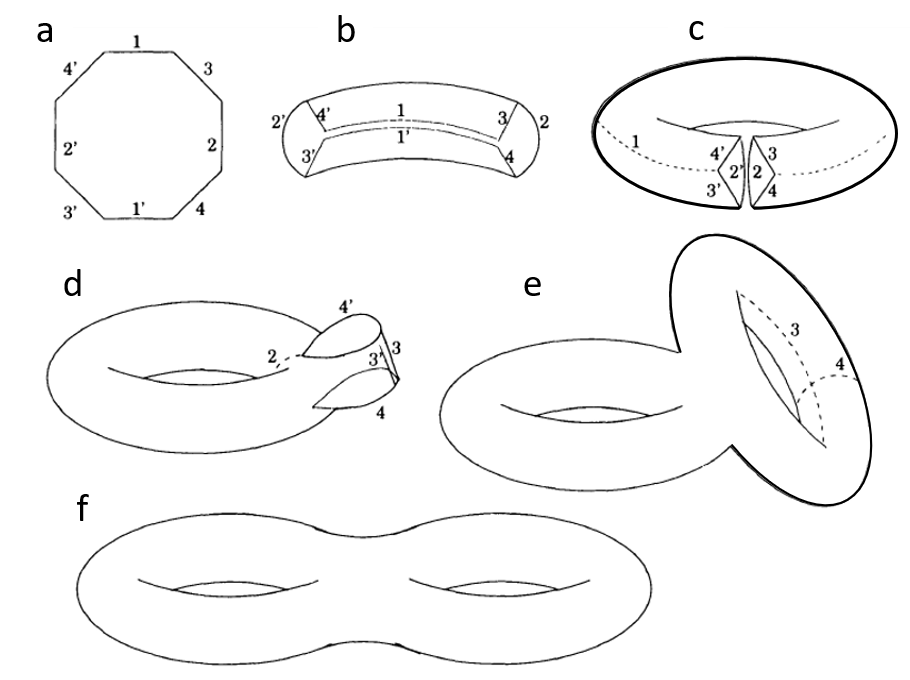,height=2.8in}
\caption{Example of topology class formed with folding and sewing operation}
\label{fig:fold}
\end{center}
\end{figure}

\begin{figure}[H]
\begin{center}
\epsfig{file=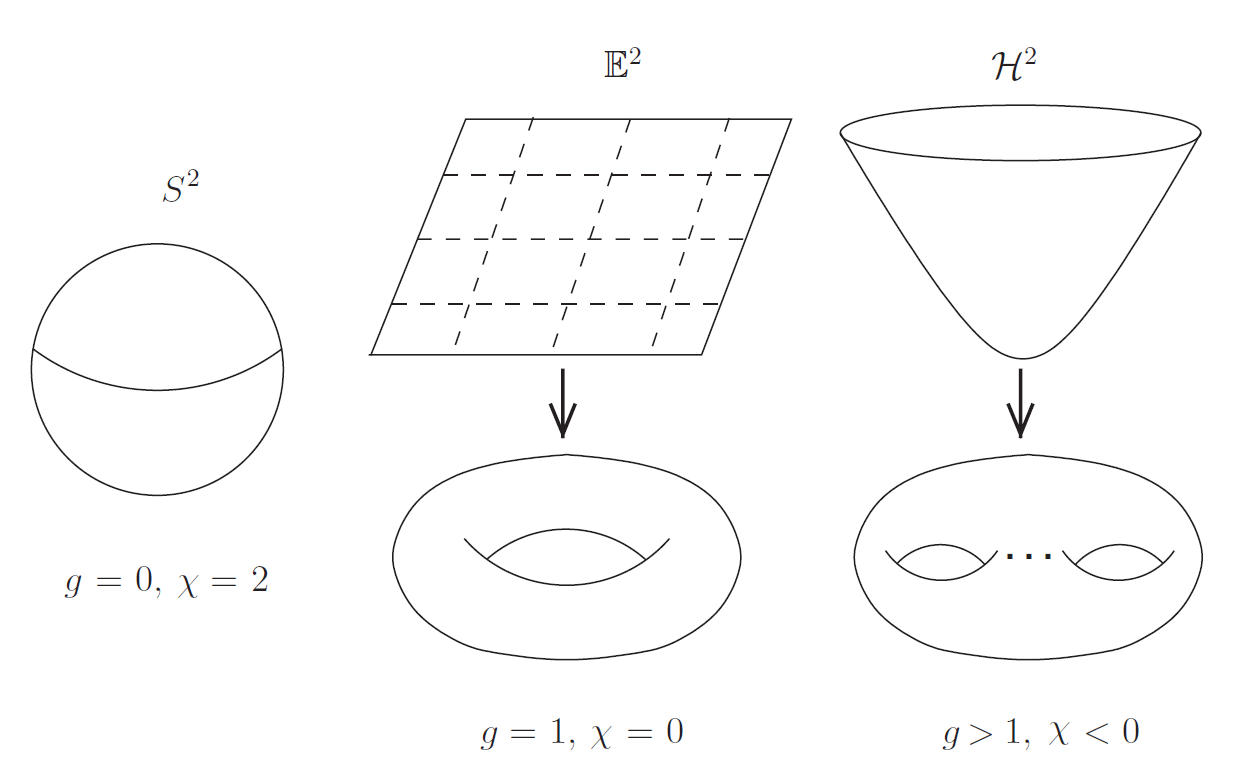,height=2.8in}
\caption{The genus trichotomy $g = 0, g = 1, g \geqslant 2 $ for oriented surfaces.}
\label{fig:genus}
\end{center}
\end{figure}

Figure~\ref{fig:fold} shows the identification of the octagon. Opposite sides of the octagon in $a$ will be identified. The sides are drawn straight for clarity. Dashed lines indicate where sides have been sewn together. First, sides 1 and 1' are identified, folding the top and bottom of the octagon away from view $b$. Sides 2 and 2' are brought together to form a torus with a diamond shaped hole, as in $c$. Next, sides 3 and 3' are stretched out and joined for $d$. The loop is lengthened along the direction of identification 3 and bent until 4 and 4' meet, forming a second torus $e$. Finally the topology is deformed to the preferred shape, seen in $f$. The number of holes in a compact surface is called its genus. Identification of a polygon of genus $g$ will clearly result in $g$ attached tori or, equivalently, a $g$-holed pacifier~\cite{Mann, Mann1997}. The sphere is already compact with $g=0$; the compact version of the plane is the one-holed torus with $g=1$, The most exciting story is that of the hyperbolic plane, which by itself can give rise to a multitude of compact geometric spaces: it can be shown that all compact geometric surfaces with $g \geqslant 2$ holes can be derived from the hyperbolic plane. To return to the basic trichotomy of positive, zero or negative curvature, we can take the Euler number $\chi= 2 − 2g$ of the surface, which is simply the quantity ‘faces − edges + vertexes’ in Euler’s formula for a triangulated surface. Then $\chi = 2$ for a sphere, as everyone knows; also $\chi = 0$ for a torus and $\chi < 0$ for the geometric surfaces with more than one hole as shown in Figure~\ref{fig:genus}~\cite{Reid}.

(3) String

\begin{figure}[H]
\begin{center}
\epsfig{file=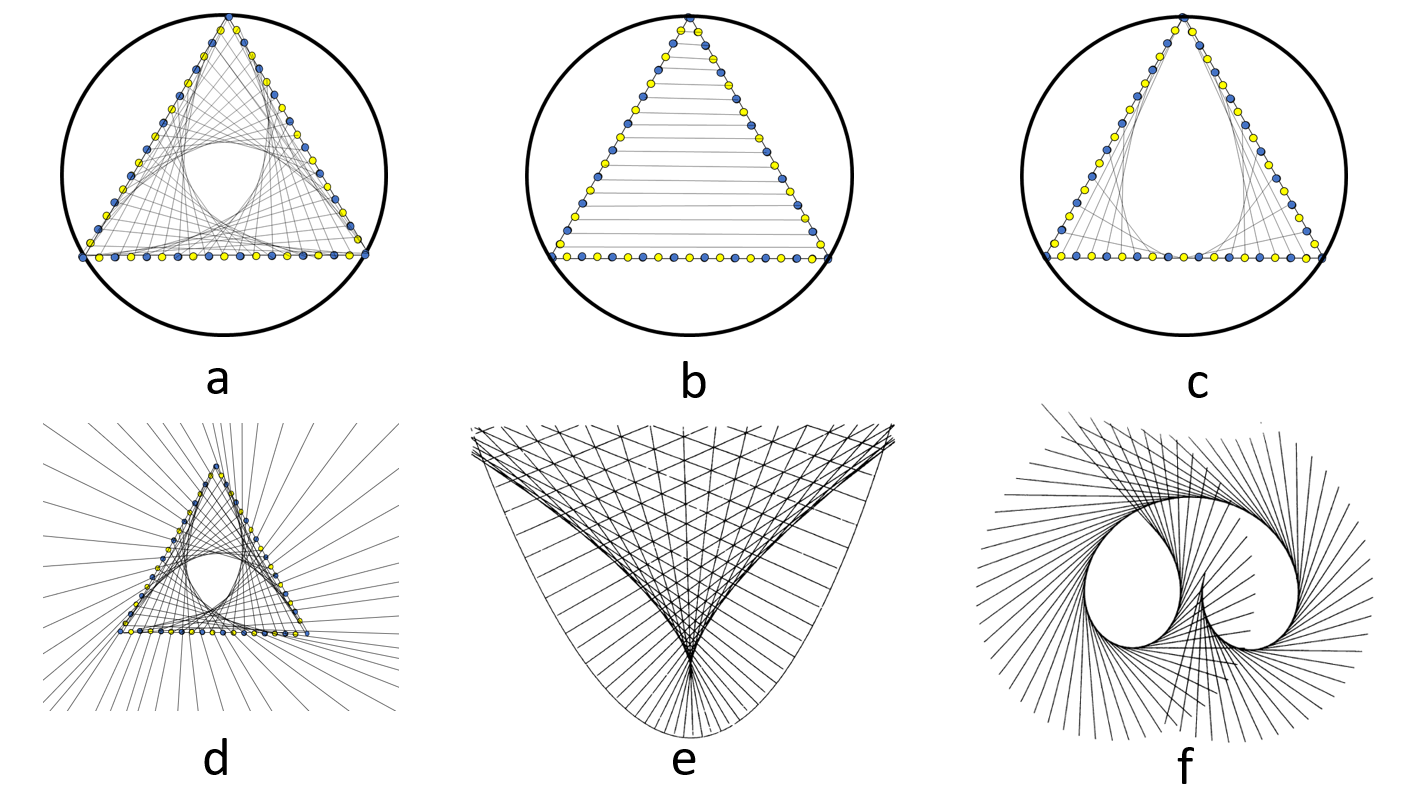,height=2.8in}
\caption{Example of string topology classes obtained via various string control code}
\label{fig:tcs}
\end{center}
\end{figure}

String operations use lines to weave plane or curve surface based on string control code. Figure~\ref{fig:tcs}  shows some string topology classes obtained via various string control code. Figure~\ref{fig:tcs}a form the topology class with three curve surfaces in a triangle. The string control code can be defined as following: 1)Form the Ying (blue) and Yang (yellow) holes in an equilateral triangle laterals. 2)Weave string line from Ying hole to adjacent  lateral Yang hole to form the curve surface between two adjacent laterals in the equilateral triangle. Since each lateral of the triangle is homeomorphic to the 1/3 arc of the circle,  the topology class with three curve surfaces can also be formed in a circle.  Figure~\ref{fig:tcs}b form the topology class with compact plane surfaces in a triangle. The string control code can be defined as following: 1)Form the Ying (blue) and Yang (yellow) holes in an equilateral triangle laterals. 2)Weave string line from Ying (Yang) holes to adjacent  lateral corresponding Ying (Yang) holes to form the plane surface between two adjacent laterals in a triangle. In the same way,  we can form the compact plane surface in a circle. Figure~\ref{fig:tcs}c form the topology class with raindrop curve surfaces in a triangle. The string control code can be defined as following: 1)Form the Ying (blue) and Yang (yellow) holes in an equilateral triangle laterals. 2)Weave string line from Ying holes only of one lateral to adjacent bottom first half lateral corresponding Ying and Yang holes to form the first curve surface and then weave string line from Ying holes only of the other lateral to adjacent bottom second half lateral corresponding Ying and Yang holes to form the second curve surface and together these two curve surface form a raindrop curve surface in an equilateral triangle. In the same way,  we can form a raindrop curve surface in a circle. Figure~\ref{fig:tcs}d form the topology class with three curve surfaces out of the triangle and extend to the whole space. The string control code can be defined as following: 1)Form the Ying (blue) and Yang (yellow) holes in an equilateral triangle laterals. 2)Weave semi straight string line from Ying holes to adjacent  lateral Yang hole and then extend to infinite space to form the topology class with three curve surfaces out of the triangle and extend to the whole space. In the same way,  we can form the topology class with three curve surfaces out of the circle and extend to the whole space. Figure~\ref{fig:tcs}e form the topology class of normals to a parabola and their envelope. Figure~\ref{fig:tcs}f form the topology class of tangent developable of a helix. The string control code of Figure~\ref{fig:tcs}e and Figure~\ref{fig:tcs}f are the definition of normal and tangent lines and can be found in reference ~\cite{Bruce}.

\begin{figure}[H]
\begin{center}
\epsfig{file=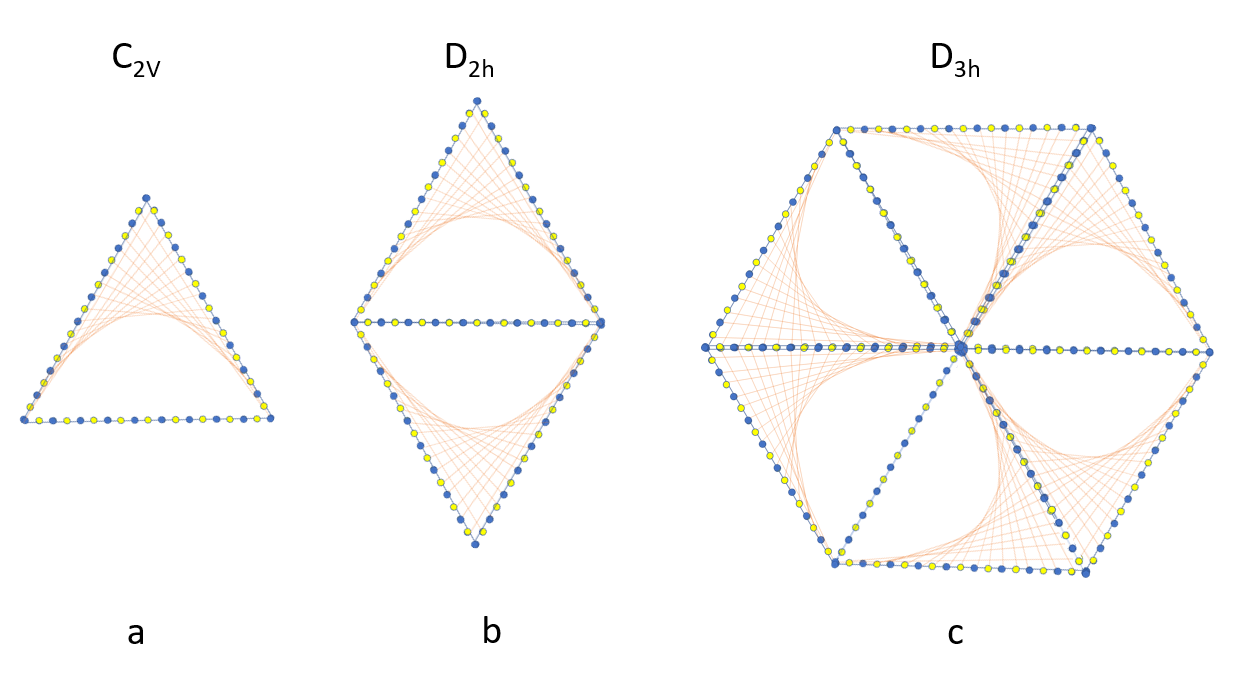,height=2.8in}
\caption{Example of one basic  triangle string topology class form the 2D quadrilateral with $D_{2h} $ and the 2D hexagon with $D_{3h} $ symmetry using Yin-Yang string control code}
\label{fig:tri-c2}
\end{center}
\end{figure}

Figure~\ref{fig:tri-c2} describe how one basic  triangle string topology class form the 2D space with point group symmetries. Figure~\ref{fig:tri-c2}a has $ C_{2V}$ symmetry and can be obtained via the string control code we defined  in Figure~\ref{fig:tcs}a. Figure~\ref{fig:tri-c2}b has $ D_{2h}$ symmetry and can be obtained by adding two Figure~\ref{fig:tri-c2}a  $ C_{2V}$ symmetry triangle together. Figure~\ref{fig:tri-c2}c has $ D_{3h}$ symmetry and can be obtained by adding three Figure~\ref{fig:tri-c2}b  $ D_{2h}$ symmetry quadrilateral together. Figure~\ref{fig:tri-c2}c can be an basic hexagon unit to form graphene like 2D space with  $ D_{3h}$ symmetry. This probably can be used to explain the unconventional superconductivity in magic-angle graphene superlattices discussed in references ~\cite{Cao2018A,Cao2018B}. The slight tweak angle can be calculated by the broken of the symmetry of the two overlapped graphene sheets with $ D_{3h}$ symmetries.

\begin{figure}[H]
\begin{center}
\epsfig{file=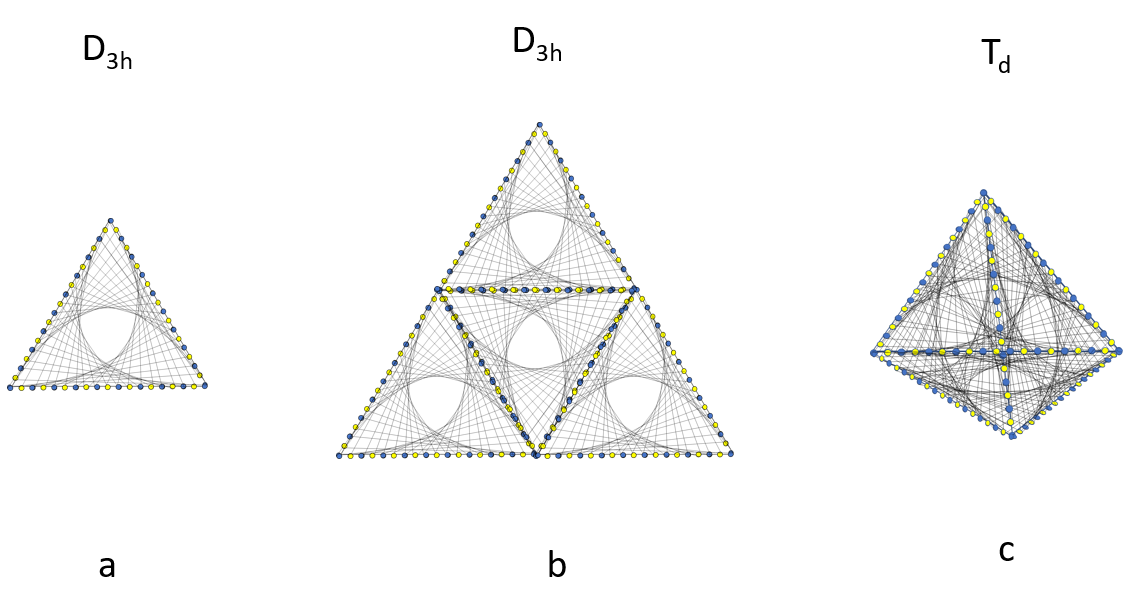,height=2.8in}
\caption{Example of one basic  triangle string topology class form the 3D tetrahedron  with $T_d $ symmetry using Yin-Yang string control code}
\label{fig:tri-c3}
\end{center}
\end{figure}

Figure~\ref{fig:tri-c3} describe how one basic  triangle string topology class form the 3D space with point group symmetries. Figure~\ref{fig:tri-c3}a has $ D_{3h}$ symmetry and can be obtained via the string control code we defined  in Figure~\ref{fig:tcs}a. Figure~\ref{fig:tri-c3}b has $ D_{3h}$ symmetry and can be obtained by adding four Figure~\ref{fig:tri-c3}a  $ D_{3h}$ symmetry triangle together. Figure~\ref{fig:tri-c3}c has $ T_{d}$ symmetry and can be obtained by folding and sewing three Figure~\ref{fig:tri-c3}b  $ D_{3h}$ symmetry equilateral triangle together. Figure~\ref{fig:tri-c3}c can be an basic tetrahedron unit to form 3D space with  $ T_{d}$ symmetry. This probably can be used to explain the novel silver four atom clusters structure and their specific rules in the surface enhanced Raman spectrum that we found in our previous experiments ~\cite{Zhang1994A,Zhang1992A,Zhang1991A,Zhang1995A,Zhang1995B}

\begin{figure}[H]
\begin{center}
\epsfig{file=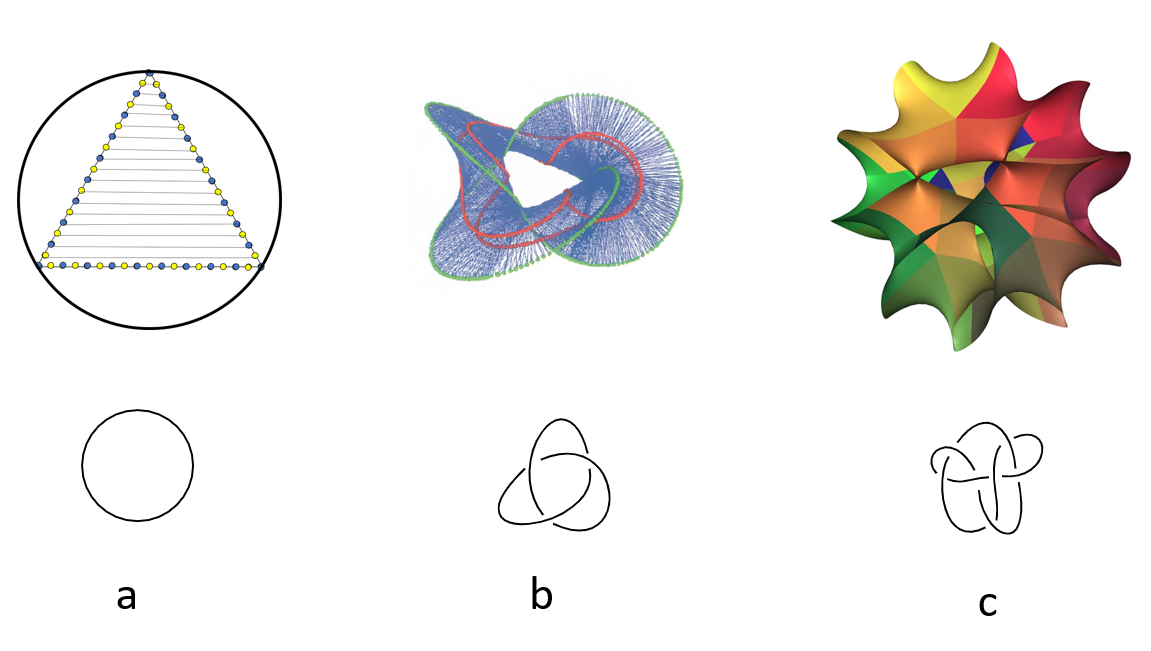,height=2.8in}
\caption{Example of relationship of duality between various string topology classes and  manifolds}
\label{fig:tri-r1}
\end{center}
\end{figure}

Figure~\ref{fig:tri-r1} shows how a 2D compact space can be changed to high dimension manifold via topology class operations.  Figure~\ref{fig:tri-r1}a bottom is simply a circle,  through the string operation we described in Figure~\ref{fig:tcs}a, it is easily become a 2D compact space as shown in the top. Now we follow the topological class operation steps described in section ~\ref{subsubsec:Homeomorphic} cut the circle first, tie a knot, and then connect the two ends as they were, the circle is easy to become  a knotted trefoil as shown in Figure~\ref{fig:tri-r1}b bottom. Accordingly the Figure~\ref{fig:tri-r1}a top 2D  compact manifold become the Figure~\ref{fig:tri-r1}b top 3D trefoil manifold. In the same ways, we can get the high dimension Calabi–Yau manifold as shown in Figure~\ref{fig:tri-r1}c ~\cite{Calabi,Yau1978, Yau2010}. 

\begin{figure}[H]
\begin{center}
\epsfig{file=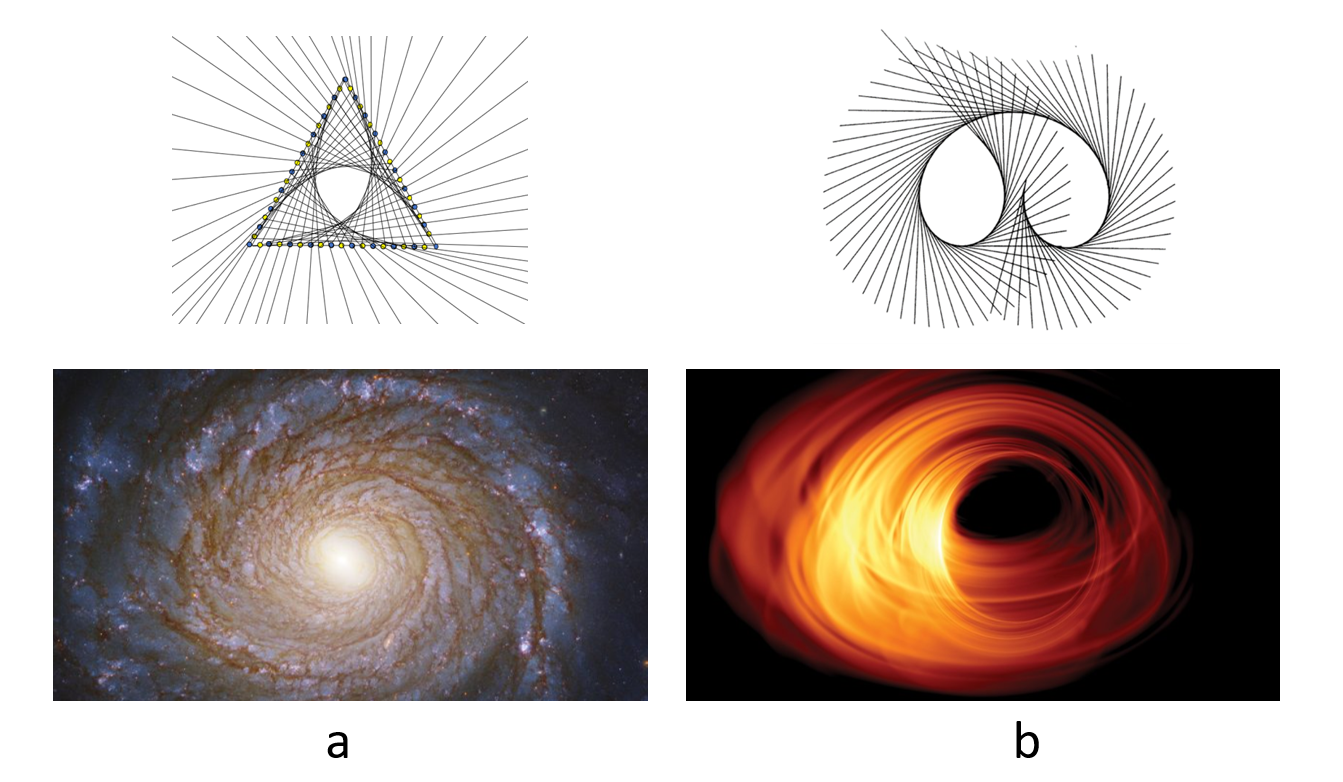,height=2.8in}
\caption{Example of relationship of duality between various string topology classes and  black holes singularities}
\label{fig:tri-r2}
\end{center}
\end{figure}

In a similar way , the 2D string topology classes can also be connected to  black holes singularities as shown in Figure~\ref{fig:tri-r2}. Although the mathematics of the black holes singularities~\cite{Chandrasekhar,Poisson} can be more complicated than this simplified thinking. The duality between various  topology classes control codes and  black holes singularities will also be discussed in section~\ref{subsubsec:Duality} and section~\ref{subsubsec:Topological trinity principle} 

Topological class obtained via string option here for the first time to be listed among other topological classes. We mainly focus on the options on triangle because it is the smallest unit that can form the 2D plane and 3D space as described in Figure~\ref{fig:tri-c2} and Figure~\ref{fig:tri-c3}. It can be considered as one of the clues for the trinity principles that will be discussed in section~\ref{subsubsec:Topological trinity principle}.
\\
\\

(4) Projection

\begin{figure}[H]
\begin{center}
\epsfig{file=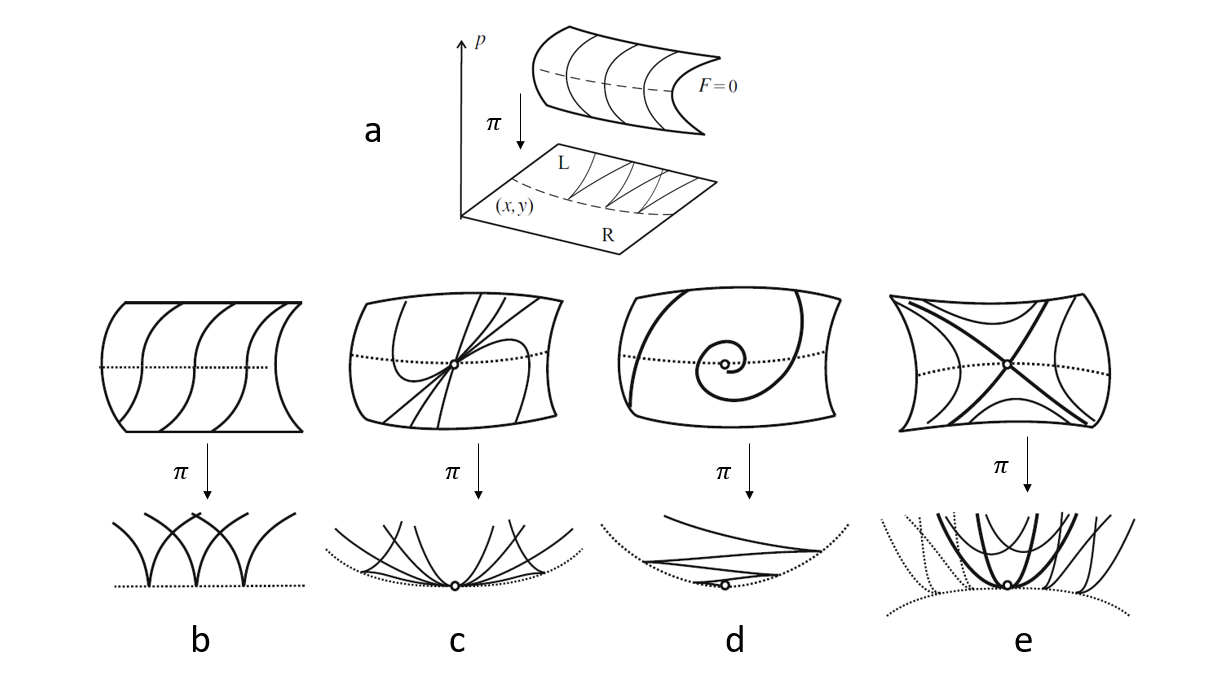,height=2.8in}
\caption{Example of topology class formed with projection}
\label{fig:pj}
\end{center}
\end{figure}

A projection operation is the process or technique of reproducing a spatial object upon a plane or curved surface. Figure~\ref{fig:pj} shows an example of topology class obtained via the projection operation. Opposite to the string operation where we can get high 3D image via the operation of low 2D circle as shown in Figure~\ref{fig:tri-r1}, here we can get low 2D singularities image via the projection operation of high 3D curved surface. Figure~\ref{fig:pj}a is the isotropic surface $F$ in projectivized tangent bundle, integral curves of the field $X$ (top) and integral curves of the equation $F(x, y, p) = 0$ (down). The dashed lines represent the criminant (top) and the discriminant curve (down). Figure~\ref{fig:pj}b-e from the top to the bottom: integral curves of the field $\dot{x}= F_p, \dot{y}= pF_p, \dot{p} = −(F_x + pF_y) $  on the isotropic surface $F$ and isotropic lines, obtained by the projection $ \pi: F  \to S$. The dashed lines represent the criminant (top) and the discriminant curve (down)~\cite{Pavlova}

(5) Iteration and Transformation

\begin{figure}[H]
\begin{center}
\epsfig{file=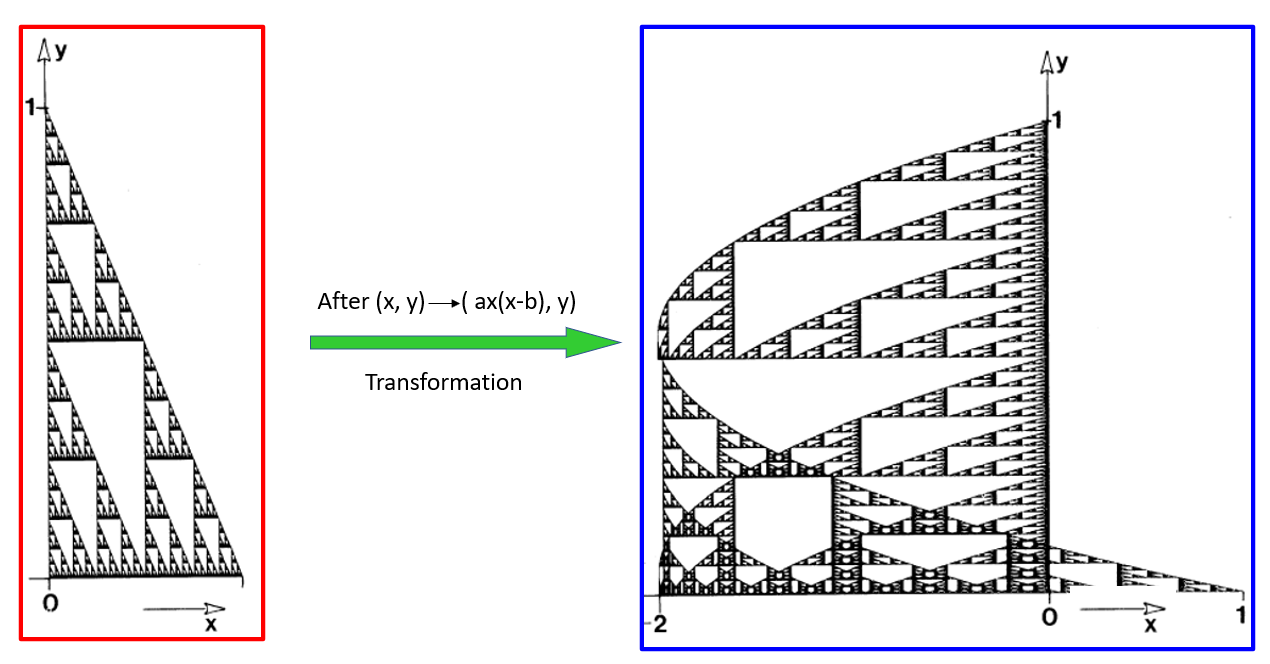,height=2.8in}
\caption{Example of topology class formed with iteration and transformation}
\label{fig:itf}
\end{center}
\end{figure}

Iteration operation is a procedure in which repetition of a sequence of operations yields results successively closer to a desired result. Transformation operation is the operation of changing  one configuration or expression into another in accordance with a mathematical rule. Figure~\ref{fig:itf} shows an example of topology class obtained via the iteration and transformation operation. Figure~\ref{fig:itf} left is a Sierpinski triangle fractal obtained through the iteration operation. Figure~\ref{fig:itf} right is this Sierpinski triangle fractal before and after the polynomial transformation $x \mapsto ax(x-b)$ is applied to the $x$-axis. Here $ a $ and $b$ are the real constants~\cite{Barnsley}. Figure~\ref{fig:tcm}e is obtained via the transformation $ x^2 = v^3$ from the trefoil knot ~\cite{Seade}.

\subsubsection{Topological Space and Time}
\label{subsubsec:Space}

In topology and related branches of mathematics, a topological space may be defined as a set of points, along with a set of neighbourhoods for each point, satisfying a set of axioms relating points and neighbourhoods. The definition of a topological space relies only upon set theory and is the most general notion of a mathematical space that allows for the definition of concepts such as continuity, connectedness, and convergence~\cite{Schubert,Berge}. Other spaces, such as fractal~\cite{Zhang2020A}, manifolds and metric spaces, are specializations of topological spaces with extra structures or constraints. Being so general, topological spaces are a central unifying notion and appear in virtually every branch of modern mathematics. 
The utility of the notion of a topology is shown by the fact that there are several equivalent definitions of this structure. Thus one chooses the axiomatisation suited for the application. The most commonly used is that in terms of open sets, but perhaps more intuitive is that in terms of neighbourhoods. \\ 

(1)Definition via open sets\\
A topological space is an ordered pair $(X, \tau)$, where  $X$ is a set and  $\tau$ is a collection of subsets of $X$, satisfying the following axioms~\cite{Armstrong1983}:\\
a. The empty set and $X$ itself belong to $\tau$.\\
b. Any arbitrary (finite or infinite) union of members of $\tau$ still belongs to $\tau$.\\
c. The intersection of any finite number of members of $\tau$ still belongs to $\tau$.\\
The elements of $\tau$ are called open sets and the collection $\tau$ is called a topology on $X$.\\

(2)Definition via neighbourhoods\\
Let $X$ be a set; the elements of $X$ are usually called points, though they can be any mathematical object. We allow $X$ to be empty. Let $\textbf{N}$ be a function assigning to each  $x$ (point) in $X$ a non-empty collection $\textbf{N}(x)$  of subsets of $X$. The elements of $\textbf{N}(x)$ will be called neighbourhoods of $x$ with respect to $\textbf{N}$ (or, simply, neighbourhoods of $x$). The function $\textbf{N}$ is called a neighbourhood topology if the axioms below ~\cite{Brown}are satisfied; and then  $X$ with  $\textbf{N}$ is called a topological space.\\
a. If $N$ is a neighbourhood of $x$ (i.e., $N \in \textbf{N}(x)$), then $x \in N$. In other words, each point belongs to every one of its neighbourhoods.\\
b. If $N$ is a subset of $X$ and includes a neighbourhood of $x$, then $N$  is a neighbourhood of $x$. I.e., every superset of a neighbourhood of a point $x$ in $X$is again a neighbourhood of $x$.\\
c. The intersection of two neighbourhoods of $x$ is a neighbourhood of $x$.\\
d. Any neighbourhood $N$ of $x$ includes a neighbourhood $M$ of $x$ such that $N$ is a neighbourhood of each point of $M$.\\
The first three axioms for neighbourhoods have a clear meaning. The fourth axiom has a very important use in the structure of the theory, that of linking together the neighbourhoods of different points of $X$.\\

In physics, spacetime topology is the topological structure of spacetime.  The Einstein equations~\cite{Einstein},

\begin{equation}
G_{\mu \nu}+\Lambda g_{\mu \nu} =\frac{8 \pi G}{c^4}T_{\mu \nu}
\label{equ:Einstein}
\end{equation}

where $G_{\mu \nu}$ is the Einstein tensor, $g_{\mu \nu}$ is the metric tensor, $T_{\mu \nu}$ is the stress–energy tensor,$\Lambda$ is the cosmological constant.

The Einstein tensor is defined as

\begin{equation}
G_{\mu \nu}=R_{\mu \nu}-\frac{1}{2}Rg_{\mu \nu}
\label{equ:Einstein1}
\end{equation}

where $R_{\mu \nu}$ is the Ricci curvature tensor, and $R$ is the scalar curvature. This is a symmetric second-degree tensor that depends on only the metric tensor and its first- and second derivatives.

The Einstein gravitational constant is defined as~\cite{Adler}
\begin{equation}
\kappa =\frac{8 \pi G}{c^4} \approx 2.077 \times 10^{-43}N^{-1}
\label{equ:Einstein2}
\end{equation}
where G is the Newtonian constant of gravitation and c is the speed of light in vacuum.

 The Einstein equations~\ref{equ:Einstein} and ~\ref{equ:Einstein1} are local field equations and do not contain any topological information. One can add topological degrees of freedom to $4D$ gravity by the addition of terms corresponding to various topological invariants~\cite{Chern,Nieh,Pontrjagin}.  For instance, the Chern-Simons contribution to the action looks like~\cite{Jackiw} :

\begin{equation}
S_{cs}= \int d^4x\frac{1}{2}(\epsilon_{ab}^{ij}R_{cdij})R_{abcd}
\label{equ:Chern}
\end{equation}

\subsubsection{Singularities}
\label{subsubsec:Singular}

Mathematical singularity, a point at which a given mathematical object is not defined or not "well-behaved", for example infinite or not differentiable. A complex algebraic variety $ \mathbb{X}$ is a subset of complex affine $n$-space $\mathbb{C}^n$ or of complex projective $n$-space $\mathbb{CP}^n$ defined by polynomial equations. A point $x \in \mathbb{X}$ is called a smooth point if, up to a complex analytic local coordinate change,  $ \mathbb{X}$  looks like a linear subspace near $x$. Otherwise, a point is called singular. At singular points  $ \mathbb{X}$  may have self-intersections, it may look like the vertex of a cone, or it can be much more complicated. Roughly speaking, resolution of singularities asserts that an arbitrary singular variety  $ \mathbb{X}$  can be parametrized by a smooth variety $ \mathbb{X}'$ . That is, all the points of $ \mathbb{X}'$ are smooth, and there is a subjective and proper map $ f :  \mathbb{X}'  \to   \mathbb{X} $ defined by polynomials~\cite{Kollar}. As the singularity is encoded in the construction of the mini-versaln deformation space (and/or its base-space), this space contains important information about the given germ and is a crucial source of numerical invariants. Its most important two ingredients are the tangent space and obstruction space, which became the subject of intense mathematical study.

Physical singularity, a point or region of infinite mass density at which space and time are infinitely distorted by gravitational forces and which is held to be the final state of matter falling into a black hole. The singularity in solutions of the Einstein field equations is one of two things: 1) a situation where matter is forced to be compressed to a point (a space-like singularity). 2) a situation where certain light rays come from a region with infinite curvature (a time-like singularity).  Space-like singularities are a feature of non-rotating uncharged black holes as described by the Schwarzschild metric~\cite{Schwarzschild}, while time-like singularities are those that occur in charged or rotating black hole exact solutions~\cite{Reissner, Nordstrom}. Both of them have the property of geodesic incompleteness, in which either some light-path or some particle-path cannot be extended beyond a certain proper time or affine parameter. The Penrose theorem guarantees that some sort of geodesic incompleteness occurs inside any black hole whenever matter satisfies reasonable energy conditions. The energy condition required for the black-hole singularity theorem is weak: it says that light rays are always focused together by gravity, never drawn apart, and this holds whenever the energy of matter is non-negative ~\cite{Penrose}. Hawking's singularity theorem is for the whole universe, and works backwards in time: it guarantees that the Big Bang has infinite density. This theorem is more restricted and only holds when matter obeys a stronger energy condition, called the dominant energy condition, in which the energy is larger than the pressure~\cite{Hawking}. It is still an open question whether general relativity predicts time-like singularities in the interior of realistic charged or rotating black holes, or whether these are artifacts of high-symmetry solutions and turn into space like singularities when perturbations are added. The emergence of saddle-point Van Hove singularities (VHSs)~\cite{Hove} in the density of states,accompanied by a change in Fermi surface topology constitutes an ideal ground for the emergence of different electronic phenomena, such as superconductivity, pseudo-gap, magnetism, and density waves. However, in most materials the Fermi level is too far from the VHS where the change of electronic topology takes place, making it difficult to reach with standard chemical doping or gating techniques. 

Technological singularity is a hypothetical moment in time when any physically conceivable level of technological advancement is attained instantaneously. Artificial intelligence and neurotechnology are the main areaes for the technological singularity studies~\cite{Armstrong,Shanahan}. People argue that if a technological singularity did indeed occur, there would be seismic. Most likely, the singularity will happen before we realise it and before we can do anything about it. As we take each step forward we will feel that all is OK until we take one step too far and there will be no turning back—the singularity will be upon us. From our studies, we believe that like in the number theory, almost all real numbers are irrational numbers,  almost all real worlds are singularities; there exist topological duality connections between singularities and their control codes as we discussed in sections ~\ref{subsubsec:Class}, ~\ref{subsubsec:Duality} and ~\ref{subsubsec:Topological trinity principle}; the technological singularity  can be realized through the devices that have inner singularities as been studied before~\cite{Zhang2015I, Zhang2018F}.

\subsubsection{Duality}
\label{subsubsec:Duality}
Duality is the quality or state of having two different or opposite parts or elements. In physics, the duality represents a major advance in the understanding of string theory and quantum gravity~\cite{deHaro,Savit}.   This is because it provides a non-perturbative formulation of string theory with certain boundary conditions and because it is the most successful realization of the holographic principle, an idea in quantum gravity originally proposed by Gerard 't Hooft  by studying analogies between string theory and nuclear physics ~\cite{tHooft} and promoted by Leonard Susskind who made early contributions to the idea of holography in quantum gravity~\cite{Susskind}.

The anti-de Sitter/conformal field theory  (AdS/CFT) correspondence was first proposed by Juan Maldacena~\cite{Maldacena}.  It is closely related to another duality conjectured by Igor Klebanov and Alexander  Polyakov in 2002~\cite{Klebanov}.  This duality states that certain "higher spin gauge theories" on anti-de Sitter space are equivalent to conformal field theories with O(N) symmetry. Here the theory in the bulk is a type of gauge theory describing particles of arbitrarily high spin. It is similar to string theory, where the excited modes of vibrating strings correspond to particles with higher spin, and it may help to better understand the string theoretic versions of AdS/CFT and possibly even prove the correspondence. In 2010, Simone Giombi and Xi Yin obtained further evidence for this duality by computing quantities called three-point functions~\cite{Yin}.

In society science, the duality of two opposite elements Yin and Yang are the key for changes. The Book of Changes (or I Ching, as it is often known)shows how the  two opposite elements Yin and Yang compensate and conflict between each other, mutual root between each other to make the myriad things change.  It is one of the treasures of world literature, and a central text in the history of Chinese civilization. Chinese medicine, geomancy, the arts of the war, and countless other arts are based on its teachings~\cite{Balkin}.

\subsubsection{Topology Partition}
Partition is an important method in science and technology. It breaks down the complicated problem to fundamental parts so that each parts can have clear vision and easy solutions.

In physics,  a partition function describes the statistical properties of a system in thermodynamic equilibrium. Partition functions are functions of the thermodynamic state variables, such as the temperature and volume. Most of the aggregate thermodynamic variables of the system, such as the total energy, free energy, entropy, and pressure, can be expressed in terms of the partition function or its derivatives.In our previous studies, the method of multi-scale fractal theory dividing large scale into many small scales
is similar to the method of dividing the system into many macroscopically small and microscopically large subsystems in thermodynamics, the partition function is introduced to reveals the geometric
characteristics of the fractal body~\cite{Zhang2020A}. The partition function is also an important tool for topological quantum field theory, as well as for traditional quantum field theory. In the latter it plays a key part in the determination of the perturbative structure, while in the former it yields important topological invariants~\cite{Gegenberg,Witten,Witten1989}. 

In mathematics, the partition topology is a topology that can be induced on any set X by partitioning X into disjoint subsets P; these subsets form the basis for the topology~\cite{Todorcevic}.

\subsubsection{Topological dynamics}

In Newton's theory the motion of a dynamical system was described by a system of differential equations. H.Poincare created a new branch of mathematics on the qualitative theory of differential equations. As a
result of Poincare's qualitative approach, the focus of the theory of dynamics shifted away from the differential equations. In George David Birkhoff's treatise on dynamical systems, he discussed many dynamical phenomena in the context of transformation groups acting on general metric spaces~\cite{Birkhoff}. Since then,topological dynamics become a branch of the theory of dynamical systems in which qualitative, asymptotic properties of dynamical systems are studied from the viewpoint of general topology. The study of the topological dynamics for continuous actions of non-compact groups on compact spaces becomes a venerable topic~\cite{Akin,Akin2008,Ellis,Aoki, Asaoka}. The central object of study in topological dynamics is a topological dynamical system, i.e. a topological space, together with a continuous transformation, a continuous flow, or more generally, a semigroup of continuous transformations of that space. The origins of topological dynamics lie in the study of asymptotic properties of trajectories of systems of autonomous ordinary differential equations, in particular, the behavior of limit sets and various manifestations of "repetitiveness" of the motion, such as periodic trajectories, recurrence and minimality, stability, non-wandering points.  Unlike the theory of smooth dynamical systems, where the main object of study is a smooth manifold with a diffeomorphism or a smooth flow, phase spaces considered in topological dynamics are general metric spaces. This necessitates development of entirely different techniques but allows an extra degree of flexibility even in the smooth setting, because invariant subsets of a manifold are frequently very complicated topologically; additionally, shift spaces arising via symbolic representations can be considered on an equal footing with more geometric actions. Topological dynamics has intimate connections with ergodic theory of dynamical systems, and many fundamental concepts of the latter have topological analogues such as  topological entropy. The topological entropy of a topological dynamical system is a nonnegative extended real number that is a measure of the complexity of the system. Topological entropy was first introduced in 1965 by Adler, Konheim and McAndrew. Their definition was modelled after the definition of the Kolmogorov–Sinai, or metric entropy. Later, Dinaburg and Rufus Bowen gave a different, weaker definition reminiscent of the Hausdorff dimension. The second definition clarified the meaning of the topological entropy: for a system given by an iterated function, the topological entropy represents the exponential growth rate of the number of distinguishable orbits of the iterates. An important variational principle relates the notions of topological and measure-theoretic entropy.

In physics, the topological entanglement entropy or topological entropy ~\cite{Hamma2005, Kitaev2006,Levin,Islam}, usually denoted by $\gamma$ , is a number characterizing many-body states that possess topological order. A non-zero topological entanglement entropy reflects the presence of long range quantum entanglements in a many-body quantum state. So the topological entanglement entropy links topological order with pattern of long range quantum entanglements. Given a topologically ordered state, the topological entropy can be extracted from the asymptotic behavior of the Von Neumann entropy measuring the quantum entanglement between a spatial block and the rest of the system. The entanglement entropy of a simply connected region of boundary length $L$, within an infinite two-dimensional topologically ordered state, has the following form for large $L$

\begin{equation}
S_L  \longrightarrow   \alpha  L -\gamma + O(L^{-\nu}),       \nu > 0
\label{equ:Tee}
\end{equation}

where $-\gamma$  is the topological entanglement entropy.
\\

\subsection{Application of Intuitive Concept of Topology}

If we use vertices to represent atoms, edges to represent bonds between atoms, and a graph to represent molecular structure, then this graph is called a molecular graph. Generally speaking, we call the quantity that does not depend on the labeling method of vertices in the molecular graph or its matrix representation as the invariant of the molecular graph~\cite{Xin}. For example,the invariant of the molecular graph can be the number of the vertices, the number of edges, the number of the vertices of adjacency matrix and so on and in addition, the characteristic polynomials of molecular graphs and their maps are also invariant of molecular graphs. 

Although molecular graphs can be used to express the topological properties of molecules, in essence, molecular graphs are non-numerical mathematical objects. The various measurable properties of molecules are usually expressed in numerical values. Therefore, in order to link the topological properties of molecules with the various measurable properties of molecules, the information obtained in the molecular graphs must first be converted into a quantity that can be expressed numerically.The various invariants of the molecular graph are the quantities that can play this role.In other words, the invariants of the molecular graphs can not only quantitatively express the structure of the molecule, moreover, it can be used to correlate the relationship between the structure and performance of molecules. Usually, the invariant of the molecular graph with this effect is called the molecular topological index. 

The first molecular topological index recognized by the chemical community is the Wiener topological index~\cite{Wiener1947A}.It was proposed by the chemist H. Weiner in 1947. Wiener's topological index is established on the basis of the invariant of the topological distance on the molecular graph. After the Wiener topological index was proposed, it has been widely used and researched~\cite{Wiener1947B,Rouvray1986A, Rouvray1986B, Rouvray1987A,Rouvray1987B,Rouvray1988}. Among the more than 40 topological indexes currently available, the Weiner topological index is still the most important one.

Following Wiener’s pioneering work, another important topological index in the history of topological index development was proposed by Japanese chemist Hosoya in 1971~\cite{Hosoya1971}. Usually called Z topological index. The most important property of this topological index is that it is closely related to the characteristic polynomial of the molecular graph. Z topological index is not only used in chemistry, but also in physics and mathematics~\cite{Hosoya1972,Hosoya1976,Hosoya1980,Hosoya1985,Gutman1978A,Gutman1987}.

In 1975, Randic first proposed the molecular connectivity index $^1 \chi$~\cite{Randic1975},following the development by Kier and Hall etal.~\cite{Kier1976A,Kier1976B,Kier1978,Kier1983}, a complete series of molecular connectivity indexes was formed. The molecular connectivity index is established on the basis of the concept of vertex degree and valence of molecular graph. Molecular connectivity index has very successful applications in chemistry, physics and molecular biology. It is a molecular topological index widely used, researched and developed~\cite{Rouvray1988,Randic1984,Randic1988,Jurs}. Based on the molecular connectivity index, we have proposed the bonding parameter topological index~\cite{Zhang1989A, Zhang1990A, Zhang1990B,Zhang1990C}. And apply it to high-temperature superconductors and surface adsorption systems. In the following sections, we will mainly focus on the molecular connectivity topological index development and bonding parameter topological index and its applications.

\subsubsection{Randic molecular topological index}

\begin{figure}[H]
\begin{center}
\epsfig{file=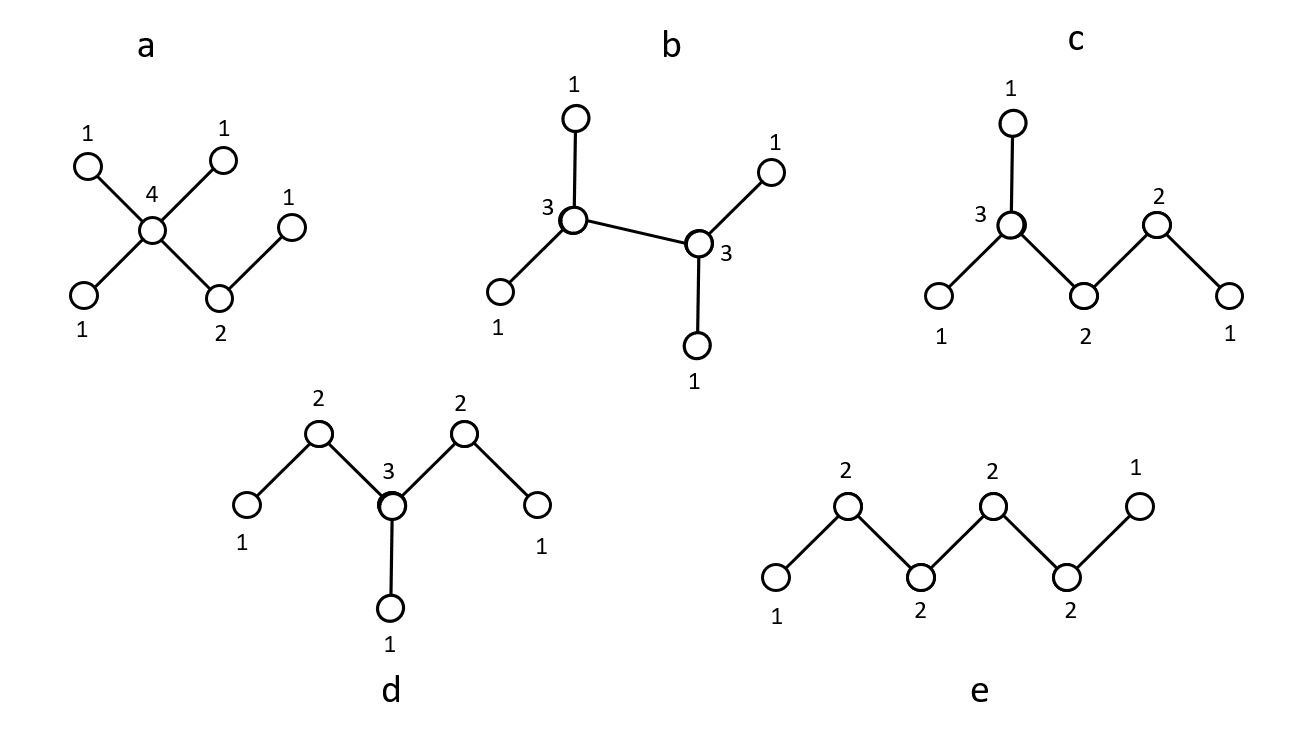,height=2.8in}
\caption{Molecular cluster with different degree of branching}
\label{fig:mg1}
\end{center}
\end{figure}

A lot of experimental facts show that in the series of isomeric organic compounds, the properties of the compounds are also different due to the different degree of branching. However, the concept of degree of branching is a more qualitative one. For example, the molecular diagram of the isomers of hexane is shown in Figure~\ref{fig:mg1}. Qualitatively speaking, the degree of branching can be arranged as follows:

\begin{equation}
e<d<c<b<a
\label{equ:mt1}
\end{equation}

Randie proposed a method to quantitatively describe the degree of branching of alkane compound series~\cite{Randic1975}. It can be seen from the molecular diagram of alkane through the carbon skeleton that the degree of branching on each carbon atom is related to the number of adjacent carbon atoms. In other words, it is related to the degree of the vertex on the molecular graph. In Figure~\ref{fig:mg1}, the degree of each vertex of each molecular graph is marked with a number on the graph. In this way, each carbon-carbon bond (or each edge on the molecular graph) can be characterized by a pair of numbers. The five edges shown in Figure~\ref{fig:mg1}a can be represented by the following numbers respectively:\\

$ (1,4); (1,4);(1,4);(2,4); (1,2)$\\

In hydrocarbons. The maximum number of atoms adjacent to a carbon atom is 4 and the minimum number is 1. Therefore, there are only 10 types of carbon-carbon bonds in hydrocarbons:\\

$ (1,1); (1,2);(1,3);(1,4)$\\
$ (2,2); (2,3);(2,4)$\\
$ (3,3); (3,4)$\\
$ (4,4)$\\

If a calculation method can be found, starting from the two vertex degrees related to each edge, calculate a certain value $C_k$ for each edge, and then sum the values $C_k$ calculated in this way for all edges in the molecular graph , which is

\begin{equation}
\sum\limits_{k} C_k \nonumber
\label{equ:mtn1}
\end{equation}

Then, the value obtained in this way can quantitatively describe the characteristics of the molecular map. Express the degree of branching of the isomeric compound series. In 1975, Randie proposed a calculation method to meet the above requirements and established a new topological index, namely Randie molecular topological index $\chi$

In Randie’s original work, $p_i$ and $p_j$ are used to represent the degrees of two vertices $v_i$ and $v_j$, respectively. For the edge $e_k$ formed by the two vertices of  $v_i$ and $v_j$, the square root of the reciprocal of the product of the degrees of the two vertices can be used,  corresponding value $C_k$, namely

\begin{equation}
C_k = (p_ip_j)^{-\frac{1}{2}}
\label{equ:mt2}
\end{equation}

In other words, each edge $C_k$ in the molecular graph can be marked with the value $C_k$ calculated in this way, and $C_k$ is called the edge index. Randie defined the sum of the edge indices $C_k$ of all edges in the molecular graph as his molecular topological index $\chi$, namely

\begin{equation}
\chi = \sum (p_ip_j)^{-\frac{1}{2}}
\label{equ:mt3}
\end{equation}

Randie puts forward an important purpose of his topological index $\chi$, in how to fully express the degree of molecular branching and the influence of such branching on the properties of molecules, so in his original literature this topological index $\chi$ is also called molecular branch topology index. It should be pointed out that because the definition of topological index  includes the vertex degree closely related to the valence of carbon, this lays the foundation for further development of this topological index to enable it to deal with a wider range of topics such as unsaturated compounds and organic compounds containing heteroatoms. At present, the topics dealt with by topological index $\chi$ far exceed those of saturated economic branching.

\subsubsection{Molecular Connectivity Index Series}

\begin{figure}[H]
\begin{center}
\epsfig{file=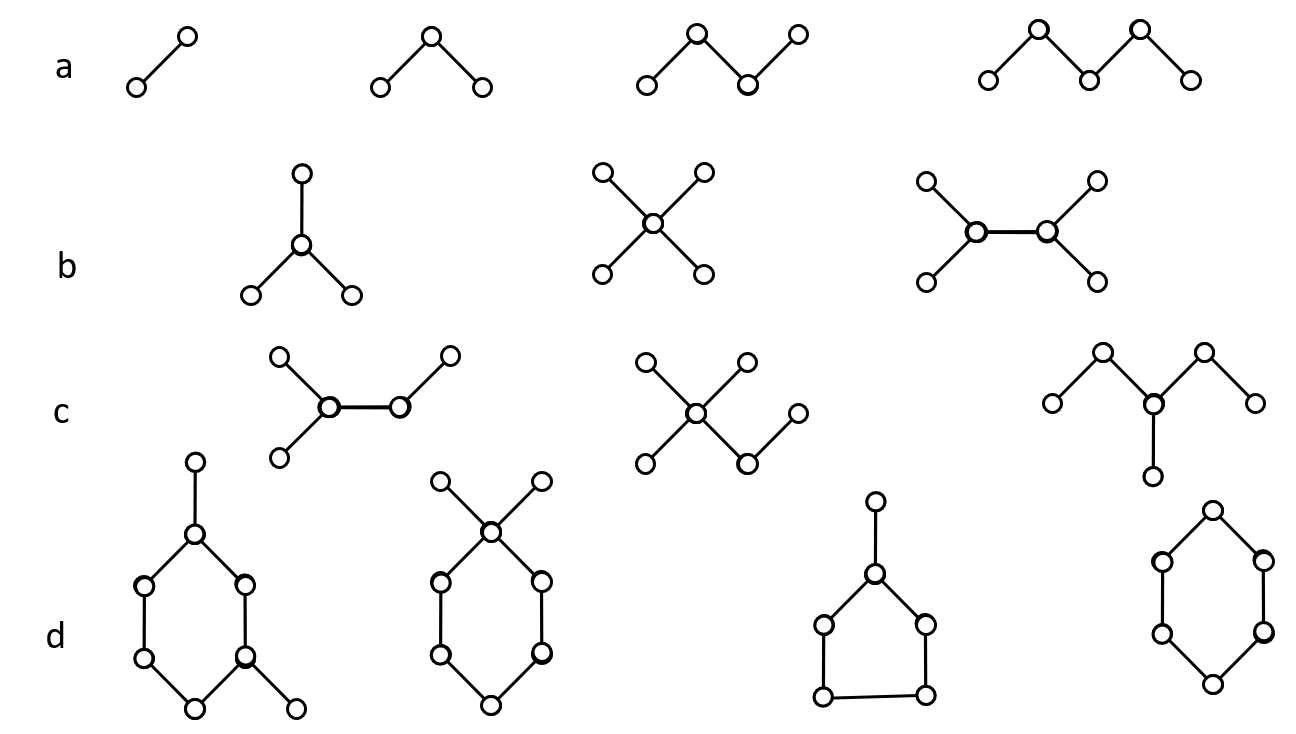,height=2.8in}
\caption{Four types of molecular graph}
\label{fig:mg2}
\end{center}
\end{figure}

After Randie proposed its topological index $\chi$, Kier and Hallet al. further developed it, thus establishing a series of molecular connectivity indexes~\cite{Kier1976A,Kier1976B,Kier1978,Kier1983}.  It is usually represented by the  $^m\chi_t$  symbol, where $m = 0,1, 2, 3 \cdots$,  usually $^0\chi,^1\chi, ^2\chi, ^3\chi \cdots$ are called zero-order, first-order, second-order, and third-order molecular connectivity indexes, respectively. And $t$ represents the type of subgraph contained in the molecular graph $G$. The connected subgraphs in the molecular graph $G$ can be divided into the following four types:

1. The first type of subgraph is called path, which is represented by the symbol p, that is, t = p. The path is formed by successively connecting different sides, as shown in Figure~\ref{fig:mg2}a. The degree of the vertices in this type of subgraph is not greater than 2.

2. The second type of subgraph is called Cluster, which is represented by the symbol c, ie t=c, as shown in Figure~\ref{fig:mg2}b. In this type of subgraph To include a vertex of degree 3 or 4, but not a vertex of degree 

3. The third type of subgraph is called path/cluster, which is represented by the symbol PC, ie t=pc, as shown in Figure~\ref{fig:mg2}c. In this type of subgraph, in addition to vertices of degree 3 or 4, vertices of degree 2 must be included.

4. The fourth type of subgraph is called a chain, which is represented by the symbol CH, that is, t=CH, as shown in Figure~\ref{fig:mg2}d. At least one ring must be included in this type of subgraph.

\begin{figure}[H]
\begin{center}
\epsfig{file=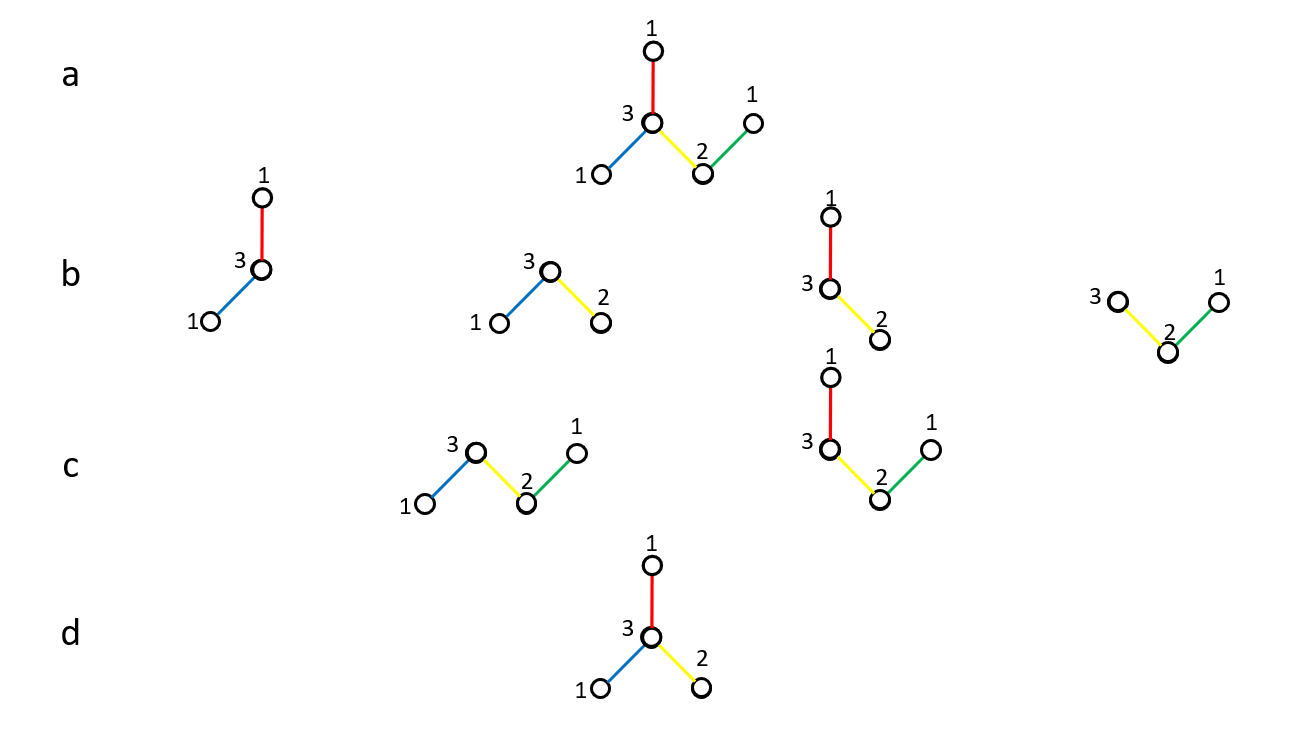,height=3.2in}
\caption{Molecular graphs with different types subgraphs}
\label{fig:mg3}
\end{center}
\end{figure}

As mentioned earlier, in the initial expression of Randie's topological index ~\ref{equ:mt3}, $\chi$ is defined by each edge in the molecular graph.  The corresponding edge index $C_k$. It is obtained by the summation, in the molecular connectivity index series, it is equivalent to the first-order molecular connectivity index $^m\chi_t$ with $m =l$. When calculating the first-level topological index, we divide the molecular graph into such subgraphs, that is, each subgraph is equivalent to a  edge, and then calculate the edge index for each subgraph according to the formula ~\ref{equ:mt2}, Finally, $^1\chi$ is calculated from the formula ~\ref{equ:mt3}. Now we further consider the secondary molecular bonding index $^2\chi$. When calculating $^2\chi$, we first divide the molecular graph into such subgraphs, and each subgraph is composed of two adjacent edges. For example, for the molecular graph G on Figure ~\ref{fig:mg3}a, there may be 4 such subgraphs, as shown in Figure ~\ref{fig:mg3}b; secondly, for each subgraph $G_k$ in Figure ~\ref{fig:mg3}b, use the following formula to calculate its corresponding exponential index $^2C_k$, namely

\begin{equation}
^2C_k = (p_ip_jp_k)^{-\frac{1}{2}}
\label{equ:mt4}
\end{equation}

Among them $p_i, p_j and p_k$ are the degrees of the three vertices that constitute this subgraph $G_k$. Obviously formula ~\ref{equ:mt4} is a direct derivation of formula ~\ref{equ:mt2}; finally for all such subgraphs, by summing, you can get the $^2\chi$, that is.

\begin{equation}
^2\chi = \sum\limits_{k}^{n_2} (p_ip_jp_k)_k^{-\frac{1}{2}}
\label{equ:mt5}
\end{equation}

Among them, $n_2$ represents the number of subgraphs $G_k$, formed by two adjacent edges. For example, for the molecular diagram of Figure ~\ref{fig:mg3}a, we can use formula~\ref{equ:mt5} calculates its second-level molecular connectivity index to be considered as

\begin{equation}
^2\chi = \frac{1}{\sqrt{1 \cdot 3 \cdot 1}}+ \frac{1}{\sqrt{1 \cdot 3 \cdot 2}}+\frac{1}{\sqrt{1 \cdot 3 \cdot 2}} + \frac{1}{\sqrt{3 \cdot 2 \cdot 1}}=1.558 \nonumber
\label{equ:mtn1}
\end{equation}

As can be seen from Figure ~\ref{fig:mg3}b, these subgraphs are all path-type subgraphs, that is, $t=p$, so $^2\chi_p$ can be used to express the second-order molecular connectivity index of this molecular graph.
In the calculation of the three-level molecular connectivity index $^3\chi_p$, the molecular graph is first divided into such subgraphs, each of which is connected by three sides. For example, the molecular graph on Figure Figure ~\ref{fig:mg3}a can be divided into three such subgraphs, as shown in the Figure ~\ref{fig:mg3}c and d; secondly, use the following formula to calculate the index $^3C_k$ of each such subgraph $G_k$, namely

\begin{equation}
^3C_k = (p_ip_jp_kp_l)^{-\frac{1}{2}}
\label{equ:mt6}
\end{equation}

Among them, $p_i, p_j,p_k and p_l$  are the degrees of the 4 vertices on each subgraph; it should be pointed out that when calculating the three-level molecular connectivity index $^3\chi_p$, due to the appearance of various subgraphs such as paths and clusters, that is $t=p$, C and CH. For the molecular diagram on Figure ~\ref{fig:mg3}a, they belong to $t=p$ There are two subgraphs, as shown in Figure ~\ref{fig:mg3}c. Add the indices $^3C_k$ of these two subgraphs to get the molecular connectivity index $^3\chi_p$, namely,

\begin{equation}
^3\chi_p = \frac{1}{\sqrt{1 \cdot 3 \cdot 2 \cdot 1}}+ \frac{1}{\sqrt{1 \cdot 3 \cdot 2 \cdot 1}}=0.816 \nonumber
\label{equ:mtn2}
\end{equation}

However, there is only one subgraph belonging to $t=c$, as shown in Figure ~\ref{fig:mg3}d. From its index  $^3C_k$,  the molecular connectivity index $^3\chi_c$ can be obtained, namely,

\begin{equation}
^3\chi_c = \frac{1}{\sqrt{1 \cdot 3 \cdot 2 \cdot 1}} =0.408 \nonumber
\label{equ:mtn3}
\end{equation}

It can be seen from this example that for the three-level molecular connectivity index, it is necessary to include $^3\chi_p,^3\chi_c $ and $ ^3\chi_{CH} $, three types of indices. According to a similar method, the molecular connectivity index above the fourth level can also be calculated. $^4\chi_p,^4\chi_c, ^4\chi_{CH} $ and $ ^4\chi_{pc} $ etc.

To sum up, the molecular connectivity index $^m\chi_t $ is determined by the $t$ type subgraph with $m$ edges, which can be generally expressed as

\begin{equation}
^m\chi_t = \sum\limits_{k=1}^{n_m} {^mC_k}
\label{equ:mt7}
\end{equation}

\begin{equation}
^mC_k = \prod_{\substack{i=1}}^{m+1}(p_i)_k^{-\frac{1}{2}}
\label{equ:mt8}
\end{equation}

Where $n_m$ is the number of $t$type subgraphs containing $m$ edges, and $p_i$ is the degree of vertex $i$. In the general expressions ~\ref{equ:mt7} and ~\ref{equ:mt8} of the series of molecular connectivity index, relative to Randie's original topological index ~\ref{equ:mt3}, the following several aspects of expansion is done :

1. Not only the contribution of each edge in the molecular graph to the topological index is considered, but also the contribution of two adjacent edges, three connected edges, and four connected edges to the topological index of the subgraphs are further considered, thus introducing the first-level, second-level Level, third-level and fourth-level molecular connectivity index $^1\chi,^2\chi, ^3\chi$ and $^4\chi $ etc. The essence of this expansion is not only to consider the contribution of each bond in the molecule, but also to further consider the contribution when there is an interaction between two bonds, the contribution when there is an interaction between three bonds, etc.

2. The summation in the formula ~\ref{equ:mt7} is performed on the subgraphs of  $m$ and  $t$ given in the molecular diagram, instead of only the molecular diagram in the formula ~\ref{equ:mt3} on each side. Therefore, the molecular connectivity index $m$ further distinguishes the topological indexes of different types of $t$ subgraphs in the molecular graph. For example, in the above example, there are 2 and 3 for the three-level molecular connectivity index. The essence of this derivation is that for the molecular connectivity index of the same series $m$, the contribution of different subgraph types can be further distinguished.

3. For a given subgraph of  $m$ and  $t$, as a direct derivation of equation ~\ref{equ:mt2}, use equation ~\ref{equ:mt8} to calculate the corresponding subgraph Index  $^mC_k$

In various isomers with the same carbon number $n$, if their topological index $\chi $ (or $^1\chi $) is the same, but at least some other topological index $^m\chi_t $ values ​​are different, so use the molecular connectivity index series $^m\chi_t $ to characterize molecular structure, compared with only Randie topological index $\chi $, its resolution power is increased.

\subsubsection{Valence Connectivity Index}
When applying the above-mentioned molecular connectivity index $^m\chi_t $  to organic compounds containing heteroatoms and unsaturated molecules containing multiple bonds, difficulties are encountered. In the case of multiple bonds, such as butane and butadiene:

\begin{figure}[H]
\begin{center}
\epsfig{file=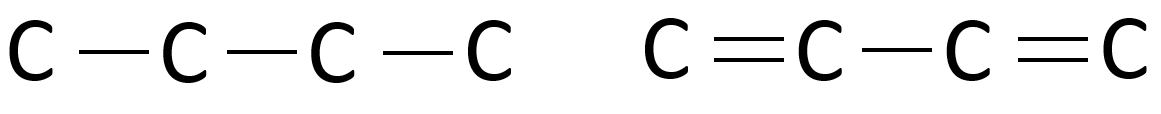,height=0.3in}
\end{center}
\end{figure}

According to the above-mentioned vertical calculation method, the $ p_i $ value of the two is the same, so the $^m\chi_t $  value is the same, that is, the difference in the properties of the two molecules cannot be reflected. For another example, for the following two organic compounds containing OH groups;

\begin{figure}[H]
\begin{center}
\epsfig{file=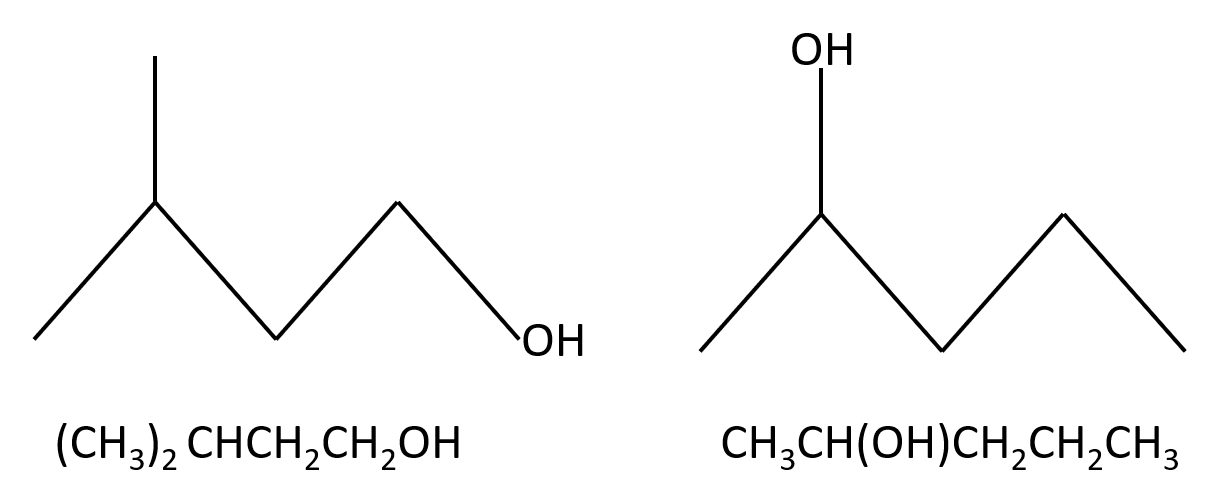,height=1.6in}
\end{center}
\end{figure}

They also have the same molecular graph, and according to the above calculation method of $^1\chi $, they also have the same double value, which cannot reflect the different properties of the two. In order to consider organic compounds containing heteroatoms and multiple bonds, Kier et al. improved the molecular connectivity index $^m\chi_t $ and proposed a valence connectivity index $^m\chi_t ^{\upsilon}$

In the $^m\chi_t ^{\upsilon}$ index, $p_i $ is the degree of the first vertex. In the molecular diagram of the carbon atom skeleton, $p_i $ is the number of C-C bonds formed by a carbon atom with other carbon atoms. It is closely related to the valence of the carbon atom and can only take the values ​​1, 2, 3, and 4. Accordingly, Kier et al. proposed to replace $p_i $ with the following quantity $\delta_i $ which is closely related to the valence of the atom on the vertex, namely~\cite{Kier1976A,Kier1976B,Kier1978,Kier1983}.

\begin{equation}
\delta_i = z_i^{\upsilon}-h_i
\label{equ:vt1}
\end{equation}

Among them, $z_i^{\upsilon}$ is the number of valence electrons on the i-th atom, and $h_i$ is the number of hydrogen atoms associated with the i-th atom. Using $\delta_i $, Kier et al. defined a valence connectivity index $^m\chi_t ^{\upsilon}$ as follows:

\begin{equation}
^m\chi_t^{\upsilon}=\sum\limits_{k=1}^{n_m} \prod_{\substack{i=1}}^{m+1}(\delta_i)_k^{-\frac{1}{2}}
\label{equ:vt2}
\end{equation}

Take the unsaturated molecule with multiple bonds as an example, we use the above $\delta_i $ value to calculate $^m\chi_t ^{\upsilon}$. For the above examples of butane and butadiene, their $\delta_i $ values ​​are as follows:

\begin{figure}[H]
\begin{center}
\epsfig{file=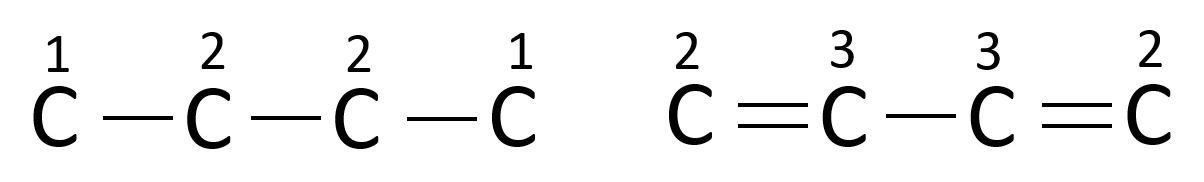,height=0.4in}
\end{center}
\end{figure}

The $^1\chi_t ^{\upsilon}$ value calculated from this is obviously different for these two molecules. As for whether the double bond is calculated as one side or as two sides when calculating $^m\chi_t ^{\upsilon}$ using the formula~\ref{equ:vt2}, it is still undecided, you can choose one of them. For example, for the following molecules;

\begin{figure}[H]
\begin{center}
\epsfig{file=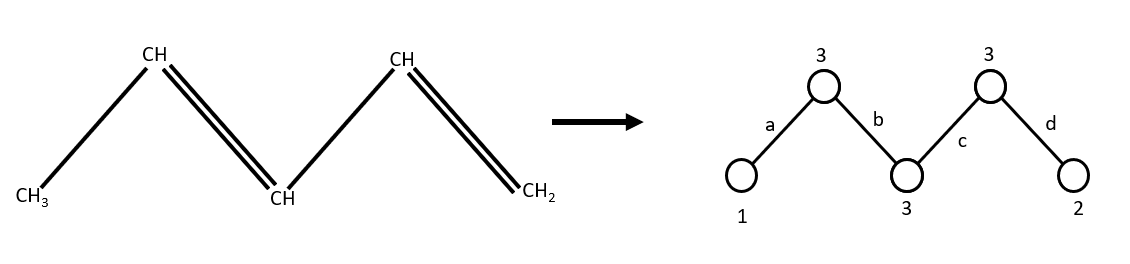,height=1.2in}
\end{center}
\end{figure}

There are two algorithms. Marray uses the method of calculating as one edge. The b and c edges give the same value in the calculation of  $^1\chi_t ^{\upsilon}$, even though b is a double bond and c is a single bond. Randie calculates the double bond as two sides, that is, calculate the  $^1\chi_t ^{\upsilon}$, value of this molecule with the following formula:

\begin{equation}
^1\chi^{\upsilon}=(1 \cdot 3)^{-\frac{1}{2}}+2(3 \cdot 3)^{-\frac{1}{2}}+(3 \cdot 3)^{-\frac{1}{2}}+2(3 \cdot 2)^{-\frac{1}{2}}\nonumber
\label{equ:vt3}
\end{equation}

The processing method of multiple bonds can also be extended to multiple bonds containing heteroatoms, such as c = o, c and N, and X = O, etc ~\cite{Xin}.

\subsubsection{Balaban Connectivity Index}
The molecular graph can be represented by the adjacency matrix $ A $ and the distance matrix $D$. The $D$ matrix has more information than the $ A $ matrix. The sum of the elements in each row or column of $ A $ matrix is ​​called the degree of the vertex, which constitutes the invariant of the molecular graph. The molecular connectivity index $^m\chi_t ^{\upsilon}$ is established based on the vertex degree~\cite{Balaban1982,Balaban1983,Balaban1988,Balaban1991}.

The result obtained by adding the elements of one row or one column of the distance matrix  $D$ is called the distance sum corresponding to the vertex, denoted by $s_j, j=l, 2,...n$. The sum of distance of each vertex $ s_j$ is similar to vertex degree $ p_i$. It is also an invariant of the molecular graph, sometimes it is called distance degree. In organic molecules, for the molecular graph of the carbon atom skeleton, each vertex degree $ p_i$ can only take a finite integer value of 1, 2, 3, and 4, but there is no such limitation for the distance degree $ s_j$. If $q $ is used to represent the number of adjacent vertices, it is equal to the number of edges in a simple molecular graph of saturated molecules, then

\begin{equation} 
\begin{aligned}
\overline{S_j} &=\frac{s_j}{q}
\end{aligned}
\label{equ:bt1}
\end{equation}

Defined as the average sum of distance. For a simple molecular graph of saturated acyclic molecules, Balaban uses a similar expression to the Randie molecular connectivity index $\chi$, replacing the vertex degree  $ p_i$ with the average sum of distance $\overline{S_j}$, and constructs a new topological index $J$, namely

\begin{equation} 
\begin{aligned}
J &=  \sum_i (\overline{S_i} \cdot \overline{S_j})^{-\frac{1}{2}}
\end{aligned}
\label{equ:bt2}
\end{equation}

The summation is performed according to the edges in the molecular graph. This topological index is usually called the connectivity index of the average distance sum, and we call it the Balaban connectivity index for short

\subsubsection{Bonding Parameter Topological Index and Application}

As mentioned above, the molecular connectivity index $\chi$ proposed by Randie has been developed from three aspects:\\
1. Kier et al. established the molecular connectivity index series $^1\chi,^2\chi,^3\chi, \cdots $ namely $^m\chi_t$ ~\cite{Kier1976A,Kier1976B,Kier1978,Kier1983}\\
2. Kier et al. used $\delta_i$ instead of $p_i$ to propose a series of valence connectivity indexes  $^m\chi_t ^{\upsilon}$ ~\cite{Kier1976A,Kier1976B,Kier1978,Kier1983}\\
3. Balaban uses distance $s_i$ instead of vertex $p_i$ to establish a high-resolution topological index $J$.~\cite{Balaban1982,Balaban1983,Balaban1988,Balaban1991}\\

Through the development of these aspects, not only the molecular connectivity index can describe the structure of the molecule more effectively, but also the application of related molecular structure and physical properties, chemical properties and biological activities of the molecule has also shown greater performance. In order to enable the molecular connectivity index to reflect more molecular structure information, especially electronic structure information, so that it can not only be used for organic compound systems containing heteroatoms, but also for inorganic compounds, such as high temperature superconductor and surface adsorption system, we propose a bonding parameter connectivity index $H_t$. Its general form can be expressed as:

\begin{equation}
H_t= (\sum\limits_{i}\frac{1}{e^{(X_i-Y_i)}\sqrt{p_iq_i}})^2=(\sum\limits_{i}\frac{1}{e^{\bigtriangleup_i}\sqrt{p_iq_i}})^2
\label{equ:ht}
\end{equation}

Among them, $p_i$ and $q_i$ represent the degree of the vertices at both ends of the i-th side in the molecular graph, and $X$ and  $Y$  are the bonding parameters of the atoms corresponding to the two vertices.  The number, $i$ is the sum of all edges in the molecular graph, and $t$ represents the type of vertex atomic bond parameters. Currently, the bond parameter types we have considered are the following:\\
$t = 1: X_i,Y_i$ respectively represent the first ionization potential and electron affinity energy of element atoms.\\
$t = 2: X_i,Y_i$ respectively represent the electronegativity of element atoms $(X_i > Y_i)$\\
$t = 3: X_i,Y_i$ respectively represents the charge-radius ratio of the element atom $z/r_{cov}(X_i> Y_i)$, $z$ is the number of valence electrons of the atom, and $r_{cov} $ is the covalent radius of the atom;\\
$t = 4: X_i-Y_i$ which represents the bond length of the i-th side.\\

In specific applications, we use the expanded form of the bonding parameter connectivity index:

\begin{equation}
H_{tn} = (\sum\limits_{i}\frac{1}{(1 + \bigtriangleup_i + \frac{1}{2!}\bigtriangleup_i^2 + \frac{1}{3!}\bigtriangleup_i^3 + \dots + \frac{1}{n!}\bigtriangleup_i^n +\dots )\sqrt{p_iq_i}})^2
\label{equ:htn}
\end{equation}

Where$ \bigtriangleup_i= X_i-Y_i$, $n$ means that $e^{(Xi-Yi)}$ expands to the nth power of $ \bigtriangleup_i$. For example, when $n = 0$, we can get:

\begin{equation} 
\begin{aligned}
H_{t0} &= ( \sum_i \frac{1}{\sqrt{p_i \cdot q_i }})^2 
\end{aligned}
\label{equ:ht0}
\end{equation}

It is the square of the Ramdic molecular connectivity index $\chi$ ~\cite{Randic1975}. 

When $n=1,t= 1$, we can get:

\begin{equation} 
\begin{aligned}
H_{11} &= ( \sum_i \frac{1}{(1 + X_i-Y_i)\sqrt{p_i \cdot q_i }})^2 
\end{aligned}
\label{equ:h11}
\end{equation}

When $n=1, t = 3$, we can get:

\begin{equation} 
\begin{aligned}
H_{31} &= ( \sum_i \frac{1}{(1 + \bigtriangleup_i)\sqrt{p_i \cdot q_i }})^2 
\end{aligned}
\label{equ:h31}
\end{equation}

At present, we have applied the bonding parameter connectivity index H to $XY_n (n=1,2, 3, 4)$ molecular systems containing heteroatoms~\cite{Zhang1989A}, high-temperature superconductors~\cite{Zhang1990A} and surface adsorption systems~\cite{Zhang1990B,Zhang1990C}, and obtained some valuable results. In the following,  we will describe two examples.

(1) Correlation of the chemical properties of isomorphic heteroatom molecules $XY_n (n=2, 3, 4)$ with $H_{11}$

Table~\ref{tab:atom} lists the value of the first ionization potential X and electron affinity energy Y of some commonly used atoms. Tables~\ref{tab:h1}, ~\ref{tab:h2} and ~\ref{tab:h3} list the molecular configuration, molecular formula, topological index of $H_{11}$ bonding parameters and generate enthalpy data. For the molecular configuration of a plane triangle and a regular tetrahedron, we only consider the cluster type marked with red and blue bond to calculate $H_{11}$ value.

\begin{figure}[H]
\begin{center}
\epsfig{file=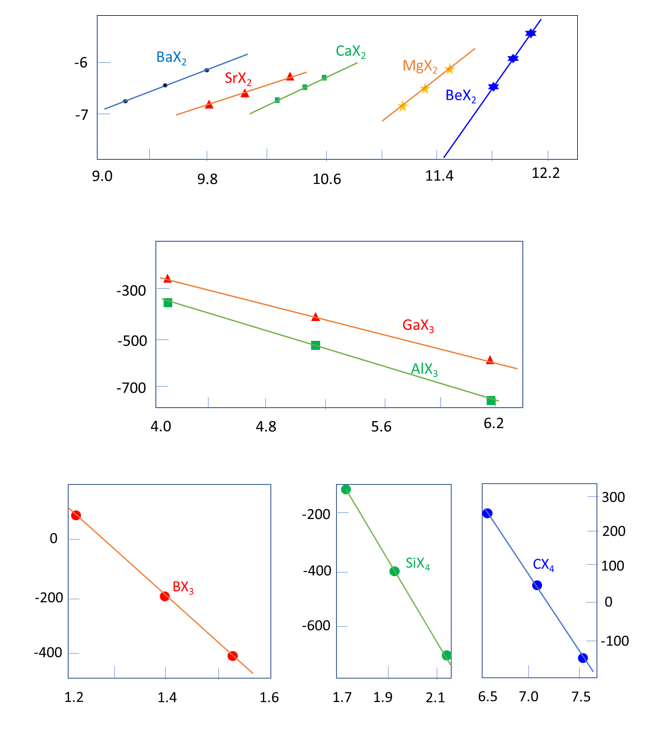,height=3.6in}
\caption{Liner regression line for the plot of  $ln(1/H_{11})$ against $ln(1/\Delta H_f)$ or $H_{11}$ against $\Delta H_f$ for  isomorphic heteroatom molecules listed in  Tables~\ref{tab:h1}, ~\ref{tab:h2} and ~\ref{tab:h3} }
\label{fig:mx-ti}
\end{center}
\end{figure}

Figure~\ref{fig:mx-ti} shows the linear relationship between $H_{11}$ and generate enthalpy  $ \Delta H_f $. From the figure, we can see that, except for the fluoride, $H_{11}$ has an excellent linear relationship with the formation of the compound. For example, in the case of Figure ~\ref{fig:mx-ti}a, the correlation coefficient is as high as 0.9999.  There are two reasons why the fluoride deviates from the linear relationship. First, fluoride may have different molecular structure compare to other compound. Second, the Y value of F atoms given in Table~\ref{tab:atom} has a large deviation. The Y values ​​of the F atom that we obtained by using three different molecular configurations of $ H_{11} - \Delta H_f $ diagrams that shown in Figure~\ref{fig:mx-ti} are all around $410 kJ mol^{-1}$

\begin{table}[H]
\begin{center}

\begin{tabular}{p{3.8cm}p{3.8cm}p{3.8cm}} \hline 
& \multicolumn{2}{l}{X and Y Value of Element Atom} \\ \cline{2-3} 
Element Atom  & $X$ & $Y$ \\ \hline

$B$ &  807 & 35   \\ 
$Al$ &  584  & 54  \\ 
$Ga$ &  585  & 55   \\ 
$In$ &  565  & 76 \\ 
$Tl$ &  596  & 123 \\
$C$ &  1093 & 119   \\ 
$Si$ &  793  & 140  \\ 
$Ge$ &  768  & 138  \\ 
$Sn$ &  715  & 148 \\ 
$Pb$ &  722  & 179 \\
$Li$ &  526 &   \\ 
$Na$ &  502  &   \\ 
$K$ &  425  &   \\ 
$Rb$ &  409  &  \\ 
$Cs$ &  382  &  \\
$Be$ &  906 &   \\ 
$Mg$ &  744  &   \\ 
$Ca$ &  596  &   \\ 
$Sr$ &  556 &  \\ 
$Ba$ &  509  &  \\
$N$ &    & -20   \\ 
$P$ &    & 82  \\ 
$As$ &   & 78 \\ 
$Sb$ &   & 65\\ 
$Bi$ &   & -27\\
$O$ &    & 149   \\ 
$S$ &    & 206  \\ 
$Se$ &   & 210 \\ 
$Te$ &   & 220\\ 
$Po$ &   & 196\\
$F$ &    & 340  \\ 
$Cl$ &    &365  \\ 
$Br$ &   & 344 \\ 
$I$ &   & 315\\ 
$H$ &   & 78 \\ \hline

\end{tabular}
\caption{The value of the first ionization potential X and electron affinity energy Y of some element atoms}
\label{tab:atom}
\end{center}
\end{table}

\begin{table}[H]
\begin{center}

\begin{tabular}{p{3cm}p{3cm}p{3cm}p{3cm}} \hline 
Configuration &  Formula  & $ln(1/H_{11}) $ & $ln(|1/\bigtriangleup H_f|)$ \\ \hline 
  &  $BeF_2$   & $11.988$  & $-6.93$ \\
  &  $BeCl_2$   & $11.897$  & $-6.19$ \\
  &  $BeBr_2$   & $11.973$  & $-5.91$ \\
  &  $BeI_2$   & $12.74$  & $-5.36$ \\
  &  $MgF_2$   & $11.315$  & $-7.02$ \\
  &  $MgCl_2$   & $11.187$  & $-6.46$ \\
  &  $MgBr_2$   & $11.295$  & $-6.26$ \\
  &  $MgI_2$   & $11.434$  & $-5.90$ \\
  &  $CaF_2$   & $10.405$  & $-7.11$ \\
\parbox[c]{2em} {\includegraphics[width=0.8in]{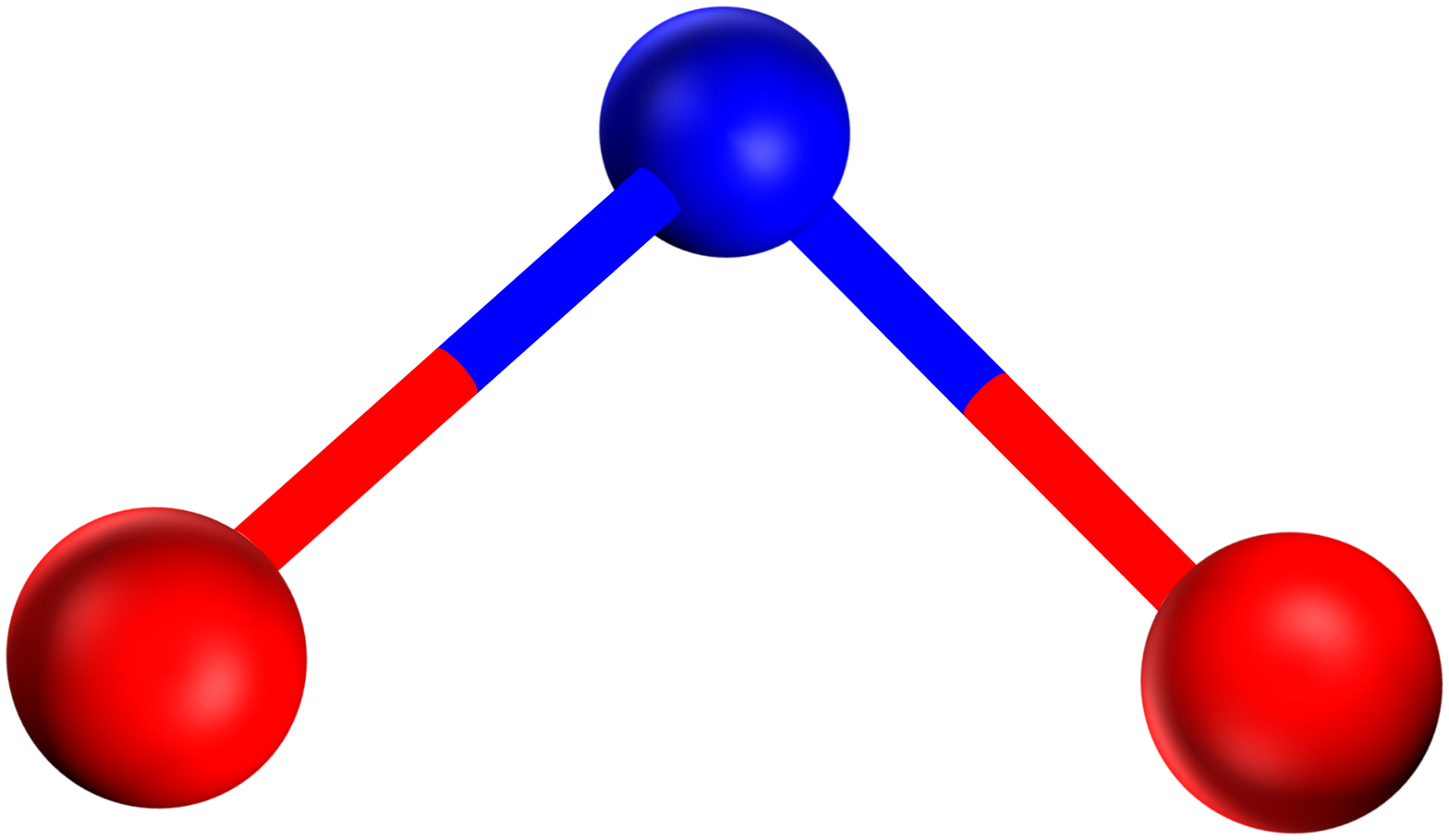}}
 &   $CaCl_2$   & $10.200$  & $-6.68$ \\
  &  $CaBr_2$   & $10.374$  & $-6.53$ \\
  &  $CaI_2$   & $10.591$  & $-6.28$ \\
  &  $SrF_2$   & $10.067$  & $-7.10$ \\
  &  $SrCl_2$   & $9.822$  & $-6.72$ \\
  &  $SrBr_2$   & $10.029$  & $-6.58$ \\
  &  $SrI_2$   & $10.285$  & $-6.32$ \\
  &  $BaF_2$   & $9.578$  & $-7.09$ \\
  &  $BaCl_2$   & $9.260$  & $-6.75$ \\
  &  $BaBr_2$   & $9.531$  & $-6.63$ \\
  &  $BaI_2$   & $9.853$  & $-6.40$ \\
\hline

\end{tabular}
\caption{List of configuration , formula, $ln(1/H_{11}) $,generate enthalpy $ln(|1/\bigtriangleup H_f|)$ of $XY_2$ molecules }
\label{tab:h1}
\end{center}
\end{table}

\begin{table}[H]
\begin{center}

\begin{tabular}{p{3cm}p{3cm}p{3cm}p{3cm}} \hline 
Configuration &  Formula  & $H_{11}(10^{-5}) $ & $\bigtriangleup H_f(kJ mol^{-1} )$ \\ \hline 
  &  $ BF_3 $   & $1.370$  & $-1137$ \\
  &  $BCl_3$   & $1.529$  & $-404$ \\
  &  $BBr_3$   & $1.393$  & $-206$ \\
  &  $BI_3$   & $1.234$  & $71$ \\
  &  $AlF_3$   & $4.998$  & $-1504$ \\
\parbox[c]{1em} {\includegraphics[width=1in]{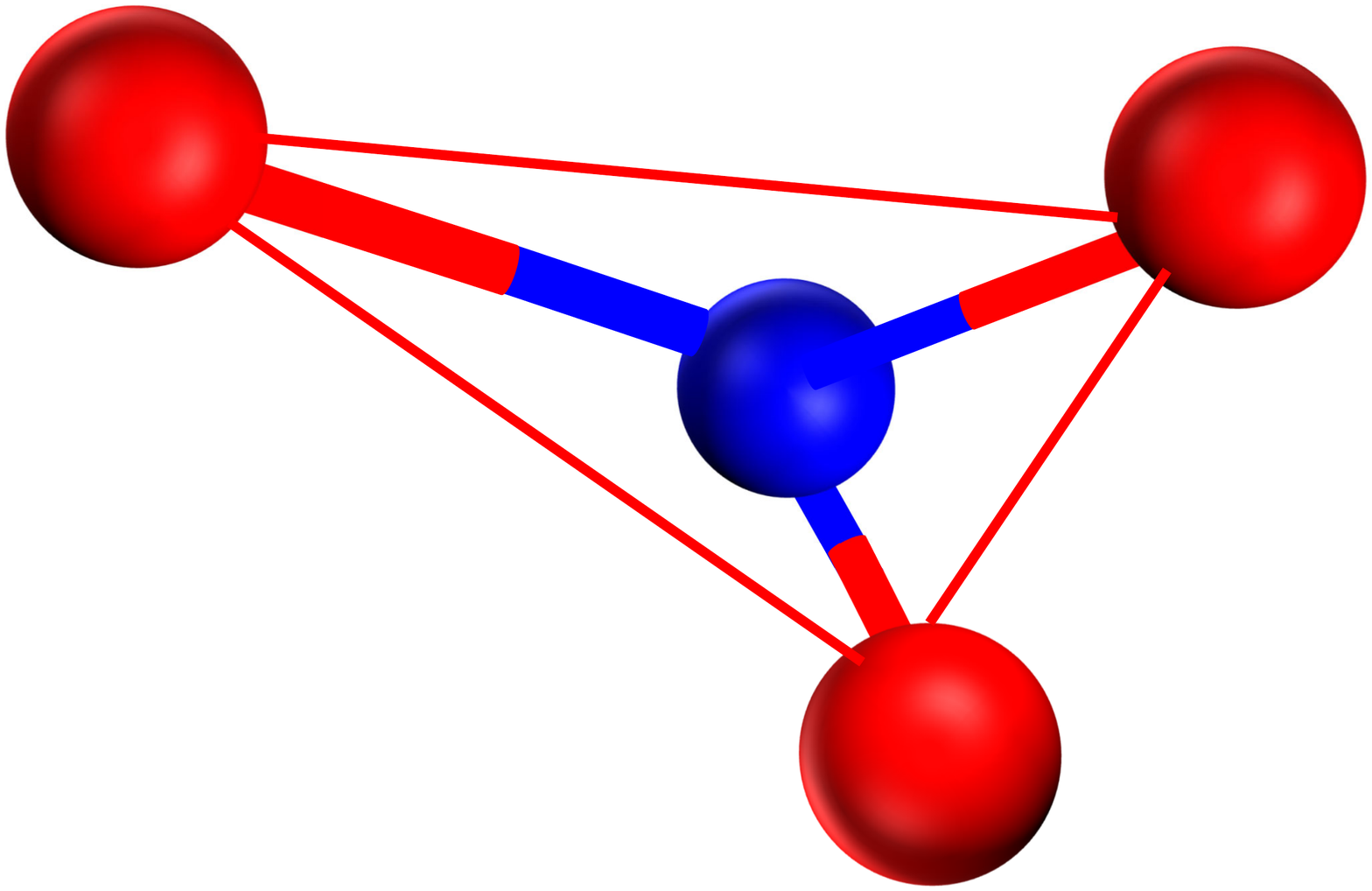}}
&   $AlCl_3$   & $6.198$  & $-704$ \\
  &  $AlBr_3$   & $5.165$  & $-516$ \\
  &  $AlI_3$   & $4.115$  & $-315$ \\
  &  $CaF_3$   & $4.957$  & $-1160$ \\
  &  $CaCl_3$   & $6.142$  & $-525$ \\
  &  $CaBr_3$   & $5.123$  & $-387$ \\
  &  $caI_3$   & $4.085$  & $-239$ \\
  \hline

\end{tabular}
\caption{List of configuration , formula, $H_{11}$,generate enthalpy  $ \Delta H_f $ of $XY_3$ molecules}
\label{tab:h2}
\end{center}
\end{table}

\begin{table}[H]
\begin{center}

\begin{tabular}{p{3cm}p{3cm}p{3cm}p{3cm}} \hline 
Configuration &  Formula  & $H_{11}(10^{-6}) $ & $\bigtriangleup H_f(kJ mol^{-1} )$\\ \hline 
  &  $SiF_4$   & $1.941$  & $-1615$ \\
  &  $SiCl_4$   & $11.988$  & $-657$ \\
  &  $SiBr_4$   & $11.988$  & $-399$ \\
  &  $SiI_4$   & $11.988$  & $-189$ \\
\parbox[c]{1em} {\includegraphics[width=1in]{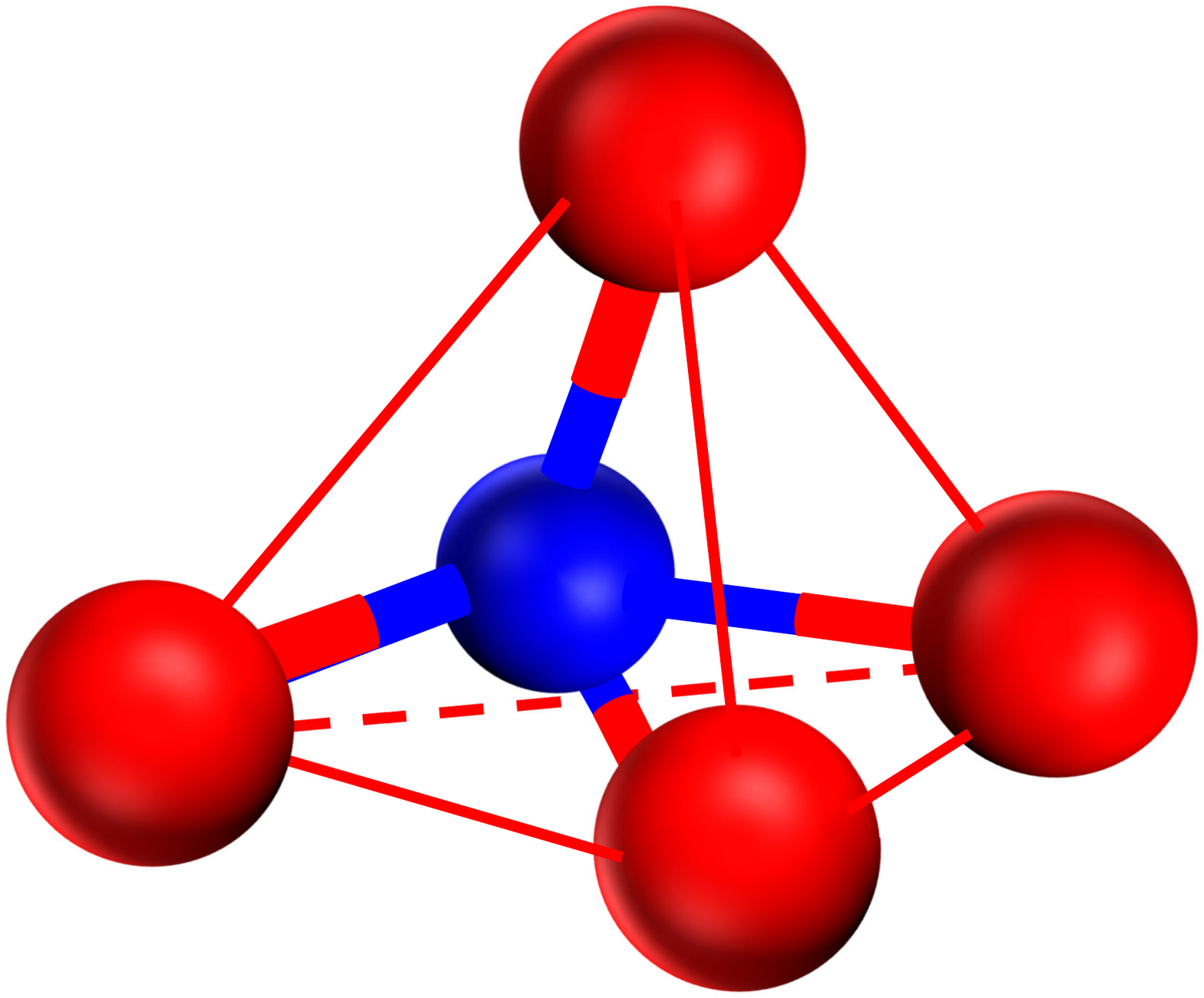}}
  &  $CF_4$   & $11.988$  & $-925$ \\
  &  $CCl_4$   & $11.988$  & $-135$ \\
  &  $CBr_4$   & $11.988$  & $19$ \\
  &  $CI_4$   & $11.988$  & $262$ \\ 
\hline
\end{tabular}

\caption{List of configuration , formula, $H_{11}$,generate enthalpy  $ \Delta H_f $ of $XY_4$ molecules}
\label{tab:h3}
\end{center}
\end{table}

(2) Correlation of Y series single-phase $Ba_2YCu_3O_{9-\delta}$ critical temperature of oxide superconductor with $H_{31}$

\begin{figure}[H]
\begin{center}
\epsfig{file=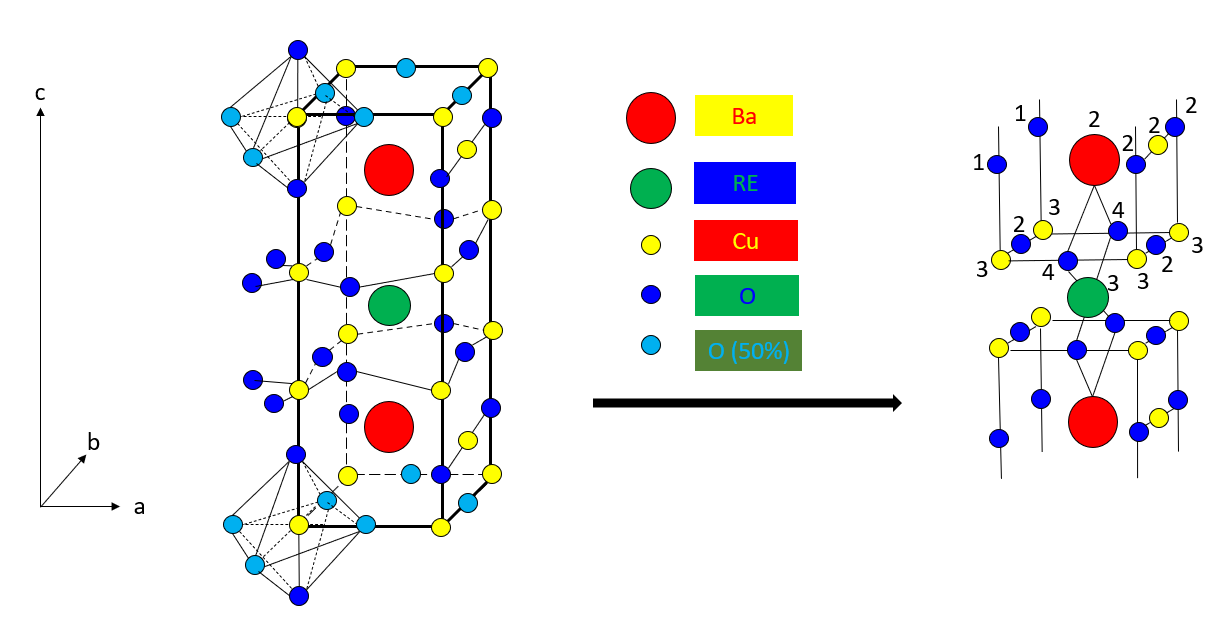,height=2.8in}
\caption{Structure of single phase $Ba_2RECu_3O_{9-\delta}$ superconductors and corresponding chemical  bonding model in one cell }
\label{fig:sup-ti}
\end{center}
\end{figure}

In exploring the relationship between the critical temperature of high-temperature superconductors and their structure, in addition to the microscopic theoretical methods based on quantum mechanics, how to make full use of the accumulated experimental data to establish a relationship between the critical temperature $T_c$ and the structural parameters of the superconductor with the semi-empirical method of quantitative relations should also be given full attention. The topological property of a molecule is the geometric property that the molecule remains unchanged during the continuous deformation process under the influence of internal and external factors, and it is one of the inherent properties of the molecule. Molecular topological index is an index that digitizes molecular structure. Its essence is the topological invariant of molecular graph. The purpose of here is to use our proposed bonding parameter topological index ~\cite{Zhang1989A} to study the relationship between the critical temperature of high-temperature superconductors and the properties of their constituent elements.
Single-phase Ba is only a perovskite structure with oxygen defects, as shown in Figure ~\ref{fig:sup-ti}(where RE stands for rare earth elements). Its space group $Pmmm$ and lattice constants are listed in Table~\ref{tab:Super1}. The $Ba$ and $RE$ in the $C$ axis are arranged in an orderly manner to extend the $C$  axis three times. The key to conduction lies in the two $CuO$ planes, in which the copper is in a distorted flat and square coordination. Experiments show that the interaction between the two  $CuO$  layers of this type of superconductor is very small~\cite{Cava}, and they have only a weak interaction with the $RE$ atoms sandwiched between them~\cite{Zhang1993A,Zhang1994B}. Based on the above experimental facts, we proposed a chemical bonding model of the interaction between atoms in a unit cell of this type of superconductor. As shown in Figure ~\ref{fig:sup-ti}: in the direction of the  $C$ axis, since the interaction of the two $CuO$ layers is very small, it can be considered that the $Cu$ and $O$ atoms between the two $CuO$ planes do not interact. The two $CuO$ planes only weakly interact through the bonding of the $RE$ atom between them and the oxygen atom closest to $RE$. Since there is a layer of $CuO$ plane, the ordered arrangement of  $Ba$ and $RE$ atoms in the $C$ axis direction cannot be formed by the direct bonding of the two. It is formed by the bridge action of oxygen atoms in the  $CuO$ plane. In order to give a quantitative concept, let the distance between the two  $CuO$ planes in the $C$ axis direction be twice the distance between other atoms. Therefore, in a unit cell, only the distance between two atoms (for $Cu, O$ atoms) can be bonded. Here, 1.1733 is the maximum value of the lattice constant $C$ in the $C$-axis direction in Table~\ref{tab:Super1}.

From equation~\ref{equ:h31} and the symmetry of Figure ~\ref{fig:sup-ti}, the topological index of the $H_{31}$ bonding parameter can be obtained as

\begin{equation} 
\begin{aligned}
H_{31} &= 4[\frac{1}{(1 + \bigtriangleup_{O-Cu})}(\frac{2}{\sqrt{3 \cdot 1 }}+\frac{2}{\sqrt{2 \cdot 2 }}+\frac{4}{\sqrt{3 \cdot 4 }}+\frac{6}{\sqrt{3 \cdot 2 }}) \\
& +\frac{1}{(1 + \bigtriangleup_{O-Ba})}\frac{2}{\sqrt{4 \cdot 2 }}+\frac{1}{(1 + \bigtriangleup_{O-X})}\frac{2}{\sqrt{4 \cdot 3 }}]^2 \\
 &= 4 (\frac{A}{1 + \bigtriangleup_{O-Cu}}+\frac{B}{1 + \bigtriangleup_{O-Ba}}+\frac{C}{1 + \bigtriangleup_{O-X}})^2
\end{aligned}
\label{equ:sct1}
\end{equation}

Where $X$ stands for $RE$ atom, $A = 5.7589, B=0.7071. C=0.4082$. The $H_{31}$ bonding parameter topological index of the single-phase $Ba_2RECu_3O_{9-\delta}$ series superconductor calculated according to equation~\ref{equ:sct1} is listed in In Table~\ref{tab:Super2}. Using the regression of $H_{31}$ and the zero resistance critical temperature T of the corresponding superconductor, the semi-empirical formula for the critical temperature of this type of superconductor is obtained as

\begin{equation} 
\begin{aligned}
T_c &= aH_{31} + b \\
 &= 4a (\frac{A}{1 + \bigtriangleup_{O-Cu}}+\frac{B}{1 + \bigtriangleup_{O-Ba}}+\frac{C}{1 + \bigtriangleup_{O-X}})^2 +b
\end{aligned}
\label{equ:sct2}
\end{equation}

Among them, the values ​​of $a$ and $b$ are as follows: when $X = Y, Yb, Sm, a = 1588.24, b=-4355.94$, when$ X=Eu, Py, Gd, Er, H, a = 44 19 .35, b=-12295.51$ The critical temperature values ​​of this type of superconductor obtained by equation~\ref{equ:sct2} are listed in Table~\ref{tab:Super2}. Using two sets of values ​​of $a$ and $b$, the critical temperature of this type of superconductor containing other rare earth elements was calculated respectively, and the calculation results are also listed in Table~\ref{tab:Super2}.

\begin{table}[H]
\begin{center}

\begin{tabular}{p{3.8cm}p{2.6cm}p{2.6cm}p{2.6cm}} \hline 
& \multicolumn{3}{c}{Lattice Constants (nm)} \\ \cline{2-4} 
Superconductors  & $a$ & $b$ & $c$ \\ \hline

$Ba_2YCu_3O_{9-\delta}$ &  0.3881  & 0.3881  & 1.1661 \\ 
$Ba_2SmCu_3O_{9-\delta}$ &  0.3845  & 0.3899  & 1.1702 \\ 
$Ba_2EnCu_3O_{9-\delta}$&  0.3850  & 0.3901  & 1.1733 \\ 
$Ba_2DyCu_3O_{9-\delta}$ &  0.3810  & 0.3873  & 1.1652 \\ 
$Ba_2GdCu_3O_{9-\delta}$ &  0.3832  & 0.3893 & 1.1696 \\ 
$Ba_2HoCu_3O_{9-\delta}$&  0.3810 & 0.3924  & 1.1648\\ 
$Ba_2ErCu_3O_{9-\delta}$ &  0.3827  & 0.3894  & 1.2694 \\ 
$Ba_2YbCu_3O_{9-\delta}$ &  -  & -  & - \\ \hline

\end{tabular}
\caption{Lattice constants of single-phase $Ba_2RECu_3O_{9-\delta}$. }
\label{tab:Super1}
\end{center}
\end{table}

\begin{table}[H]
\begin{center}

\begin{tabular}{p{3.8cm}p{2.6cm}p{2.6cm}p{2.6cm}} \hline 
& \multicolumn{2}{r}{$T_C(K)$} \\ \cline{3-4} 
Superconductors  & $H_{31}$ & Cale & Exp \\ \hline

$Ba_2YCu_3O_{9-\delta}$ &  2.8003  & 91.6  & 91.8 \\ 
$Ba_2SmCu_3O_{9-\delta}$ &  2.7987  & 89.06  & 89.0 \\ 
$Ba_2EnCu_3O_{9-\delta}$&  2.7998  & 77.8  & 77.8 \\ 
$Ba_2DyCu_3O_{9-\delta}$ &  2.8024  & 89.3  & 91.5 \\ 
$Ba_2GdCu_3O_{9-\delta}$ &  2.8013  & 84.4 & 83.6 \\ 
$Ba_2HoCu_3O_{9-\delta}$&  2.8029 & 91.5  & 91.5\\ 
$Ba_2ErCu_3O_{9-\delta}$ &  2.8034  & 93.7  & 83.5 \\ 
$Ba_2YbCu_3O_{9-\delta}$ &  2.7952  & 83.5 & 83.5 \\ 
$Ba_2CeCu_3O_{9-\delta}$ &  2.7998  & 90.8  & - \\ 
$Ba_2LuCu_3O_{9-\delta}$ &  2.8045  & 98.3  & - \\ 
$Ba_2TbCu_3O_{9-\delta}$&  2.8019  & 94.0  & - \\ 
$Ba_2LaCu_3O_{9-\delta}$ &  2.79668  & 85.9  & - \\ 
$Ba_2PrCu_3O_{9-\delta}$ &  2.7998  & 90.8 & - \\ 
$Ba_2NdCu_3O_{9-\delta}$&  2.7998 & 90.8  & -\\ 
$Ba_2TmCu_3O_{9-\delta}$ &  2.8045  & 98.3  & - \\ \hline

\end{tabular}
\caption{$H_{31}$ and critical temperature of single-phase $Ba_2RECu_3O_{9-\delta}$. }
\label{tab:Super2}
\end{center}
\end{table}

Rouray and Kumazaki further studied the relationship between the fractal properties of the sintered molecular configuration and the topological index, and obtained encouraging results ~\cite{Rouvray1991}. Perhaps there is a certain connection between fractal and topological properties. A very meaningful subject to be further studied.

\subsection{Other Topological Theories in Chemical Physics}

In addition to the molecular topological index theory introduced above, there are many applications of topological theories in chemical physics. Here we briefly introduce two.

\subsubsection{Topological Theory of Molecular Potential Energy Surface}

Since the 1980s, people have gradually realized that molecular potential energy surfaces, as hypersurfaces in multi-dimensional spaces, have topological properties, and hope that through the study of the topological properties of molecular potential energy surfaces, they can explore a method that can be directly obtained  some important information on the molecular potential energy surface without quantum theoretical calculations. The representative work in this area is a series of work by P. G. Mezey~\cite{Mezey1977,Mezey1980,Mezey1981A,Mezey1981B,Mezey1982,Mezey1983,Mezey1985A,Mezey1985B,Mezey1986,Mezey1988A,Mezey1988B, Mezey1989}. The mathematical tools he uses are mainly topological space theory in topology, fundamental groups and homology groups. The object of processing is the molecular potential energy surface and it is regarded as an n-dimensional space hypersurface with algebraic structure. In addition to gaining a deeper understanding of the important properties of the potential energy surface, his work has also enabled us to have a further topological understanding of traditional chemical concepts such as molecular structure and chemical reaction mechanisms. At the same time, it played a role in the development of computer-aided synthesis design methods.

\subsubsection{Molecular Quantum Topological Theory}

As we all know, molecules are composed of atoms, but how should the exact boundaries between adjacent atoms in a molecule be divided? When a chemical reaction occurs in a molecule, it must be accompanied by the formation or breakage of a chemical bond. If you study the fine process of molecular reaction dynamics, when will the chemical bond be broken? How does a certain atom in a molecule move as a unit during a chemical reaction? What is the law of quantum mechanics? This type of chemical problem has been vague and unresolved for a long time. In recent years, the quantum topology theory proposed and developed by Bader et al. has given clear answers to these questions.

The content of quantum topology theory is to combine topology and quantum mechanics to discuss the structure and reaction of molecules. The mathematical tools of this theory are topology and catastrophe theory, the physical method is quantum mechanics, and the object of study is molecular structure.
And chemical reaction. Therefore, this is a marginal theory combining mathematics, physics and chemistry. This theoretical method is unique and the viewpoint is novel, which is quite eye-catching in theoretical physics and theoretical chemistry.

Quantum topology theory takes the charge density in molecules as the basic research object. The charge density of the molecule, especially the electron density determine the basic properties of molecules. According to the Rohenkerg-Kohn theorem, electron density is the basic physical quantity for understanding the properties of molecules. Quantum topology theory, density functional theory and quantum hydrodynamics together constitute the three main branches of modern electron density theory. In 1975, Bader first used the orthogonal trajectory (OT)  to describe the charge density of molecules~\cite{Bader1975}. Bader and coworkers systematically established quantum topology theory and developed quantum mechanics of subspace and further studied the relationship between the topological properties of the charge density in the molecule and the topological properties of the molecular potential energy surface~\cite{Bader1980,Bader1982A,Bader1982B,Bader1987, Bader1988,Bader2007, Bader2008}.

\section{Logical Topological  Properties and  Characterizations in Layer 2}

In topology and related areas of physics, a logical topological property or topological invariant is a property of topological results obtained via the logical deriving from first physical principals. In this section we will first start with the  molecular topological properties and characterization , second the condense matter topological properties and characterization and  end with topological quantum computing discussions

\subsection{Molecular Topological Properties Characterization}
We discuss our previous ab-initio theoretical studies of molecular and electronic structures of neutral silver bromide clusters ~\cite{Zhang2000B} from the point view of topology in this section.

\subsubsection{Molecular Topology Invariant Characterization and Application}
\label{subsubsec:Molecular Topology Invariant}

\begin{figure}[H]
\begin{center}
\epsfig{file=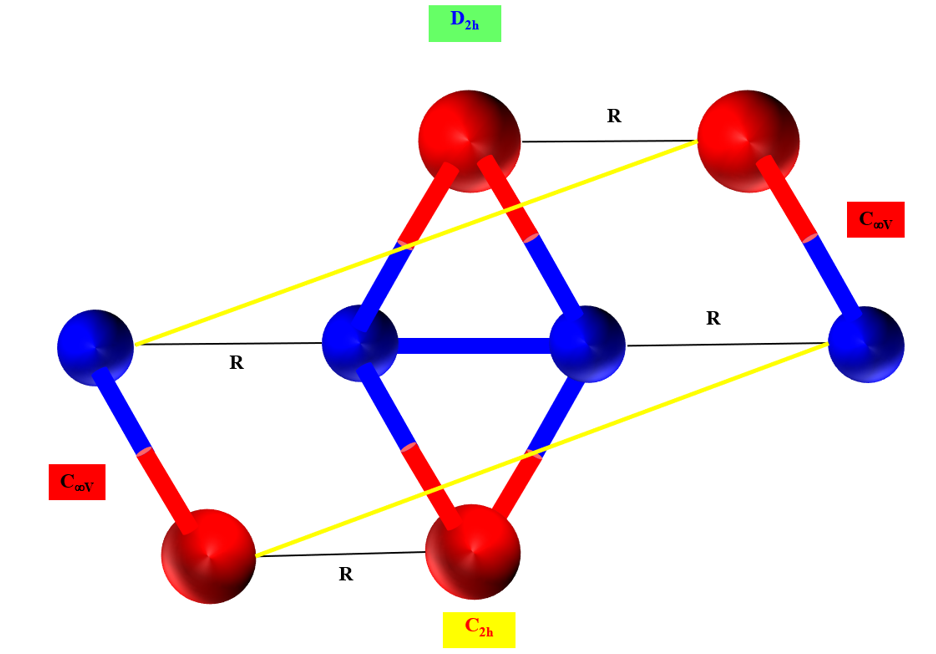,height=2.8in}
\caption{Topological transformation of two $C_{\infty V}$ monomers form $C_{2h}$ Dimer and $D_{2h}$ Dimer as $R$ changed within range $0 \leqslant R \leqslant \infty$ }
\label{fig:Di}
\end{center}
\end{figure}

Figure ~\ref{fig:Di} shows 2  $C_{\infty V}$ monomers can form a $D_{2h}$ dimer along a molecular topology invariant the $C_{2h}$ symmetry dimer ~\cite{Zhang2000B,Zhang2019B,Zhang2019C}. This follows the following physical principal: the stable state with lowest energy level forms along its intrinsic symmetry. Here the stable state is $D_{2h}$ dimer, the intrinsic symmetry is the $C_{2h}$. The topological transformation of two  $C_{\infty V}$ monomers  are  homeomorphic to the  $C_{2h}$ symmetry dimer. There are a lot of applications of the molecular topology invariant, for example, the $C_{2h}$ symmetry dimer here. It is the key to form the stable state  $D_{2h}$ dimer. During the topological continuous deformation transformation, when the symmetry broken  from $C_{2h}$ to $D_{2h}$, catastrophe phenomena happened ~\cite{Zhang1992D} , in the case of silver bromide dimerization,  the silver-silver bonding formed and thus form the silver bromide $D_{2h}$ dimer ~\cite{Zhang2000B,Zhang2019B,Zhang2019C,Zhang2000A,Zhang2000C,Zhang1997A}.  It can also be used to form the molecular orbital energy level correlation diagram.

\begin{figure}[H]
\begin{center}
\epsfig{file=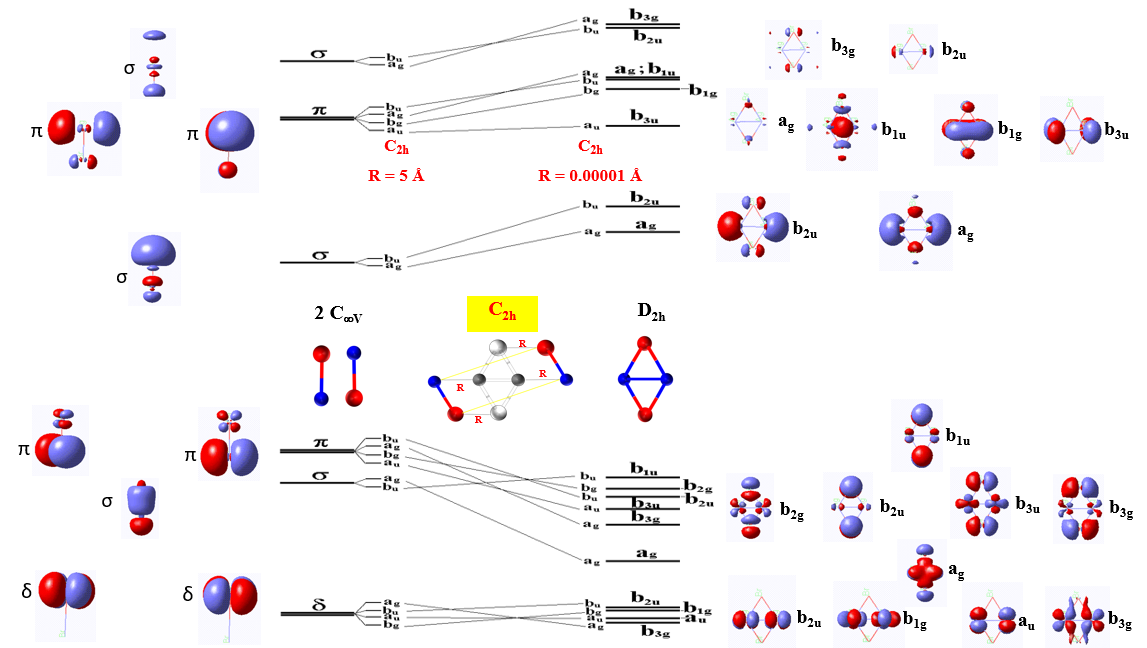,height=3.0in}
\caption{Molecular orbital energy level correlation diagram for dimerization obtained by the topological transformation mapping of two $C_{\infty V}$ monomers symmetry configuration  starting from $C_{2h}$ dimer symmetry configuration to $D_{2h}$ dimer symmetry configuration as $R$ changed from $R = \infty$ to $R = 0$ . Here ab initial calculation show that when $R= 5 \AA $, the  $C_{2h}$ dimer energy level already match with two $C_{\infty V}$ monomers energy level at $R = \infty$,  when $R=0.00001 \AA $,  the  $C_{2h}$ dimer energy level already match with  $D_{2h}$  dimer energy level at $R=0$ } 
\label{fig:Di-ec}
\end{center}
\end{figure}

Figure ~\ref{fig:Di-ec} shows molecular orbital energy level correlation diagram for dimerization. It was obtained via the ab initial calculation of the orbital energy level of molecular topology invariant $C_{2h}$ symmetry dimer at different R value ~\cite{Zhang2000B,Zhang2000C}. For example, when $R = \infty$, the energy level of $C_{2h}$ symmetry dimer should equal to the sum of two $C_{\infty V}$ monomer energy level. This is exactly what we see in our calculation. All the molecular orbitals  and orbital energy level of the two $C_{\infty V}$ monomers are listed in the left, their symmetry with energy level from low to high $\delta, \sigma, \pi, \sigma, \pi, \sigma $ are marked at the left of the orbitals  and the top of each orbital energy level.   All the molecular orbitals and orbital energy level of  $D_{2h}$ dimer  are listed in the right, their symmetry with energy level from low to high $b_{3g}, a_u, b_{1g},b_{2u}, a_g,b_{3g}, b_{3u},b_{2u}, b_{2g}, b_{1u}, a_g, b_{2u}, b_{3u}, b_{1g}, b_{1u}, a_g, b_{2u}, b_{3g}$ are marked at the right of each orbitals  and the top of each orbital energy level. In the middle, two sets of orbitals symmetry and orbital energy level of $C_{2h}$  dimer are listed. The first sets value are calculated for  $C_{2h}$  dimer at $R= 5 \AA $, it was found that the $C_{2h}$  dimer energy levels are close enough to the sum of the energy levels of two $C_{\infty V}$ monomers. Short correlation lines are used to connect between the two molecular orbital energy levels of sum of two $C_{\infty V}$ monomers and $C_{2h}$  dimer at $R= 5 \AA $. The second sets value are calculated for  $C_{2h}$  dimer at  $R=0.00001 \AA $, it was found that the $C_{2h}$  dimer energy levels are close enough to the molecular orbital energy levels of  $D_{2h}$ dimer. Long correlation lines are used to connect between the two sets of molecular orbital energy levels of  $C_{2h}$  dimer at $R= 5 \AA $ and $R=0.00001 \AA $. They are obtained via the serious ab initial calculation of molecular orbital energy levels of  $C_{2h}$  dimer from $R= 5 \AA $ to $R=0.00001 \AA $. The molecular orbital energy level change from low to high and high to low with the same symmetry can be clear traced  via this kind of serious ab initial calculation. Also in this way, we successfully set up  the  molecular orbital energy level correlation between $D_{2h}$ dimer and two $C_{\infty V}$ monomers . In our previous work~\cite{Zhang2000B,Zhang2000C}, this was called the monomer $C_{\infty V}$ and the dimer  $D_{2h}$ share a common symmetry of  $C_{2h}$. In this work, from the topology point of view, it is clear that $C_{2h}$ dimer is actually a  topology invariant during the dimerization. It is the key to find the stable $D_{2h}$ dimer and the key for the simplification of quantum computing for finding the stable dimer states as discussed in section ~\ref{subsubsec:Invariant} and will also be discussed in section ~\ref{subsec:Topological Quantum Computation}. This is the same for trimerization and tetramerization. $C_{3h}$ trimer is the  topology invariant during the trimerization, $D_{2d}$ dimer is the  topology invariant during the tetramerization as shown in Figure ~\ref{fig:dtt-e}. 

\subsubsection{Molecular Topology Class Characterization and Application}
\begin{figure}[H]
\begin{center}
\epsfig{file=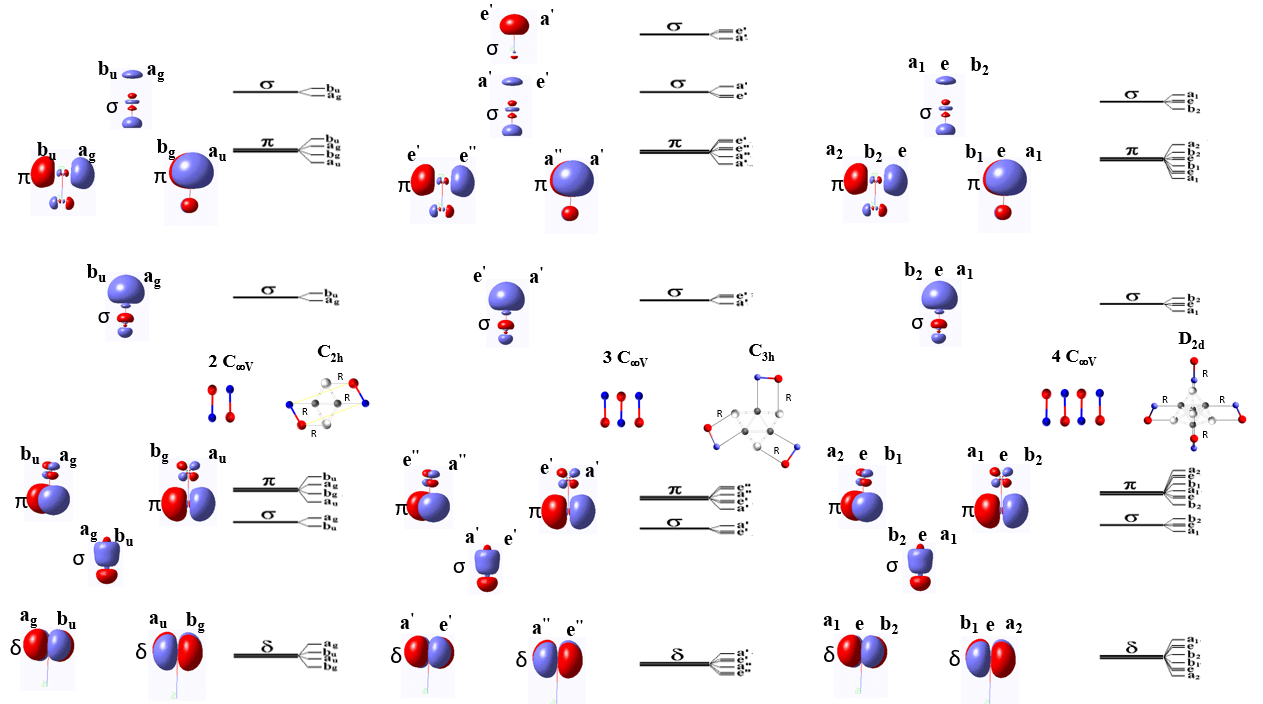,height=3.0in}
\caption{Topology class hind inside monomer molecular orbital energy level can be distinguished by the number and the symmetry of the monomer molecular clusters} 
\label{fig:dtt-e}
\end{center}
\end{figure}

In previous section, we learn that topology invariant symmetry is the key to find the stable molecular structure and the key for the simplification of quantum computing for finding the stable structure. The topology invariant symmetry is defined as the symmetry that keeps unchanged under the continue deformation -topological transformation. In a given system,  there may exist many these kinds of topology invariant symmetry, we call it topology class of symmetry in the system. Figure ~\ref{fig:dtt-e} shows the example of the topology class of symmetry that can exist in a four monomers system that far away from each other. Let us focus on the change of the monomer molecular orbital energy levels of each topology class. To be simple, we can consider the lowest energy level with $ \sigma  $ symmetry. If the four monomers get close, they collapse to two $C_{2h}$ dimers, the $ \sigma  $ symmetry will split to two with $a_g, b_u$ symmetry;If the four monomers get close, they collapse to one $C_{\infty V}$ monomer and one $C_{3h}$ trimer, the $ \sigma  $ symmetry will keep unchanged for the monomer and split to three with $a', e'$ symmetry, the $ e'$ is energy degenerate state with two levels;If the four monomers get close, they collapse to one $D_{2d}$ tetramer, the $ \sigma $ symmetry will split to four with $a_1, e, b_2$ symmetry, the $ e$ is energy degenerate state with two levels. Now let's increase the monomer number from four to infinity, you can image 
that when there infinity monomers get close to form a solid, the $ \sigma $ symmetry will split to infinity to form a energy band with certain symmetry depending on the symmetry of the solid that formed. It has been conjectured that the energy levels of generic integrable quantum systems have the same statistical properties as random numbers from a Poisson process~\cite{Bogomolny}. Statistical properties of eigenvalues of energies of quantum systems are described by standard random matrix ensembles depending only on system symmetries. The zeros of the Riemann zeta-function seem to be distributed on the 1/2-line like the eigenvalues of large random hermitian matrices.  Dyson noticed the same equations pop up in a quantum field to do with the distribution of prime numbers~\cite{Darling}. Physicists benefit because energy levels are usually hard to compute, whereas mathematicians have very efficient methods for calculating zeros of the zeta function~\cite{Wells}. There also has been some speculation about a relationship between the Lee–Yang theorem~\cite{Yang1952,Yang1952A} and the Riemann hypothesis about the Riemann zeta function~\cite{Knauf}. Why primes lie at such a delicate point of balance between randomness and order? Prime numbers also seem to be fundamental to practically all of mathematics and to the physical universe itself. Zhang announced a proof that states there are infinitely many pairs of prime numbers that differ by 70 million or less~\cite{Yitang}. This result implies the existence of an infinitely repeatable prime 2-tuple, thus establishing a theorem akin to the twin prime conjecture.

\subsubsection{Molecular Topology Probability Characterization and Application}
\begin{figure}[H]
\begin{center}
\epsfig{file=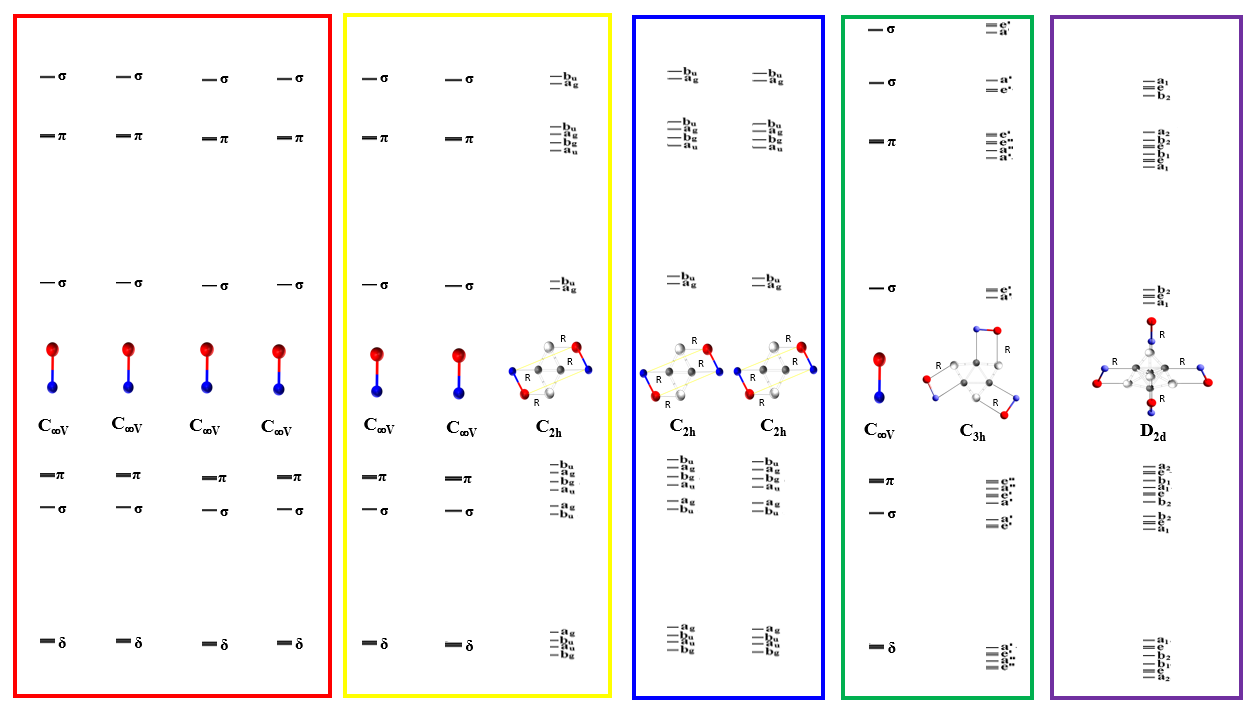,height=3.0in}
\caption{Molecular orbital energy level of the least five possible molecular topology configuration of the four molecular monomers at  $R = \infty$ }
\label{fig:trl}
\end{center}
\end{figure}

\begin{figure}[H]
\begin{center}
\epsfig{file=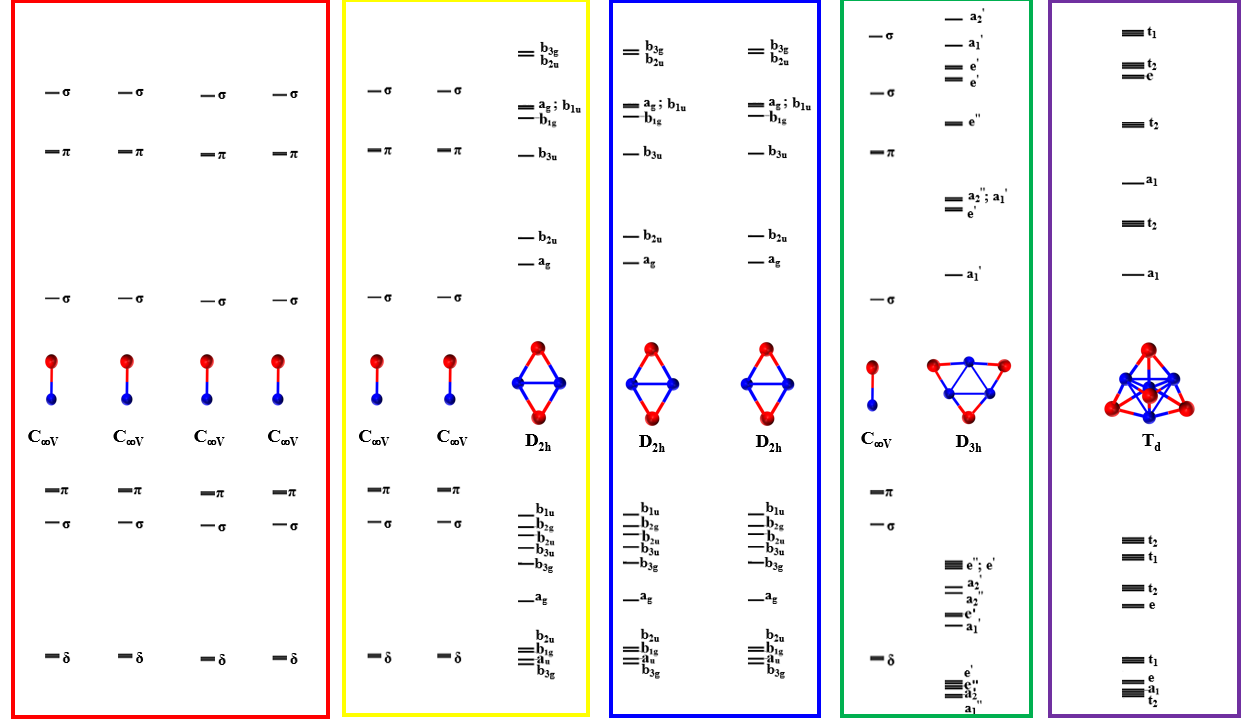,height=3.0in}
\caption{Molecular orbital energy level of the least five possible molecular topology configuration of the four molecular monomers at $R = 0$ } 
\label{fig:tr0}
\end{center}
\end{figure}

A natural question for a topological system with many topological symmetry classes is what's the probability for the  topological system collapse to a specific symmetry topology class? To be simple,a four monomers system again to be considered. There at least five possible molecular topology configurations as shown in Figure~\ref{fig:trl} at $R = \infty$. As discussed in the previous molecular topology invariant section, dependent on the environment, when $R$ change from $R = \infty$ to $R = 0$, the system can collapse to the five possible molecular symmetry topology configuration as shown in Figure~\ref{fig:tr0}. Now let's increase the monomer number from four to infinity, you can image that when there infinity monomers get close, the number of the symmetry topology class formed will be extreme large and the calculation of the probability to form each specific symmetry topology class will be extreme complicated. Symmetries arising from free probability theory~\cite{Voiculescu, Voiculescu1985} may be used to calculate  the number of the symmetry topology class formed and compare with the experiment results.  From our previous experiments, we know that the probability to detect the monomer ($p_m$), dimer ($p_d$), trimer($p_t$) and tetramer ($p_{te}$) etc  will follow the following order:

\begin{equation} 
p_m > p_d > p_t > p_{te}
\label{equ:prob}
\end{equation}

The ratio of the probability to detect the monomer ($p_m$), dimer ($p_d$), trimer($p_t$) and tetramer ($p_{te}$) etc can also be measured through the measurement of the intensity of DLD-TOF-MS of each cluster~\cite{Zhang2019B,Zhang2019C}.

\subsection{Topological Quantum Matter}

Topological quantum matter was first encountered around 1980, with the experimental discovery of the integer ~\cite{Klitzing} and later fractional~\cite{Tsui1982} quantum Hall effects (QHE) in the two dimensional electron systems in semiconductor devices, and the theoretical discovery of the entangled gapped quantum spin-liquid state of integer-spin ~\cite{Haldane1983}, which was later experimentally confirmed~\cite{Buyers}in crystals of the organic chain molecule. The common feature of these discoveries was their unexpectedness and follow the classification work of Xiao-Gang Wen~\cite{Wen,Wen1989,Wen1990,Wen2006,Wen2013}that their common feature is that they involve topologically nontrivial entangled states of matter that are fundamentally different from the previously known topologically trivial states, and this lies at the heart of their unexpected properties~\cite{Haldane}. Topological materials have become the focus of intense research in recent years, since they exhibit fundamentally new physical phenomena with potential applications for novel devices and quantum information technology. One of the hallmarks of topological materials is the existence of protected
gapless surface states, which arise due to a nontrivial topology of the bulk wave functions~\cite{Chiu,Qi}.

\subsubsection{Topological Order}
\label{subsubsec:Topological Order}

Understanding phases of matter is one of the central issues in condensed-matter physics. For a long time, we believed that all the phases and phases transitions were described by Landau symmetry breaking theory~\cite{Landau}. In 1989, it was realized that many quantum phases can contain new kinds of orders which are beyond the Landau symmetry breaking theory~\cite{Wen1989}. A quantitative theory of the new orders was developed based on robust ground-state degeneracy and the robust non-Abelian Berry’s phases of the degenerate ground states, which can be viewed as new topological non-local order parameters~\cite{Wen,Wen1990}.  The new orders were named topological order. It is a kind of order in the zero-temperature phase of matter or quantum matter. Topologically ordered states have some interesting properties, such as (1) topological degeneracy and fractional statistics or non-abelian statistics that can be used to realize a topological quantum computer; (2) perfect conducting edge states that may have important device applications; (3) emergent gauge field and Fermi statistics that suggest a quantum information origin of elementary particles (4) topological entanglement entropy that reveals the entanglement origin of topological orders. A short-range entangled quantum state can be connected to a real space direct product state through smooth local deformations. By contrast, the long range entangled states do not have this property. Note that this behavior has no relation with the correlation length of local operators. In a study of quantum matter, long-range entanglement can give rise to many new quantum phases. So long-range entanglement is a natural phenomenon that can happen in our world. They greatly expand our understanding of possible quantum phases and bring the research of quantum matter to a whole new level. To gain a systematic understanding of new quantum phases and long-range entanglement, we would like to know, what mathematical language should we use to describe long-range entanglement? Early studies suggest that tensor category and group cohomology should be a part of the mathematical framework that describes long-range entanglement~\cite{Wen2013A}. Further progresses in this direction will lead to a comprehensive understanding of long-range entanglement and topological quantum matter. However, what is really exciting in the study of quantum matter is that it might lead to a whole new point of view of our world. This is because long-range entanglement can give rise to both gauge interactions and Fermi statistics. In contrast, the geometric point of view can only lead to gauge interactions. Maybe we should use entanglement pictures to understand our world. This way, we can get both gauge interactions and fermions from a single origin- qubits. We may live in a truly quantum world and the universes communication among the billions of galaxies and planets can be built up through the long-range entanglement. So, quantum entanglement represents a new chapter in physics. To understand the meaning of long-range entanglement, imagine a quantum computer which applies a sequence of geometrically local operations to an input quantum state, producing an output product state which is completely disentangled. If the time required to complete this disentangling computation is independent of the size of the system, then we say the input state is short-range entangled; otherwise, it is long-range entangled.  Wen et al ~\cite{Wen2004,Wen2019} review the development of physical matters in four revolutions: (1) all matter is formed by particles; (2) the discovery of wave-like matter; (3) particle-like matter = wave-like matter. (4) matter and space =information (qubits), where qubits emerge as the origin of everything. In fact, it implies that information is matter, and matter is information.The energy-frequency relation $E = h \nu$ implies that matter = information. It is the richness of the orders that give rise to the richness of material world. The physical properties of a many-body state mainly come from the organization or the order of the degrees of freedom in the state.  As a result, liquids can only resist compression and have only compression wave. Since shear deformations do not cost any energy for liquids, liquids can flow freely. We see that the properties of the propagating wave are entirely determined by how the atoms are organized in the materials. Different organizations lead to different kinds of waves and different kinds of mechanical laws. To summarize, topological order and long-range entanglement give rise to new states of quantum matter. Topological order, or more generally, quantum order have many new emergent phenomena, such as emergent gauge theory, fractional charge, fractional statistics, non-abelian statistics, perfect conducting boundary, etc.

\subsubsection{Topological Materials}
\label{subsubsec:Topological Materials}

\begin{figure}[H]
\begin{center}
\epsfig{file=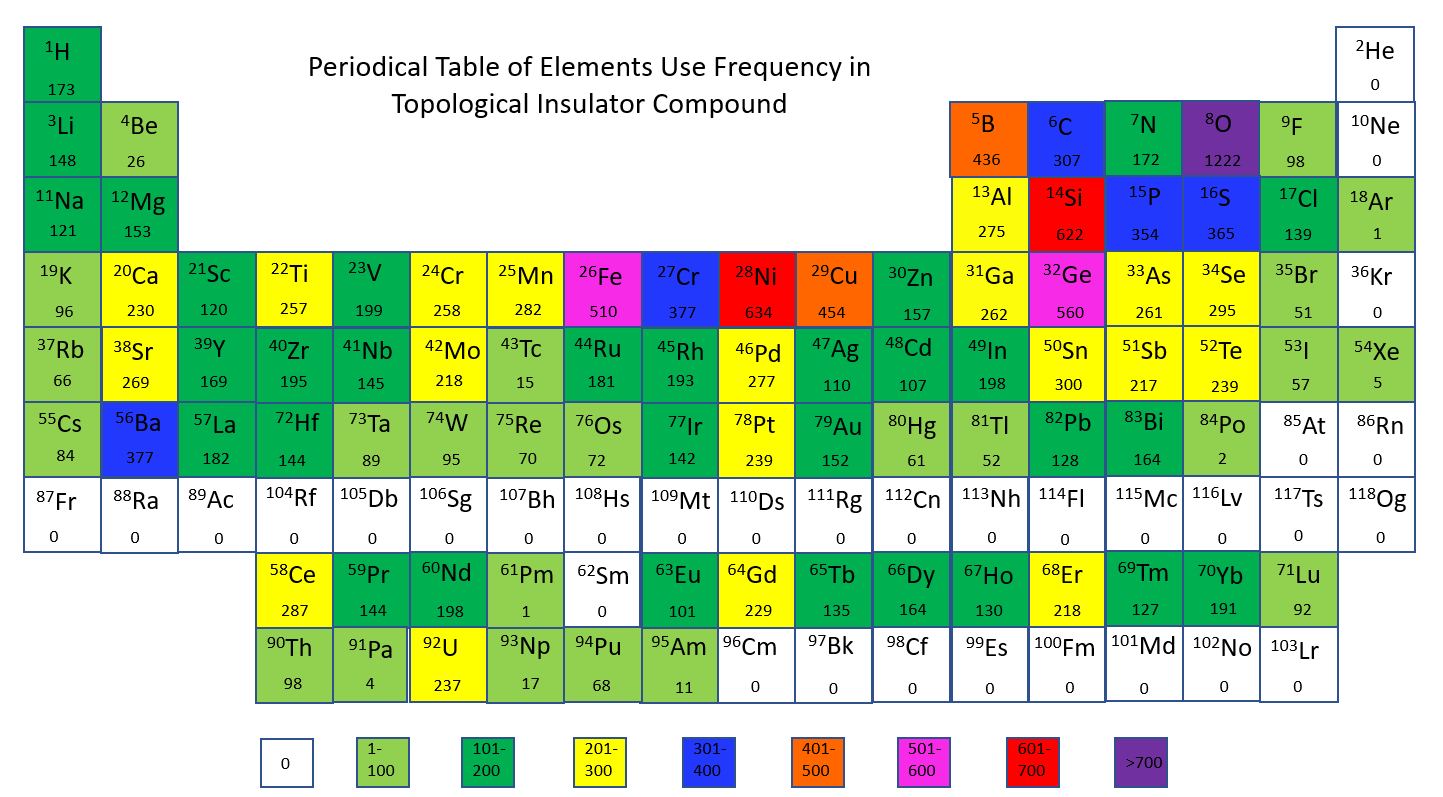,height=3.0in}
\caption{Color map of elements use frequency in topological insulator compounds}
\label{fig:pt}
\end{center}
\end{figure}

\begin{figure}[H]
\begin{center}
\epsfig{file=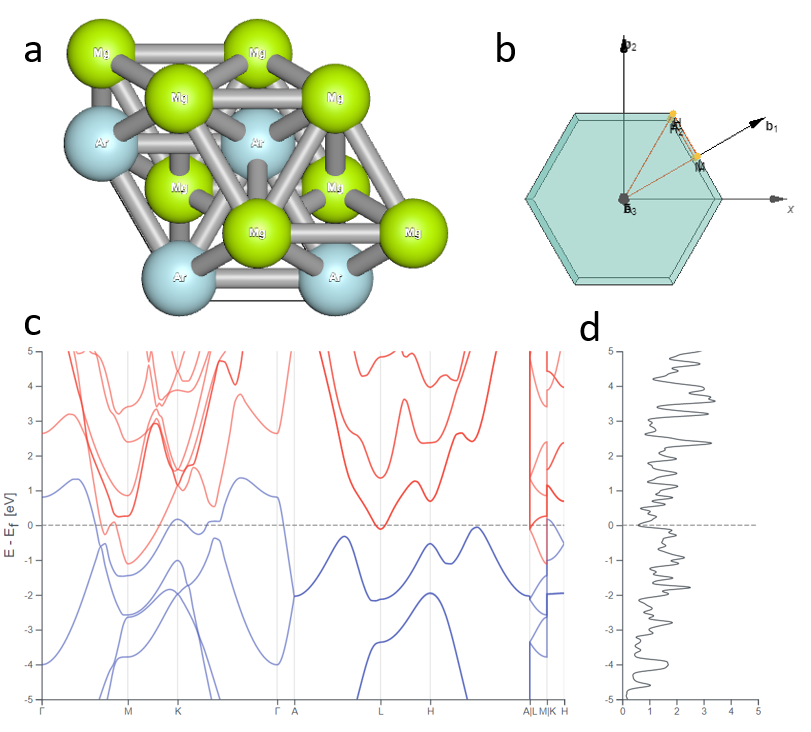,height=3.0in}
\caption{Lattice structure (a), brillouin zone(b), band structure(c) and density states(d) of $ Ar1Mg2$ TI compound}
\label{fig:arti}
\end{center}
\end{figure}

Topological insulator is an example of symmetry protected topological (SPT) phases which are protected by $U(1)$ and time reversal symmetries. Interacting bosonic topological superconductors is another example of SPT phases which is protected by time-reversal symmetry only~\cite{Wen2013}. Topological insulators are new states of quantum matter which cannot be adiabatically connected to conventional insulators and semiconductors. They are characterized by a full insulating gap in the bulk and gapless edge or surface states which are protected by time-reversal symmetry. Zhang Shoucheng was one of the founders of the field of topological insulators. He made one of the first theoretical proposals of the quantum spin Hall effect. Soon after the initial theoretical proposal, his group theoretically predicted the first realistic quantum spin Hall material in HgTe quantum wells~\cite{ShouCheng2006}. This prediction was soon confirmed experimentally~\cite{Konig}, launching worldwide research activities. Subsequently, his group predicted numerous novel topological states of matter and topological effects, including the Bi2Se3 family of 3D topological insulators~\cite{ShouCheng2009}, the topological magneto-electric effect~\cite{ShouCheng2008}, the quantum anomalous Hall effect in magnetic topological insulators, time-reversal invariant topological superconductors, and the realization of a chiral topological superconductor and of chiral Majorana fermions using the quantum anomalous Hall state in proximity with a superconductor. Most of these predicted properties have now been experimentally observed. Now all these topological materials not only have been theoretically predicted and experimentally observed in a variety of systems but also all known inorganic compounds now have been categorized using a single electron approach to trivial and topological materials~\cite{Kumar,Bradlyn, Fang, Vergniory,Tang} and published on the Web~\cite{Web}. All scientists can now search for new topological compounds on these Web pages. Figure~\ref{fig:pt} show the frequency periodic table of elements that can be the component of topological insulators (TIs)  based on current 6109 TIs that published on the Web. From the table, we can see that oxygen is the most frequently used component to form the TIs. It start from the two components compound $ B2O1$ with symmetry group $164(P-3m1)$ and topological indices $ Z_{2W,1}=0,Z_{2W,2}=0, Z_{2W,3}=1,Z_4=2$ to the six components compound $ H4Al2K1Ni1O14P3$ with symmetry group $15(C2/c)$ and topological indices $ Z_{2W,1}=0,Z_{2W,2}=0, Z_{2W,3}=0,Z_4=1$ total 1222 TI compounds~\cite{Web}.  Study the oxide TI compounds electrical, optical, magnetic, thermal and mechanical properties will definitely be a focus area for the future and their wide applications in the future devices development can be expected. On the other hand, only one argon TI compound be predicted as shown in the table. It is $ Ar1Mg2$ with symmetry group $194(P6_3/mmc)$ and topological indices $ Z_{2W,1}=0,Z_{2W,2}=0, Z_{2W,3}=0,Z_4=1,Z_{6m0}=5,Z_{12}'=5$. The lattice structure, brillouin zone, band structure and density states has been shown in Figure~\ref{fig:arti}. So far, there no experiment report of this TI compound be noticed. The study of electrical, optical, magnetic, thermal and mechanical properties of this TI compound also be expected to have wide application in the future devices development.

\begin{figure}[H]
\begin{center}
\epsfig{file=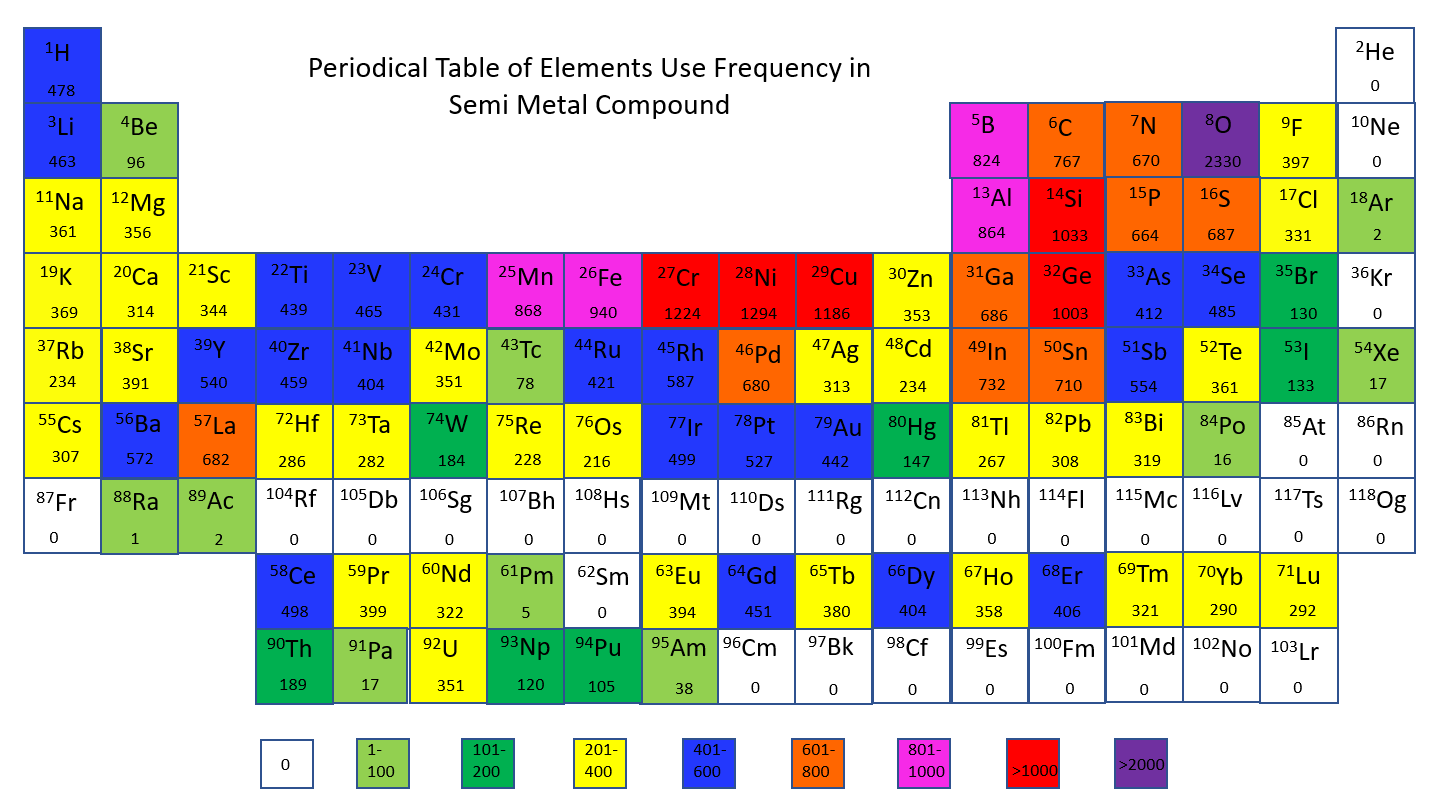,height=3.0in}
\caption{Color map of elements use frequency in semi metal compounds}
\label{fig:psm}
\end{center}
\end{figure}

Topological metals (TMs) are a kind of special metallic materials, which feature nontrivial band crossings near the Fermi energy, giving rise to peculiar quasiparticle excitations. TMs can be classified based on the characteristics of these band crossings. For example, according to the dimensionality of the crossing, TMs can be classified into nodal-point, nodal-line, and nodal-surface metals. Another important property is the type of dispersion. According to degree of the tilt of the local dispersion around the crossing, we have type-I and type-II dispersions. This leads to significant distinctions in the physical properties of the materials, owing to their contrasting Fermi surface topologies~\cite{Li2020, Hasan2015A, Hasan2015B, Hasan2017}. It has been well established that the known TMs can be classified by the dimensionality of the topologically protected band degeneracies. While Weyl and Dirac Semi Metal (SM) feature zero-dimensional points, the band crossing of nodal-line SMs forms a one-dimensional closed loop~\cite{Hasan2017,Armitage, Bansil}. Figure~\ref{fig:psm} show the frequency periodic table of elements that can be the component of SMs based on current 13895 SMs that published on the Web~\cite{Web}. From the table, we can see that oxygen again is the most frequently used component to form the SMs. It start from the two components compound $ K1O2$ with symmetry group $15(C2/c)$ to the six components compound $ H28Al1Cl1Cu1O22S2$ with symmetry group $2(P-1)$ total 2330 SM compounds~\cite{Web}. Through the compare studies of Figure~\ref{fig:pt} and Figure~\ref{fig:psm},we notice some interesting facts: 1) oxygen is most frequently used component to form both TIs and SMs. The next two elements are Silicon and Nickel, but their use frequency only half of the oxygen. 2)For the Alkali metal group, the Halogens group and the Oxygen group elements, the smaller the atomic number, the higher frequency the elements be used to form  both TIs and SMs. 3)For the Noble gases group only Argon and Xenon be used to form  both TIs and SMs.

Topological superconductors have a full pairing gap in the bulk and gapless surface states consisting of Majorana fermions. The theory of topological superconductors can be reviewed in close analogy to the theory of topological insulators~\cite{Qi}because the Bogoliubov–de Gennes (BdG) Hamiltonian for the quasiparticles of a superconductor is analogous to the Hamiltonian of a band insulator, with the superconducting gap corresponding to the band gap of the insulator. TR breaking superconductors are classified by an integer~\cite{Volovik1988, Read}, similar to quantum Hall insulators~\cite{Thouless}, while TR invariant superconductors are classified~\cite{Schnyder, Qi2009} by a $Z_2$ invariant in 1D and 2D, but by an integer $Z$ invariant in 3D~\cite{Schnyder,Imura}. Topological invariants count the number of protected subgap states, either bound to a defect or propagating along a boundary. In a superconductor, these are Majorana fermions, described by a real rather than a complex wave function~\cite{Majorana}. The absence of complex phase factors fundamentally modifies the random-matrix description, notably, the scattering matrix at the Fermi level is real orthogonal rather than complex unitary. Experimentally, the search for Majorana zero modes (bound to a vortex core or to the end of a superconducting wire) and Majorana edge modes (propagating along the boundary of a two-dimensional superconductor) is still at an initial stage~\cite{Wilczek2012, Alicea2013}. But much is understood from the theoretical point of view.  Some of these topological superconductors and insulators have been realized in the laboratory, many others are being searched for~\cite{Alicea2012, Beenakker,Chung2011,Stanescu2013}. Besides the TR invariant topological superconductors, the TR breaking topological superconductors have also attracted a lot of interest recently, because of their relation with non- Abelian statistics and their potential application to topological quantum computation~\cite{Ivanov,Jiang2015}. 

\subsubsection{Topological Quantum Control}
\label{subsubsec:Topological Quantum Control}

\begin{figure}[H]
\begin{center}
\epsfig{file=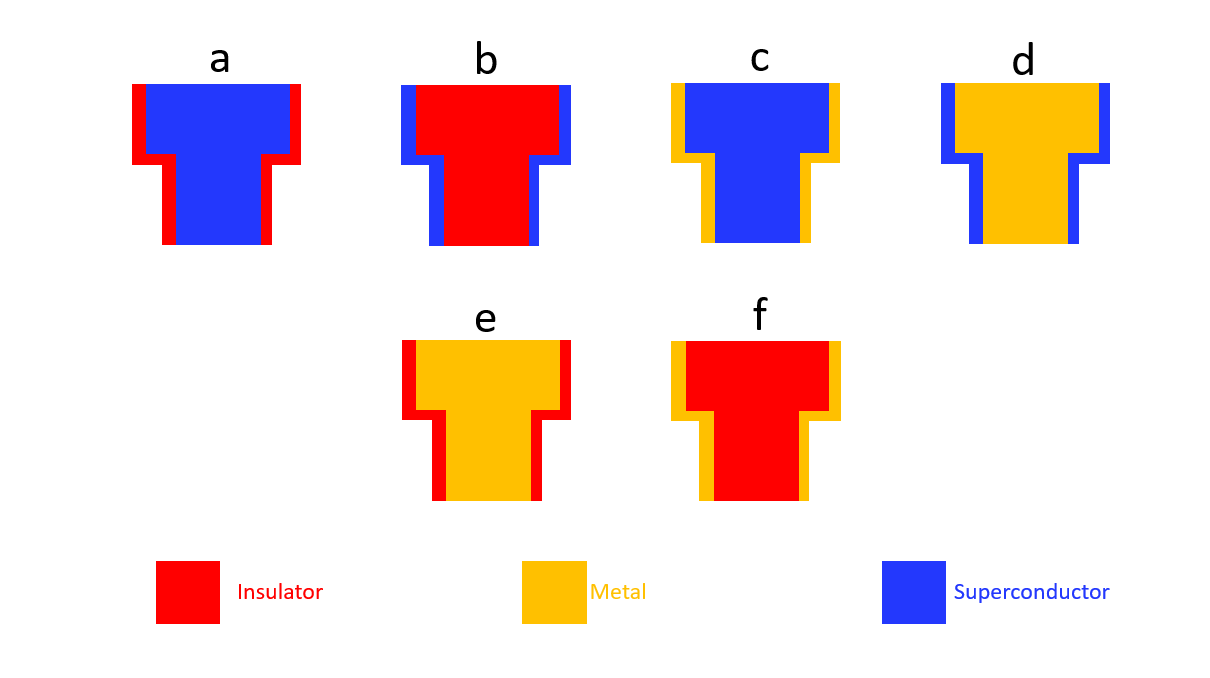,height=3.0in}
\caption{Topological classification of dual damascene interconnect structures with topological insulator, metal and superconductor materials}
\label{fig:tda}
\end{center}
\end{figure}

Microelectronics, which is the basis of information technology, will reach its physical limit, which poses a severe challenge to the development of information technology. Humans must seek new ways. While some of these most impactful technologies of the 20th century, such as the transistor and the laser, rely on quantum physics, they do not use the most extreme kinds of quantum phenomena. The ability to harness effects like quantum superposition and entanglement will usher in a new generation of transformative technologies.  And new information methods based on this strong quantum effects have emerged~\cite{Zagoskin}. This is quantum control. Open up a new way to explore new quantum phenomena, construct future information technology to achieve a theoretical breakthrough in a strong correlation system, establish a new quantum information theory and technology system, and surpass the current in terms of information volume, transmission speed, communication security and information functions; First, breakthroughs have been made in the theoretical research of strong electronic correlation systems such as high-temperature superconductors, giant magnetoresistance, ferroelectrics, etc., which will drive the development and technical realization of related basic disciplines. Quantum control is based on the strong correlation, phase coherence, and entanglement of multi-particle systems described by quantum mechanics. Through the external field modulation system structure and carrier behavior, it is the basis for the further development of quantum information processing, quantum computing and quantum communication.  The object of quantum control has evolved from a single control of a simple system to a comprehensive control of complex systems based on electrons, photons, phonons, spins, orbital electrons, quantum states, etc. Control methods have evolved from simple and low-level classic control methods widely used at present to comprehensive and deep-level advanced quantum control methods for electricity, magnetism, light, and phonons. Regulatory response refers to a variety of under the action of the external field, it produces a comprehensive effect of multiple flows, which is called a huge response effect and shows a very strong information processing capability. Eventually, a new generation of technology based on quantum control will be created. It is expected that this technology will have features that are not available in modern technologies such as good coherence~\cite{Zagoskin}, quantum sensing~\cite{Yuhua,Degen}, integrated control, fast operation, functional integration, and low power consumption. The implications of the landscape topology for practical quantum control efforts are called topological quantum control. Topologically ordered systems are intrinsically robust against local sources of decoherence, and therefore hold promise for quantum information processing~\cite{Hsieh,Jiang2011,Jiang2013}. Constructing a future device based on this idea may completely get rid of the existing known framework and construct a brand new device, and its function will be amazing~\cite{Zhang2020A,Zhang2019B,Zhang2019C}. Figure~\ref{fig:tda} shows all possible topological classification of dual damascene interconnect structures with topological insulator, metal and superconductor materials that can be used to construct the future devices connection. The topologically insulating circuit of capacitors, inductors and other devices are also proposed and tested~\cite{Jia,Jiang2015A}.

\subsection{Topological Quantum Computation}
\label{subsec:Topological Quantum Computation}

Quantum computing is the use of quantum phenomena such as superposition and entanglement to perform computation. Computers that perform quantum computations are known as quantum computers~\cite{Chuang,Kitaev2002,Preskill,Everitt,Imre}. Quantum computers are believed to be able to solve certain computational problems, such as integer factorization, substantially faster than classical computers. Quantum computing is one of the important branch of quantum information~\cite{Chuang,Wilde} that will lead the future technology development for universes communication among the billions of galaxies and planets we discussed before~\cite{Zhang2020A}. Technologies currently involved in quantum computing development mainly include (1) Photonic quantum computing~\cite{Kumar2005,Brien,Guangcan,Jianwei}, for these applications to attain their full potential, various photonic technologies are needed, including high fidelity sources of single and entangled photons, and high efficiency photon-counting detectors, both at visible and telecommunication wavelengths. (2)ion trapped quantum computing~\cite{Blinov,Blatt,Kielpinski, Haffner}, experimental approaches in ion trap quantum computing can be divided by the type of qubit, in terms of the qubit level energy splitting and the couplings required to drive quantum logic gates between the qubit states. The two primary types of trapped ion qubit architectures are optical qubits derived from a ground state and an excited metastable state separated by an optical frequency, and hyperfine qubits derived from electronic ground-state hyperfine (HF) levels separated by a microwave frequency. (3)superconductor quantum computing ~\cite{Devoret,Gambetta,Kelly, Chow},here, qubits are constructed from collective electrodynamic modes of macroscopic electrical elements, rather than microscopic degrees of freedom. An advantage of this approach is that these qubits have intrinsically large electromagnetic cross-sections, which implies they may be easily coupled together in complex topologies via simple linear electrical elements like capacitors, inductors, and transmission lines. (4)quantum annealing computing~\cite{McGeoch,Pudenz}, adiabatic quantum computation (AQC) is an alternative to the better-known gate model of quantum computation. The two models are polynomially equivalent, but otherwise quite dissimilar: one property that distinguishes AQC from the gate model is its analog nature. Quantum annealing (QA) describes a type of heuristic search algorithm for topological invariants described in sections~\ref{subsubsec:Invariant} and ~\ref{subsubsec:Molecular Topology Invariant} can be implemented to run in the “native instruction set” of an AQC platform. (5) silicon CMOS quantum computing~\cite{Charbon,Charbon2017, Sebastiano, Charbon2020}, a quantum computer comprises a cryogenic quantum processor and a classical electronic controller. When scaling up the cryogenic quantum processor to at least a few thousands, and possibly millions, of qubits required for any practical quantum algorithm, cryogenic CMOS (cryo-CMOS) electronics is required to allow feasible and compact interconnections between the controller and the quantum processor. Cryo-CMOS leverages the CMOS fabrication infrastructure while exploiting the continuous improvement of performance and miniaturization guaranteed by Moore’s law, in order to enable the fabrication of a cost-effective practical quantum computer. Designing cryo-CMOS integrated circuits requires a new set of CMOS device models, their embedding in design and verification tools, and the possibility to co-simulate the cryo-CMOS/quantum processor architecture for full-system optimization. Fault-tolerant quantum bits in surface code configurations, one of the most accepted implementations in quantum computing, operate in deep sub-Kelvin regime and require scalable classical control circuits. State-of-the-art quantum processors consist of arrays of qubits operating at a very low base temperature, typically a few tens of $mK$. The qubit states degrade naturally after a certain time, upon loss of quantum coherence. For proper operation, an error-correcting loop must be implemented by a classical controller, which, in addition of handling execution of a quantum algorithm, reads the qubit state and performs the required corrections.  (6) Spin-based quantum dots quantum computing~\cite{Eriksson,Engel,Kloeffel, Veldhorst}, essential components of spin-based quantum dot quantum computing include i) spin qubits in single electron dots, ii) qubit initialization by thermalization in a magnetic field, iii) qubit rotations performed using electron spin resonance, iv) two-qubit gates enabled by electrostatic control of exchange coupling in neighboring dots, and v) readout by spin-charge transduction. Subsequent theoretical work has shown that two-qubit gates can be sufficiently fast (sub-nanosecond) and that these same interactions can be harnessed for single-qubit rotations, albeit with some encoding overhead. The most challenging aspect of scalable spin-based quantum dot quantum computing is fast readout: spin-dependent tunneling schemes have been proposed as well as microwave-enabled, fast initialization and readout in a closed dot.(7) silicon and diamond defect spin quantum computing~\cite{Tucker,Wrachtrup, Weber,Itoh},  phosphorous donors in silicon present a unique opportunity for solid-state quantum computation. Electrons spins on isolated $Si:P$ donors have very long decoherence times of $~60 ms $  in isotopically purified $^{28}Si$ at 7{K}. The $Si:P$ donor is a self-confined, perfectly uniform single-electron quantum dot with a non-degenerate ground state. A strong Coulomb potential breaks the 6-valley degeneracy of the silicon conduction band near the donor site, yielding a substantial energy gap of $~15 meV$ to the lowest excited state as required for quantum computation.  To realize a silicon quantum computer, $P$ donors must be integrated with single-electron transistors to perform spin-to-charge state readout and top-gate arrays for accurate control of bound electron wavefunctions and nearest-neighbor exchange. Diamond crystals composed of nuclear-spin-free stable isotopes $^{12}C$ are considered to be ideal host matrixes to place spin qubits for quantum-computing and -sensing applications, because their coherent properties are not disrupted thanks to the absence of host nuclear spins. The nitrogen-vacancy (NV) center in diamond was identified as a single qubit that could be operational at room temperature. This discovery was followed immediately by studies to use NV qubits for quantum computing. (8)nuclear magnetic resonance quantum computing~\cite{Ramanathan, Vandersypen, Jones} , nuclear spins feature prominently in most condensed matter proposals for quantum computing either directly being used as computational or storage qubits, or being important sources of decoherence. the coherent control of nuclear spins has a long and successful history driven in large part by the development of nuclear magnetic resonance (NMR) techniques in biology, chemistry, physics, and medicine.The central feature of NMR that makes it amenable to quantum information processing (QIP) experiments is that, in general, the spin degrees of freedom are separable from the other degrees of freedom in the systems studied,both in the liquid and solid state. We can therefore describe the Hamiltonian of the spin system quite accurately; there is an extensive literature on methods to control nuclear spins, and the hardware to implement such control is quite precise. This readily accessible control of nuclear spins has led to liquid state NMR being used as a testbed for QIP, as well as to preliminary studies of potentially scalable approaches to QIP based on extensions of solid state NMR. The liquid state NMR QIP testbed, although it is not scalable, has permitted studies of control and QIP in Hilbert spaces larger than are presently available with other modalities, and has helped to provide concrete examples of QIP. (9) neutral atom quantum computing~\cite{Enk,lessen, Henriet}, proposals to use neutral atoms as the building blocks of a quantum computer followed closely after the first demonstration of quantum logic in ion traps. Laser cooling of ions and neutrals was initially developed as an enabling technology for precision metrology. Both systems were known to have long coherence times but also complementary features that lead to radically different approaches to, e.g. atomic clock design. Because ions are charged they can be tightly confined in deep traps and observed for very long times, but the strong Coulomb repulsion limits the number of ions that can be precisely controlled in a single trap. In contrast, neutral atoms usually interact only at very short range and can be collected in large ensembles without perturbing each other, a clear advantage for both metrology and QIP. Posssible future devices that can be used to constrcut these kinds of quantum computers will also be discussed in section~\ref{sec:Application of Topology in Integrated Circuits}. It should be pointed out that the quantum computing method and types are not limited to what we listed here.  The challenge for all these quantum computing development is the decoherence among qubits. This will introduce mistakes and errors during quantum computing. Quantum error correction ~\cite{Lidar} is a common technology solution now to solve this problem. However it will introduce big extra cost during the quantum computing technology development. Topological quantum computing is another solution without introduction of the extra costs~\cite{Wang2002,Wang2005,Wang2005A,Pachos,Stanescu}.	

A topological quantum computer is a theoretical quantum computer proposed by Alexei Kitaev in 1997~\cite{Kitaev2002,Kitaev}. It employs two-dimensional quasiparticles called anyons, whose world lines pass around one another to form braids in a three-dimensional spacetime. These braids form the logic gates that make up the computer. The advantage of a quantum computer based on quantum braids over using trapped quantum particles is that the former is much more stable. Small, cumulative perturbations can cause quantum states to decohere and introduce errors in the computation, but such small perturbations do not change the braids' topological properties.

\subsubsection{Anyons and Topological Systems}

Symmetries play a central role in physics. They dictate what one can change in a physical system without affecting any of its properties. You might have encountered symmetries like translational symmetry, where a system remains unchanged if it is spatially translated by an arbitrary distance. A system with rotational symmetry, however, is invariant under rotations. Some symmetries, like the ones mentioned above, give information about the structure of the system. Others have to do with the more fundamental physical framework that we adopt. An example for this is the invariance under Lorentz transformations in relativistic
physics. Other types of symmetries can be even more subtle. For example, it is rather self-evident
that physics should remain unchanged if we exchange two identical point-like particles. Nevertheless, this fundamental property that we call statistical symmetry gives rise to rich and beautiful physics. In three spatial dimensions it dictates the existence of bosons and fermions. These are particles with very different quantum mechanical properties. Their wave function acquires a +1 or a −1 phase, respectively, whenever two particles are exchanged. A direct consequence of this is that bosons can actually occupy the same state.
In contrast, fermions can only be stacked together with each particle occupying a different state. When one considers two spatial dimensions, a wide variety of statistical behaviours is possible. Apart from bosonic and fermionic behaviours, arbitrary phase factors, or even non-trivial unitary evolutions, can be obtained when two particles are exchanged~\cite{Leinaas}. Particles with such exotic statistics have been named anyons by Frank Wilczek~\cite{Wilczek}. For example, gases of electrons confined on thin films in the presence of sufficiently strong magnetic field and at a sufficiently low temperature give rise to the fractional quantum Hall effect~\cite{Tsui,Laughlin}. The low-energy excitations of these systems are localised quasiparticle excitations that exhibit anyonic statistics. Beyond the fractional quantum Hall effect, other two-dimensional systems have emerged which theoretically support anyons~\cite{Volovik}. These range from superconductors~\cite{Chamon} and topological insulators~\cite{ShouCheng2006, ShouCheng2009,ShouCheng2008, Hasan2010} to spin lattice models. Systems that support anyons are called topological as they inherit the topological properties of the anyonic statistical evolutions. Topological systems are usually many-particle systems that support localised excitations, so-called quasiparticles, that can exhibit anyonic behaviour. In general, they have highly entangled degenerate ground states. As a consequence local order parameters, such as the magnetisation, are not able to describe topological phases. So we need to employ non-local order parameters. Various characteristics exist that identify topological order, such as ground state degeneracy, edge states in the presence of a gapped bulk, topological entanglement entropy or the explicit detection of anyons. As topological order comes in various forms, the study and characterisation of topological systems in their generality is complex and still an open problem. Over the last years the richness in the behaviour of two-dimensional topological systems has inspired many scientists. One of the most thought-provoking ideas is to use topological systems for quantum computation. Consider, for example, the case where the exchange is not a mathematical procedure, but a physical process of moving two particles along an exchange path. The effect of this transport on the wave function should not depend on the particular shape of the path taken by the particles when they are exchanged or the speed the path is traversed. Nevertheless, the evolution might still depend on some global, topological characteristics of the path, such as the number of times the particles are exchanged. Statistical evolutions are hence topological in their nature.

\subsubsection{Quantum computation with anyons}
Quantum computation requires the encoding of quantum information and its efficient manipulation with quantum gates~\cite{Chuang}. Qubits, the quantum version of classical bits, provide an elementary encoding space. Quantum gates manipulate the qubits to eventually perform a computation. A universal quantum computer employs a sufficiently large set of gates in order to perform arbitrary quantum algorithms. In recent years, there have been two main quests for quantum computation. First, to find new algorithms, that go beyond the already discovered algorithms of searching~\cite{Grover}and factorising~\cite{Shor}. Second, to perform quantum computation that is resilient to errors. In the 1990s a surprising connection was made. It was argued by Castagnoli and Rasetti~\cite{Castagnoli}  that anyons could be employed to perform quantum computation. Kitaev~\cite{Kitaev2003}demonstrated that anyons could actually be used to perform fault-tolerant quantum computation. This was a very welcome advance as errors infest any physical realisation of quantum computation, coming from the environment or from control imperfections. Shor~\cite{Shor1995} and Steane~\cite{Steane} independently demonstrated that for sufficiently isolated quantum systems and for sufficiently precise quantum gates, quantum error correction can allow fault-tolerant computation. However, the required thresholds are too stringent and demand a large overhead in qubits and quantum gates for error correction to be realised. In contrast to this, anyonic quantum computation promises to resolve the problem of errors from the hardware level. Topological systems can serve as quantum memories or as quantum computers. They can encode information in a way that is protected from environmental perturbations. In fact, topological systems have already proven to be a serious candidate for constructing fault-tolerant quantum hard disks. The intertwining of anyons and quantum information in topological systems is performed in an unusual way. Information is encoded in the possible outcomes when bringing two anyons together. This information is not accessible when the anyons are kept apart, and hence it is protected. The exchange of anyons gives rise to statistical logical gates. In this way anyons can manipulate information with very accurate quantum gates, while keeping the information hidden at all times. If the statistical evolutions are complex enough then they can realise arbitrary quantum algorithms. Fundamental properties of anyonic quasiparticles can thus become the means to perform quantum computation. Fault-tolerance simply stems from the ability to keep these quasiparticles intact. The result is a surprisingly effective and aesthetically appealing method for performing fault-tolerant quantum computation.

We want to employ the statistical evolutions of anyons as a novel way to perform quantum computation. This promises to efficiently overcome the problem of errors that prohibit the reliable storing and manipulation of quantum information. Employing anyons for such technological tasks requires a good understanding of their properties. We also need to investigate in detail the properties of the topological systems that support anyons. Then the corresponding anyonic model supports universal quantum computation implemented just by braiding anyons~\cite{Freedman2002A,Freedman2002B}. For these models it has been shown~\cite{Simon,Burrello} that by weaving a single anyon among many static ones it is possible to perform a universal set of gates between arbitrary qubits. Then one can employ these gates to implement quantum computation following standard quantum algorithms. It is an important open problem to find a method
that efficiently overcomes probabilistic errors with a two-dimensional system. First important steps are taken in Chesi et al.~\cite{Chesi} and Hamma et al.~\cite{Hamma}. Ising anyons describe the statistical properties of Majorana fermions. The latter are under intense experimental investigation in the arena of fractional quantum Hall samples~\cite{Miller}, topological insulators~\cite{Fu}  and p-wave superconductors~\cite{Read}. It is possible to devise a more complicated scheme for universal quantum computation with purely topological means~\cite{Aguado,Mochon}. Topological degeneracy~\cite{Kells} and entanglement entropy~\cite{Chung2010,Yao}. Possible generalizations of this model have also been considered ~\cite{Yao2007,Yao2009}. The presence of interactions between vortices was established in Lahtinen et al~\cite{Lahtinen2008} and explicit demonstration of the non-Abelian statistics was obtained in Lahtinen and Pachos ~\cite{Lahtinen2009}. 

Two questions have a singular importance. The first natural question is: which physical systems can support non-Abelian anyons? Realising non-Abelian anyons in the laboratory is of fundamental and practical interest. Such exotic statistical behaviour has not yet been encountered in nature. The physical realisation of non-Abelian anyons would be the first step towards the identification of a technological platform for the realisation of topological quantum computation. The second question concerns the efficiency of topological systems in combating errors. It has been proven that the effect of coherent environmental errors in the form of local Hamiltonian perturbations can be suppressed efficiently without degrading the topologically encoded information ~\cite{Bravyi2010}. Nevertheless, there is no mechanism that can protect topological order from incoherent probabilistic errors. Topological systems nevertheless constitute a rich and versatile medium that allows imaginative proposals to be developed ~\cite{Chesi,Hamma}.

The measurement of anyons can be employed instead in order to realise a version of one-way quantum computation with static anyons~\cite{Bonderson2008,Bonderson2009}. Using this scheme, measurements of the topological charge can be used to generate the braiding transformations used in topological quantum computation without the need to physically transport anyons. Being able to bypass the anyonic transport overcomes a crucial obstacle in the implementation of topological quantum computation with
fractional quantum Hall liquids. Other electron systems are the p-wave superconductors that can support fractionally charged vortices with anyonic statistics~\cite{Read, Ivanov}. They have
been shown to be equivalent to Kitaev’s honeycomb lattice model~\cite{Chen2008,Yu2008}.  Topological insulators were discovered, which have similar properties as the integer and the fractional quantum Hall effect. Interestingly, they do not require the presence of an external magnetic field to acquire topological properties~\cite{Kane2006}. These materials are the focus of a rapidly developing field, both theoretically and experimentally~\cite{Hasan2010}. In addition, Levin and Wen proposed a family of topological models called the string net models that include the quantum doubles as special cases~\cite{Levin2005}. This generalized class provides a versatile laboratory to theoretically probe anyons. 

The second question is the concerns the stability of topological order under environmental perturbations. Topological order is stable at zero temperature, when the system is subject to local, weak and time-independent perturbations~\cite{Bravyi2010}. Such perturbations can be of the form of erroneous interaction terms added to the Hamiltonian of the system~\cite{Pastawski2009}. From the above discussion it is clear that topological systems by themselves are not able to protect topologically encoded information from temperature errors. Are there any modifications we can perform in order to achieve this goal? The aim is to safely store quantum information in a system, which is subject to finite temperature, for arbitrarily long
times and without performing continuous quantum error correction. Recently, two inspiring
schemes appeared that address this problem within the context of topological systems.
Hamma et al.~\cite{Hamma}  decorated the toric code with a scalar field that couples to anyons, thus
creating long-range attractive interactions between them. When temperature errors in the
form of anyonic excitations occur, then the attractive interaction causes them to annihilate.
Still, topological order remains intact as the ground state does not sense the presence of the
scalar field. If such a system could be designed, it would be characterised by a finite critical
temperature below which information could be reliably stored. An alternative scheme was
presented by Chesi et al.~\cite{Chesi}. It employs repulsive long-range interactions between the
anyons of the toric code. They showed that such a quantum memory is protected against
temperature fluctuations as it energetically penalises the generation of anyonic errors in the
system. These are just two examples that suggest how to combat finite temperature errors.

Beyond these two questions presented above there is a variety of problems that need to
be addressed. The better we understand the forms quantum matter can take, the better we
can encode and manipulate quantum information for technological applications, such as
performing error-free quantum computation. The theoretical and experimental investigation
of topological systems moreover provides a platform to study the physics of anyons
in their own right. We can probe and manipulate anyonic quasiparticles and study their
physics, such as transport phenomena or critical behaviour, without the need to resort to
the microscopic properties of the underlying topological system. This opens a wealth of
possibilities, where anyons can be the ingredients for fundamental research and for technological
applications.

\subsubsection{Topological quantum computation with DNA}

\begin{figure}[H]
\begin{center}
\epsfig{file=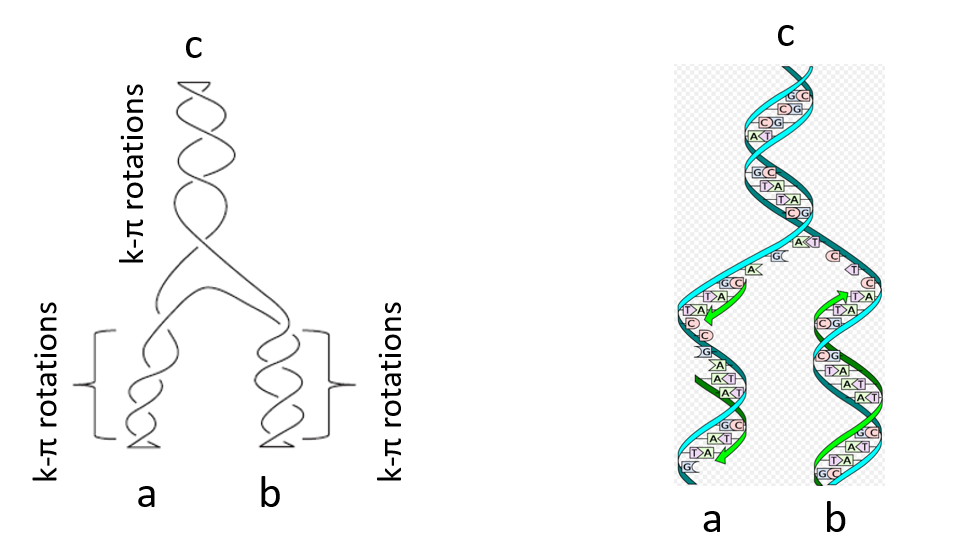,height=3.0in}
\caption{Compare of anyons and DNA quantum computation }
\label{fig:aonDNA}
\end{center}
\end{figure}

As described before, while the elements of a topological quantum computer originate in a purely mathematical realm, experiments in fractional quantum Hall systems indicate these elements may be created in the real world using semiconductors made of gallium arsenide at a temperature of near absolute zero and subjected to strong magnetic fields. One way to proceed is to get closer to the innermost structure of matter, beyond micro, into nano, even into femto. This means computing directly with molecules and atoms, maybe even with their parts. One hope is for a quantum computer, a computer of a completely different type that can make use of the strange but wonderful phenomena that occur at the atomic and intra-atomic levels. However, there are problems: just think about the wave-particle duality of quanta, about the possibility of a particle existing in several places at the same time and of several particles being in the same place at the same time, about teleportation, and so forth. There are a number of challenging difficulties from the computability point of view: for instance, that observation modifies the observed system, which means that reading some information destroys that information, or that quantum interference of particles cannot be controlled using current technology. Fortunately, for billions of years nature itself has been "computing" with molecules and cells. Thus another hope for the future is to use DNA and other proteins, manipulated by tools already known to biochemistry and genetic engineering. For instance, enzymes are one of the most intelligent gifts nature has made to us, as it has been said, and to develop wet computers with core chips consisting of bio-molecules~\cite{Calude}. Figure ~\ref{fig:aonDNA} does an analog compare between anyons with ribbons facilitates accounting for twists and exchanges and the DNA decomposition. Clockwise exchanging $k$ times anyons $a$ and $b$ can be continuously deformed to $k$ clockwise $\pi$ rotations for anyon $c$ and $k$ counterclockwise $\pi$ rotations for both anyons $a$ and $b$. These two configurations are topologically equivalent and coherence with each other~\cite{Pachos}. This process is very similar to a DNA molecular decompose to two sub DNA molecules as shown in Figure ~\ref{fig:aonDNA}.

\begin{figure}[H]
\begin{center}
\epsfig{file=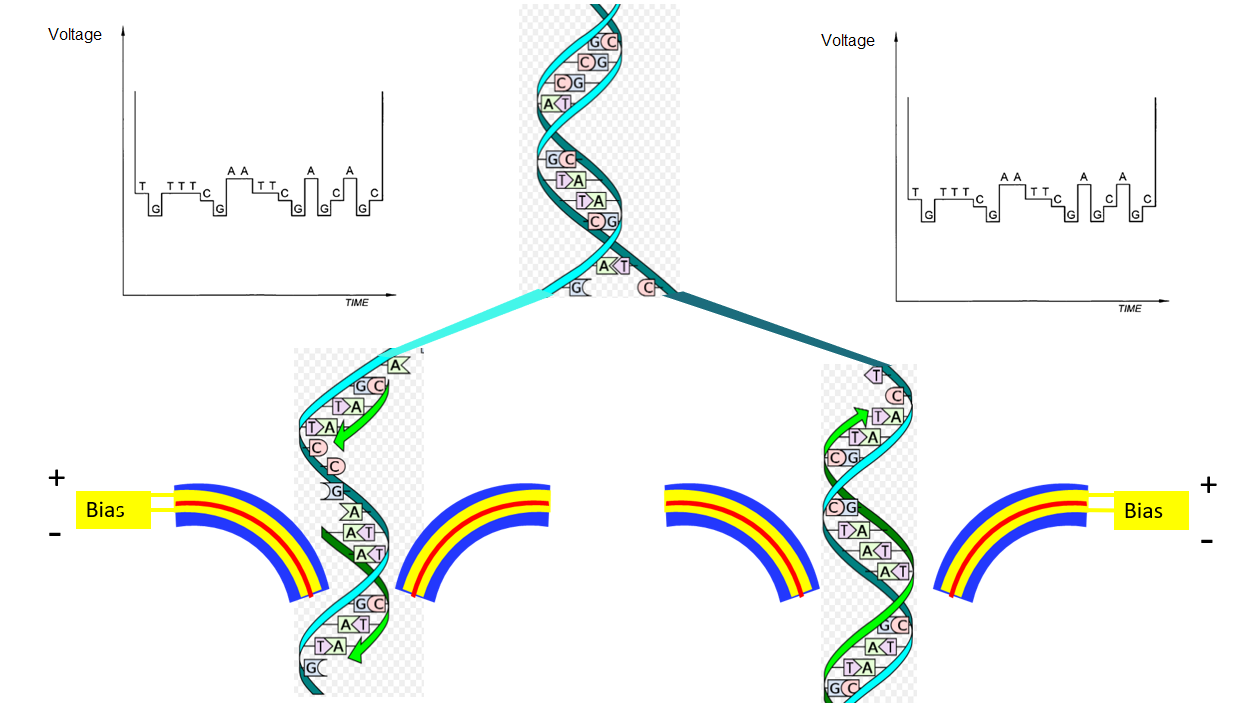,height=3.0in}
\caption{Diagram of an example to form qubit with long coherence time using DNA.}
\label{fig:tcDNA}
\end{center}
\end{figure}

As discussed before topological quantum computing can be realized with anyons. Figure ~\ref{fig:tcDNA}  gives an example of doing topological quantum computing with DNA molecules using the size controllable integrated devices we invented before ~\cite{Zhang2019C, Zhang2013C, Zhang2015H}. When a DNA molecule decompose to two sub-DNA molecules, we can use two size controllable device to record the bias as they pass through the holes, these two signals are coherence with each other since they come from the same DNA molecule. In this way, we can form a qubit with long coherence time.

\begin{figure}[H]
\begin{center}
\epsfig{file=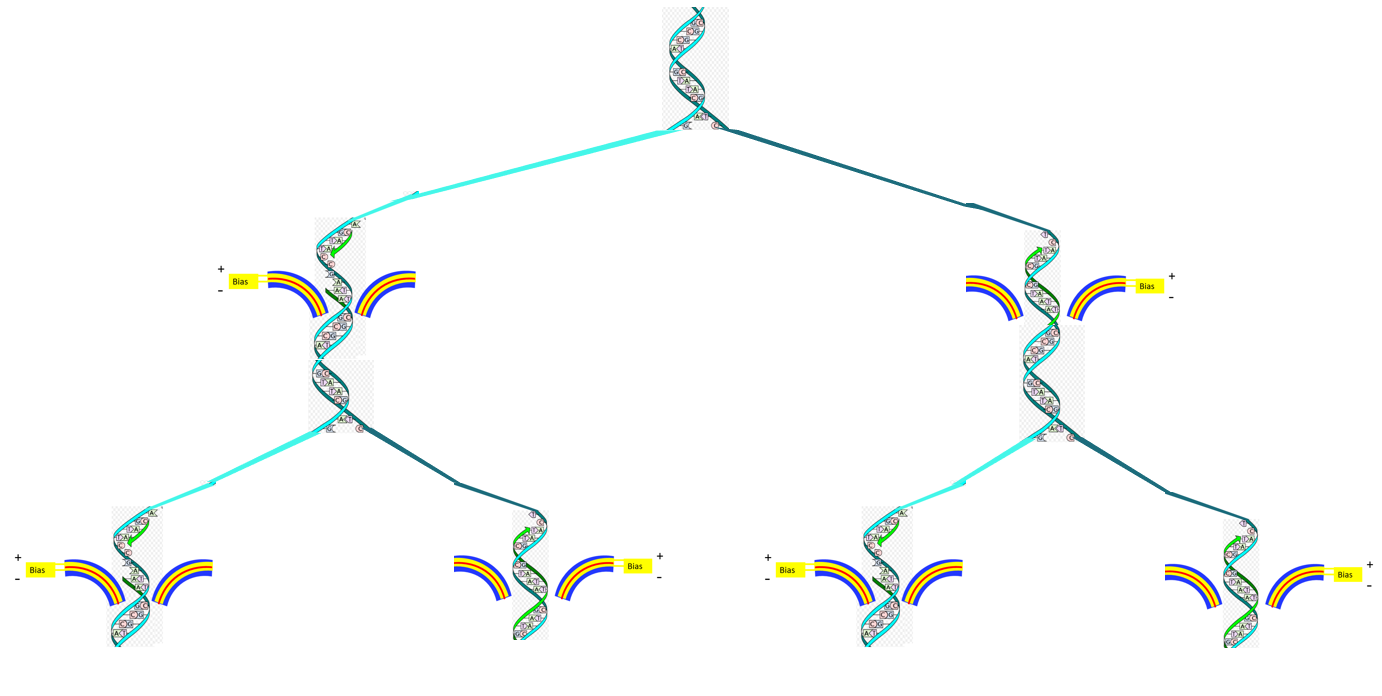,height=3.0in}
\caption{Diagram of down stream multi qubits topological quantum computing with long coherence time using DNA.}
\label{fig:dstqc}
\end{center}
\end{figure}

\begin{figure}[H]
\begin{center}
\epsfig{file=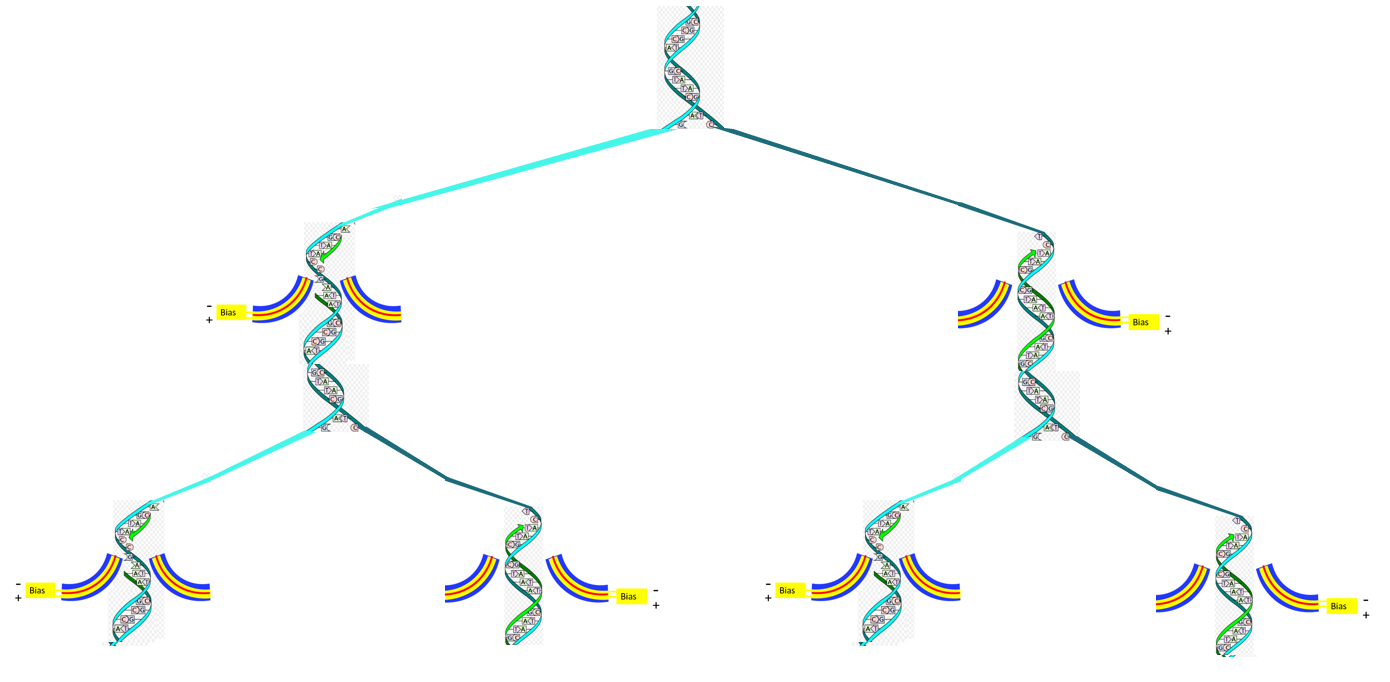,height=3.0in}
\caption{Diagram of up stream multi qubits topological quantum computing with long coherence time using DNA.}
\label{fig:ustqc}
\end{center}
\end{figure}

\begin{figure}[H]
\begin{center}
\epsfig{file=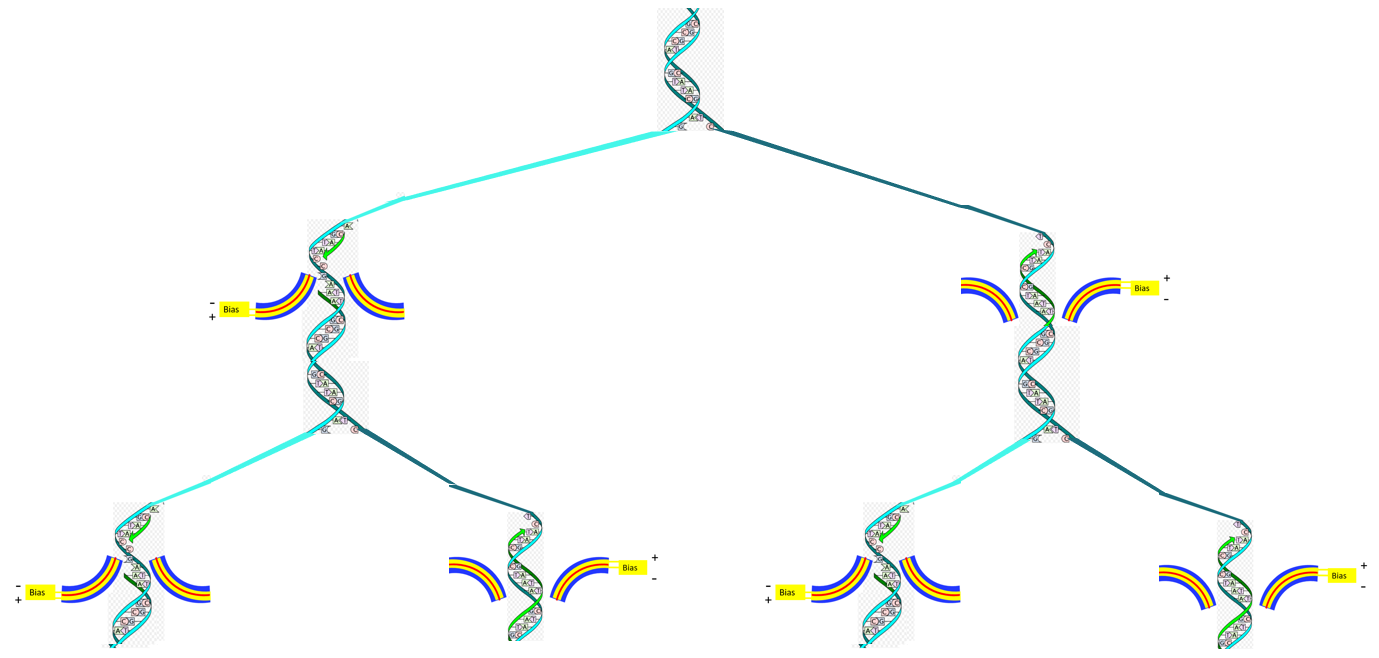,height=3.0in}
\caption{Diagram of mixing stream multi qubits topological quantum computing with long coherence time using DNA.}
\label{fig:mstqc}
\end{center}
\end{figure}

Now in the same way, we can construct multi qubits with long coherence time using DNA for topological quantum computing. They have three basic types as shown in  Figures ~\ref{fig:dstqc}, ~\ref{fig:ustqc} and ~\ref{fig:mstqc} based on the DNA molecules decomposition, composition and mixing of composition and decomposition. 

It should be pointed out that the construction of above three basic types of multi qubits can be continue to infinite numbers with Fibonacci sequence~\cite{Posamentier,Koshy2018, Koshy2019}. And the infinite number of qubits should all coherence with each other since they are from the same one DNA molecule.

\section{The nature of topological properties and its power in layer 3 }
A group of three closely related things is called trinity. The Daodejing is one of the foundational texts of Chinese thought. It explains the trinity in the power of the Way.  The Way generates the Unique;The Unique generates the Double;The Double generates the Triplet;The Triplet generates the myriad things~\cite{Ryden}. 
In this section, we will see the use of iterated function systems (IFS)to simulate the topological duality connection between the qubits represented by the control code space and singularities represented by the fractal space. This studies give us clue for the conjunction we made in section ~\ref{subsubsec:Singular} that although in the future the technology singularities will dominate the worlds, but it can still be controlled through the topological duality connections between singularities and their control codes.

\subsection{Topological power studies via IFS}
In our previous work, the fractal properties had been studied via IFS~\cite{Zhang2020A}.  Fractal space is manipulated by the manipulating space or codes that are composed of the real numbers of the transformation functions and the probabilities assigned to the functions. Here we consider the manipulating space or codes form the qubits and fractal spaces form the singularities to discuss the duality connection between qubits and singularities via IFS.  Also we use the IFS Generator program that developed by Brendan Harding and Michael Barnsley~\cite{Barnsley} to demonstrate the topological trinity principle.

\subsubsection{Topological trinity principle}
\label{subsubsec:Topological trinity principle}
\begin{figure}[H]
\begin{center}
\epsfig{file=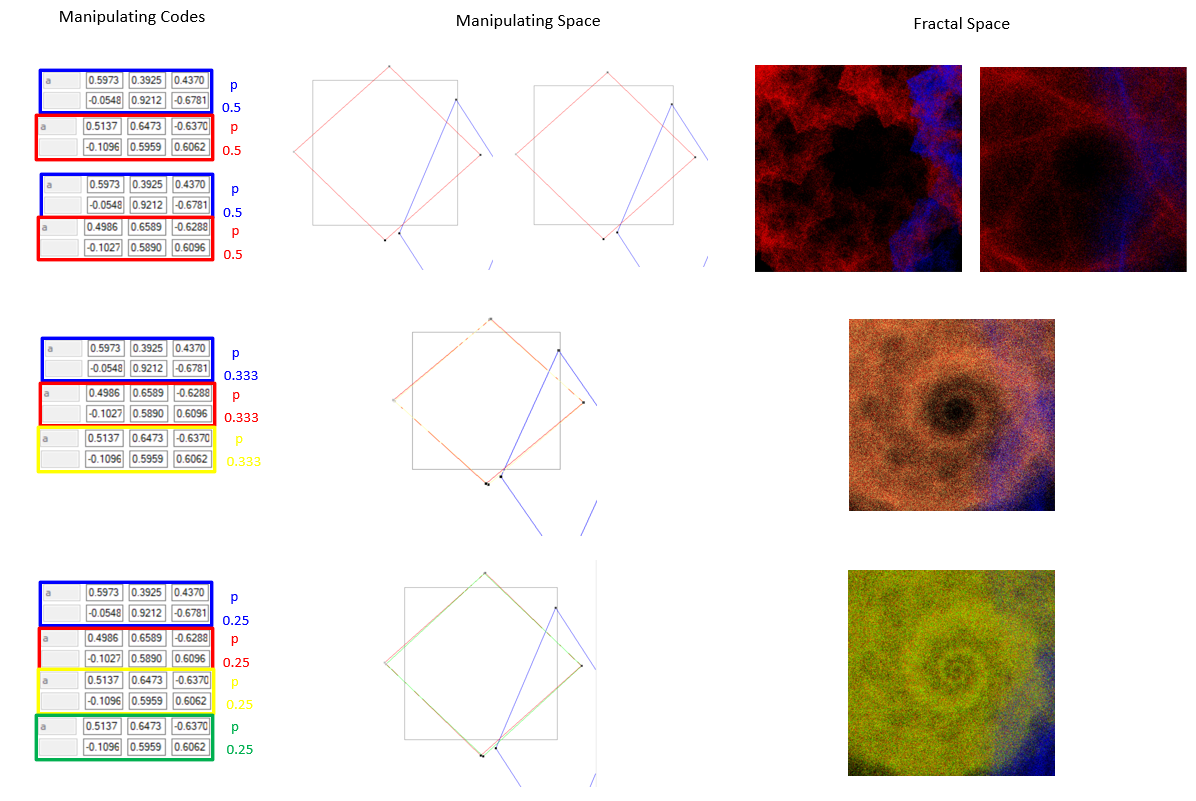,height=3.0in}
\caption{Example of trinity via IFS}
\label{fig:its0}
\end{center}
\end{figure}

 In the previous study, we conclude that the real pattern appeared in the fractal space independent to the type of transformations functions~\cite{Zhang2020A}. So here we will all use affine transformation function for our studies. Figure~\ref{fig:its0} shows the topological duality connection between manipulating codes or space qubits and the singularity black hole structure with different number of affine transformation functions. The top picture uses two affine transformation functions, we found that no matter how do you deform the qubits parameters, no singularity black hole structure can be formed. The middle picture uses three affine transformation functions, we found that with the same qubits parameters, the singularity black hole structure can be formed. The bottom picture uses four affine transformation functions, we found that with the same qubits parameters, the singularity black hole structure can be formed with rich details.Based on these studies, we propose the following topological trinity principle: To form the myriad things in fractal space via IFS, the minimal transformations functions required is three. Also these three transformations functions must meet some close relationship. This will be described in the next section.

\subsubsection{Closing principle for singularity generation}
\label{subsubsec:Closing principle for singularity generation}

\begin{figure}[H]
\begin{center}
\epsfig{file=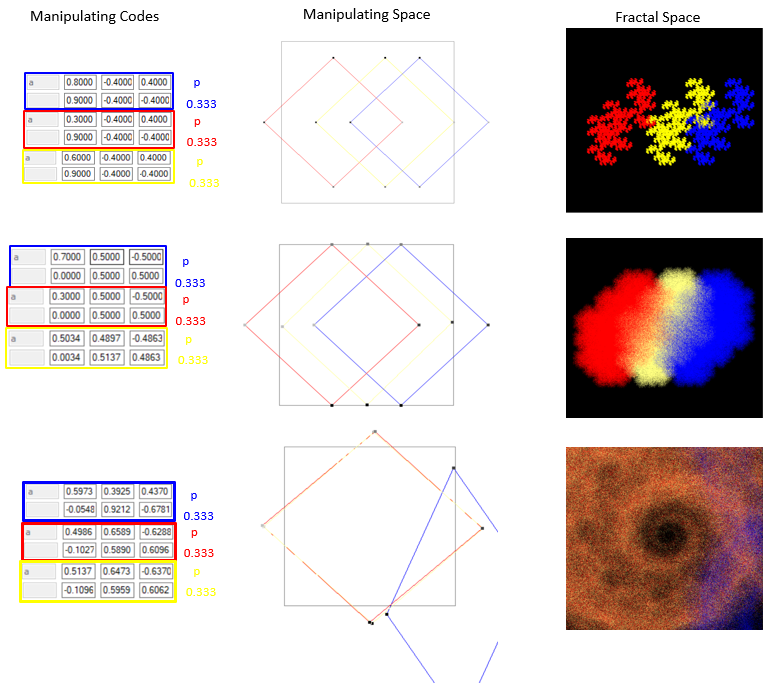,height=3.0in}
\caption{Example of fractal space structures singularity generation and the closing principals at manipulating space}
\label{fig:its1}
\end{center}
\end{figure}

Figure~\ref{fig:its1} shows an example of fractal space structures singularity generation. It was found that when the manipulating code or qubits equally distribute in the manipulating space, you can form either discrete structure such as dragons  or the fuzzy structure in the fractal space as shown in Figure~\ref{fig:its1} top and middle picture. To form the singularity structure such as black holes in fractal space, you must have two of the manipulating code or qubits close enough in the manipulating space as shown in Figure~\ref{fig:its1} bottom picture. We call this the closing principle for singularity generation.

\subsubsection{Singularity structure obtained with  same topological class of qubits}
\label{subsubsec:Singularity structure obtained with  same topological class of qubits}

\begin{figure}[H]
\begin{center}
\epsfig{file=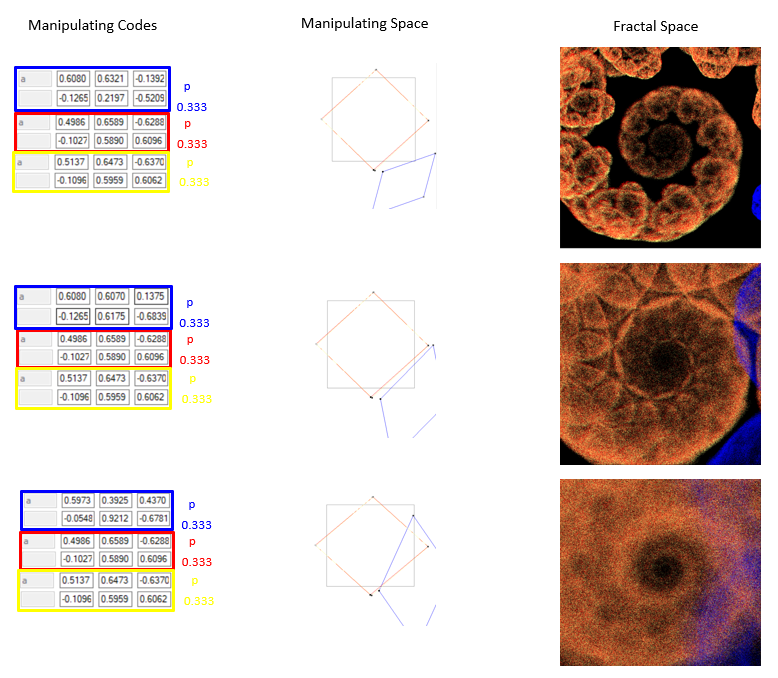,height=3.0in}
\caption{Example of fractal space structures obtained with  same topological class of manipulating space and manipulating codes}
\label{fig:its2}
\end{center}
\end{figure}

As we discussed the intuitive concept of topology in section~\ref{subsec:Intuitive} Figure~\ref{fig:tc}, the points A,B,C can inside, outside and on the circle is the topological properties. Figure~\ref{fig:its2} show the blue qubit can outside and on the red and yellow qubits circle. Figure~\ref{fig:its2} top and middle picture belong to the same topological class which are outside the circle. Figure~\ref{fig:its2} bottom picture  belong to the different topological class which is blue qubit across the red and yellow qubits circle. The black hole singularities structure in fractal space happened in this class. Is this the  sufficient condition to generate the black hole singularities structure in fractal space? This will be discussed in next section.

\subsubsection{Singularity structure obtained with  different topological class of qubits}
\label{subsubsec:Singularity structure obtained with  different topological class of qubits}

\begin{figure}[H]
\begin{center}
\epsfig{file=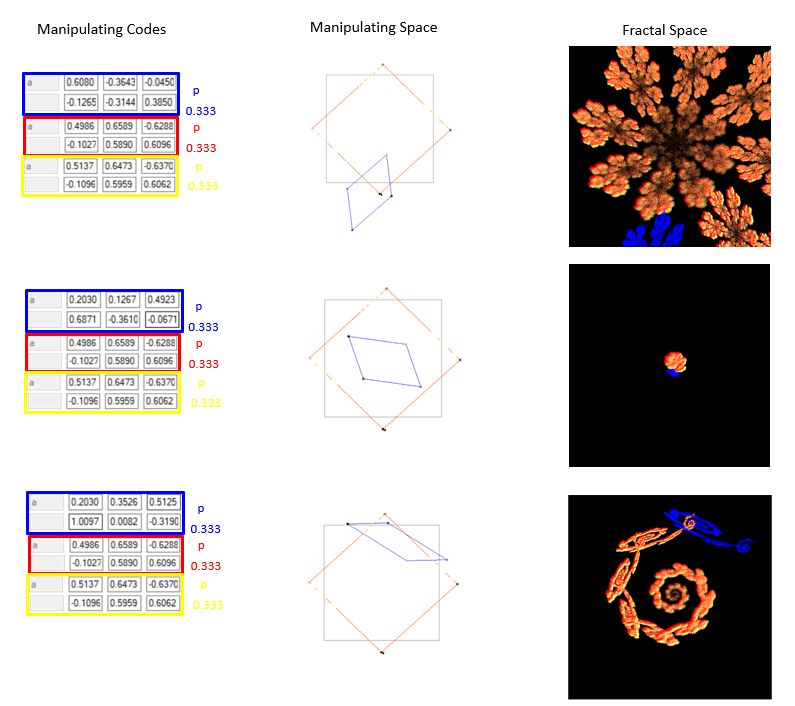,height=3.0in}
\caption{Example of fractal space structures obtained with  different topological class of manipulating space and manipulating codes with 3 iteration functions}
\label{fig:its3}
\end{center}
\end{figure}

Figure~\ref{fig:its3} show the blue qubit across and inside  the red and yellow qubits circle. Figure~\ref{fig:its3} middle picture belong to the topological class which is inside the red and yellow qubits circle. Figure~\ref{fig:its3} top and botttom picture belong to the same topological class which are across the red and yellow qubits circle. These two have the same  topological class with Figure~\ref{fig:its2} bottom picture. But their structures in the fractal space are very different. From these compare studies, we can also conclude that the closing principle and the blue qubit across the red and yellow qubits circle are ONLY the necessary condition for the black hole singularities structure generation in fractal space. They are NOT the sufficient conditions.

\subsubsection{Effects of number of iteration functions on singularity structures}
\label{subsubsec:Effects of number of iteration functions on singularity structures}

\begin{figure}[H]
\begin{center}
\epsfig{file=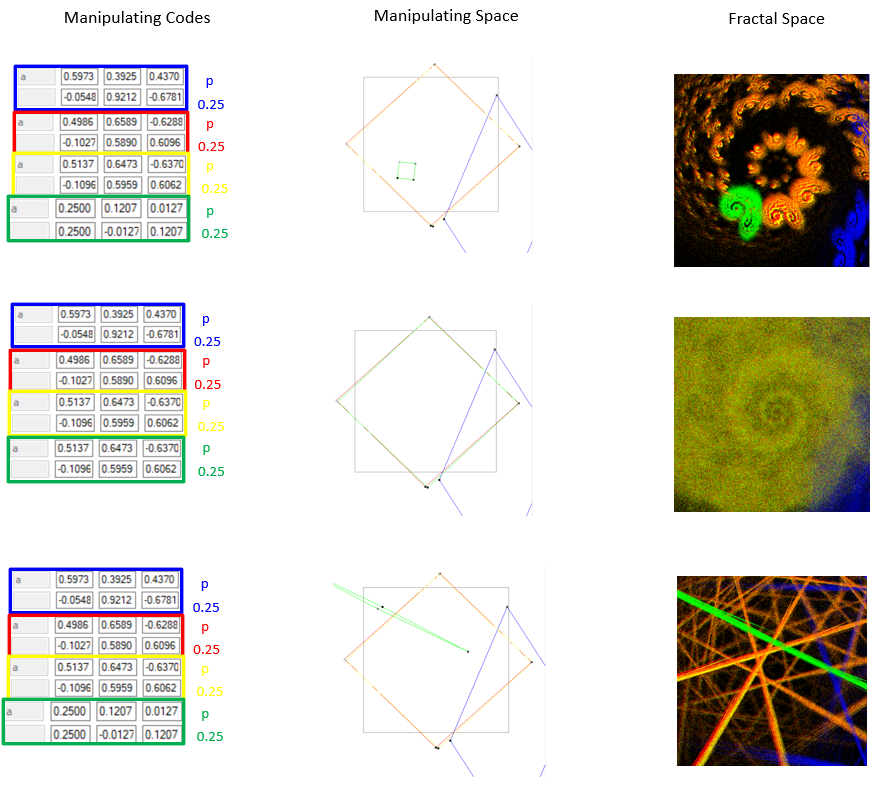,height=3.0in}
\caption{Example of fractal space structures obtained with  different topological class of manipulating space and manipulating codes with 4 iteration functions.}
\label{fig:its4}
\end{center}
\end{figure}

To study the effects of number of iteration functions on singularity structures, Figure~\ref{fig:its4} show the fractal space structures obtained with  different topological class of manipulating space qubits with 4 iteration functions. Compare to the two and three iteration functions fractal space that shown in Figure~\ref{fig:its0}, we can see that there more complex and detail structures can be observed with large number of iteration functions.

\subsection{Topological power studies via knowledge structure}

Duality and trinity are viewed as intrinsically topological objects and are recognized as common knowledge
shared among human society activity, mathematics, physics,chemistry,biology and many other kinds of nature science and technology. In this section we will continue the trinity principle studies via knowledge structure and the duality connection between symmetric trinity and arbitrary trinity.

\subsubsection{Topological power via symmetric trinity}
\begin{figure}[H]
\begin{center}
\epsfig{file=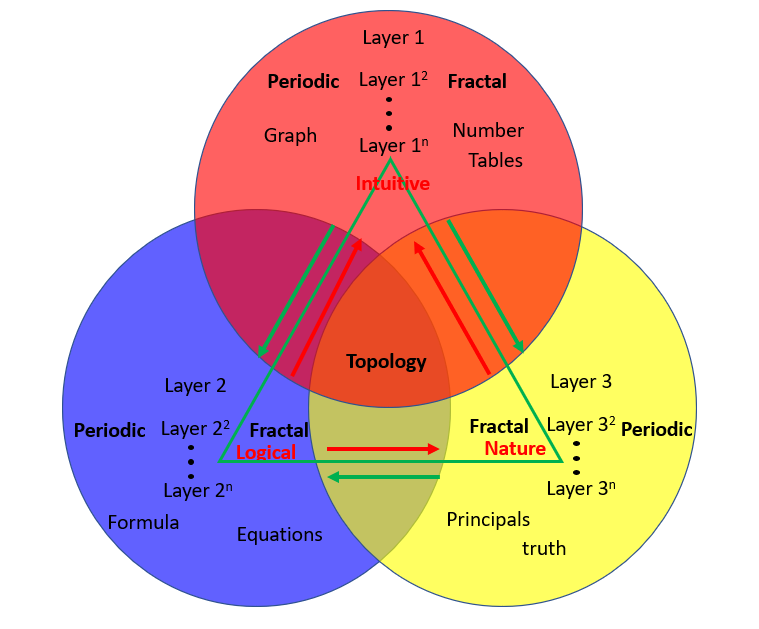,height=3.0in}
\caption{Topological properties and characterizations in three layers and their relationships: the first layer is intuitive concept,  characterizations and applications, the second layer is logical physics understanding of topological properties, characterizations and applications, and the third layer is the nature of topological properties and its power.}
\label{fig:triC}
\end{center}
\end{figure}

\begin{figure}[H]
\begin{center}
\epsfig{file=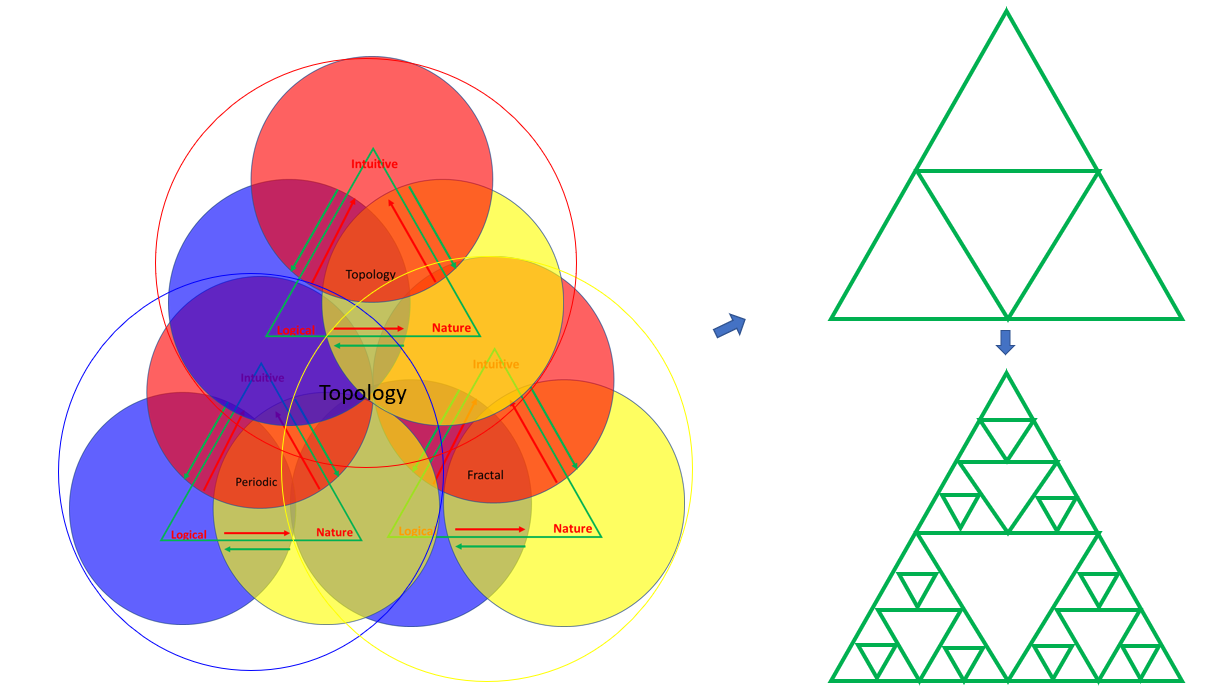,height=3.0in}
\caption{Topological properties and characterizations in three layers have multi scale characterization and can be represented by a Fractal}
\label{fig:t2f}
\end{center}
\end{figure}

In this paper, we described topological properties and characterizations in three layers as shown in Figure~\ref{fig:triC}.  The first layer is intuitive concept, characterizations and applications,the typical expression format used is graphs, numbers and tables. Even for the first layer intuitive concept, the understanding can be divided to different scale and depth, we marked to level $1$, level $1^2,...$, level $1^n$.   The second layer is logical physics understanding of topological properties, characterizations and applications, the typical expression format used is formula and equations. The second layer understanding can also be divided to different scale and depth, we marked to level $2$, level $2^2,...$, level $2^n$. And the third layer is the nature of topological properties and its power, the typical expression format used is principals and truth. The third layer understanding can also be divided to different scale and depth, we marked to level $3$, level $3^2,...$, level $3^n$. Figure~\ref{fig:triC} describe the topological properties studies, so the center of the core is the topology. In the same way, when we study the fractal properties and periodic properties, we can have fractal and periodic centered graph, put them together we get a knowledge trinity graph as shown in Figure~\ref{fig:t2f}.

If we consider the multi scale understanding, the graph can be represented by a Fractal as shown in Figure~\ref{fig:t2f}.

\subsubsection{Topological power via arbitrary trinity}
\begin{figure}[H]
\begin{center}
\epsfig{file=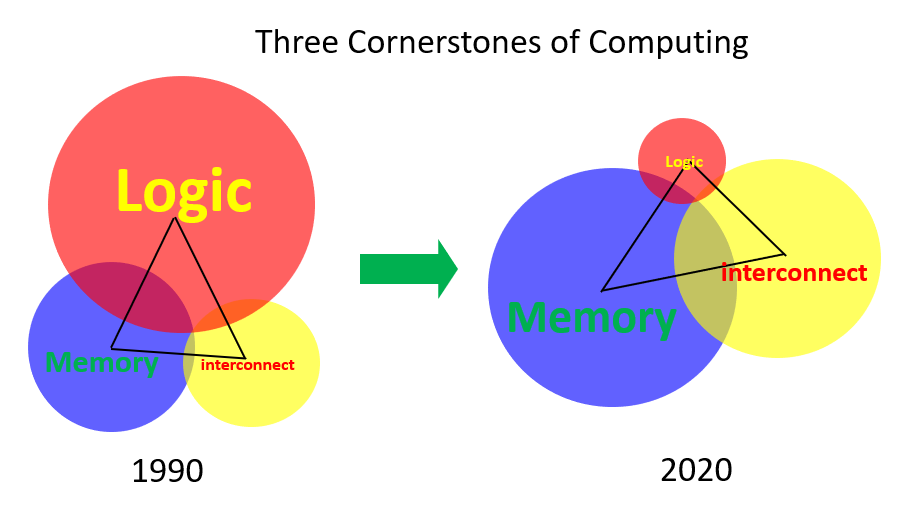,height=3.0in}
\caption{Time evolution of  the three cornerstones of computing with logic, memory and interconnect show the topological change via arbitrary trinity}
\label{fig:ct}
\end{center}
\end{figure}

It is easy for us to describe the knowledge structure use symmetric trinity in the paper, however, in the real world, the knowledge development tend to be an arbitrary trinity structure.  For example, in the electronic computing technology development, the portion of logic, memory and interconnect three cornerstone of computing ~\cite{Young} on the silicon chip area show a topological change via arbitrary trinity as shown in Figure~\ref{fig:ct} 

\subsubsection{Symmetric trinity and arbitrary trinity are mutual root of each other}

\begin{figure}[H]
\begin{center}
\epsfig{file=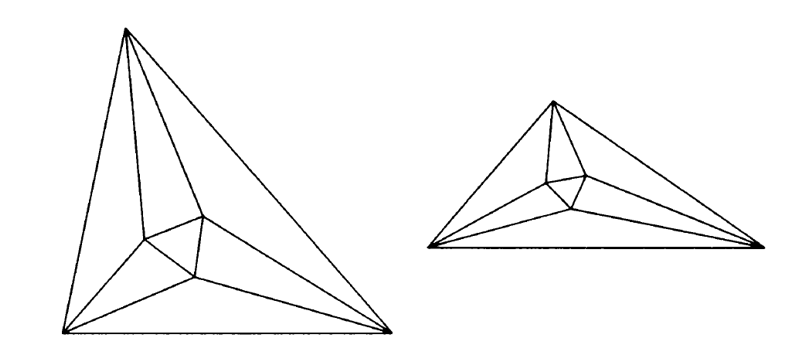,height=2.6in}
\caption{Morley's Theorem states that for any triangle, whatever the lengths of the sides, if you trisect each angle which mean you divide it into three equal angles then the lines of trisection will all intersect at the corners of an equilateral triangle - one whose sides are of equal length.}
\label{fig:tria}
\end{center}
\end{figure}

It should be pointed out that symmetric trinity and arbitrary trinity can be connected via duality as we described in section~\ref{subsubsec:Duality}.  They are compensate and conflict between each other,
mutual root between each other to form and develop the knowledge. Morley's Theorem maybe a good proof of this relationship as shown in Figure~\ref{fig:tria}.

In mathemtics, trichotomy of positive solutions are often used to solve the complicated topology and singularities problems ~\cite{Cirstea}

\section{Application of Topology in Integrated Circuits }
\label{sec:Application of Topology in Integrated Circuits}

Integrated circuits have various device architectures such as Vertical Field Effect Transistor (VFET)~\cite{Zhang2018B,Zhang2016E,Zhang2016F,Zhang2017A,Zhang2017I,Zhang2020M,Zhang2020N,Zhang2021C,Zhang2018C}, Vertical Turnal Field Effect Transistor(VTFET)~\cite{Zhang2018I,Zhang2020J,Zhang2016A,Zhang2021B,Zhang2020G},Vertical Gate All Around Vacuum Channel Transistor(VGAACT)~\cite{Zhang2017G,Zhang2020K,Zhang2021D}, Nanosheet Transistor~\cite{Zhang2017B, Zhang2018D,Zhang2017C,Zhang2019I,Zhang2020C,Zhang2020F,Zhang2020L,Zhang2020P}, Photonic Integrated Circuits~\cite{Zhang2015B, Zhang2019D,Zhang2020D}, Bio sensor~\cite{Zhang2017D,Zhang2018A}, 3D stack~\cite{Zhang2013A,Zhang2014B}, 3D package~\cite{Zhang2013B,Zhang2020O,Zhang2019E}, 3D cooling~\cite{Zhang2014A}, Radio Frequence Integrated Circuits~\cite{Zhang2019F}, Analog Integrated Circuits~\cite{Zhang2019F}, Logic Integrated Circuits~\cite{Zhang2019G}, Input/Output devices (I/O)~\cite{Zhang2017E}, Fully Deplete Silicon on Insulator device (FDSOI)~\cite{Zhang2018E}, Dynamic Random Access Memory (DRAM)~\cite{Zhang2015C}, Static Random Access Memory(SRAM)~\cite{Zhang2017F},  Phase Change Memory(PCM)~\cite{Zhang2015D}, Magnetic Random Access Memory (MRAM)~\cite{Zhang2018F}, Resistive Random Access Memory (RRAM)~\cite{Zhang2019H}, Thin Film Transistor~\cite{Zhang2018G},  Crack Stop~\cite{Zhang2017H}, Efuse~\cite{Zhang2016B}, Electical Static Diode(ESD)~\cite{Zhang2015E}, SiC device~\cite{Zhang2015F}, High Electron Mobility Transistor (HEMT)~\cite{Zhang2015G} and micro-electromechanical system (MEMS) devices~\cite{Zhang2016C,Zhang2020B} etc.

\begin{figure}[H]
\begin{center}
\epsfig{file=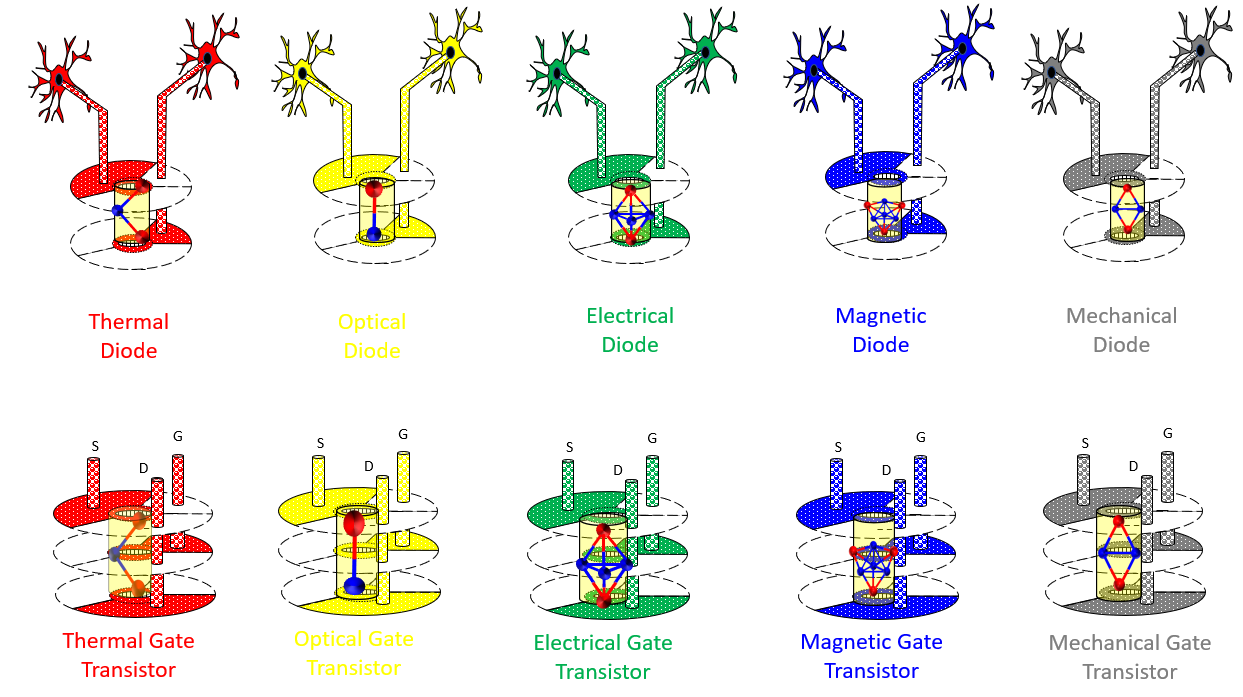,height=3.0in}
\caption{Example of form QCS individual thermal, optical, electrical, magnetic and mechanical diode or gate controlled transistor using various type molecular clusters}
\label{fig:i-di-tri}
\end{center}
\end{figure}

\begin{figure}[H]
\begin{center}
\epsfig{file=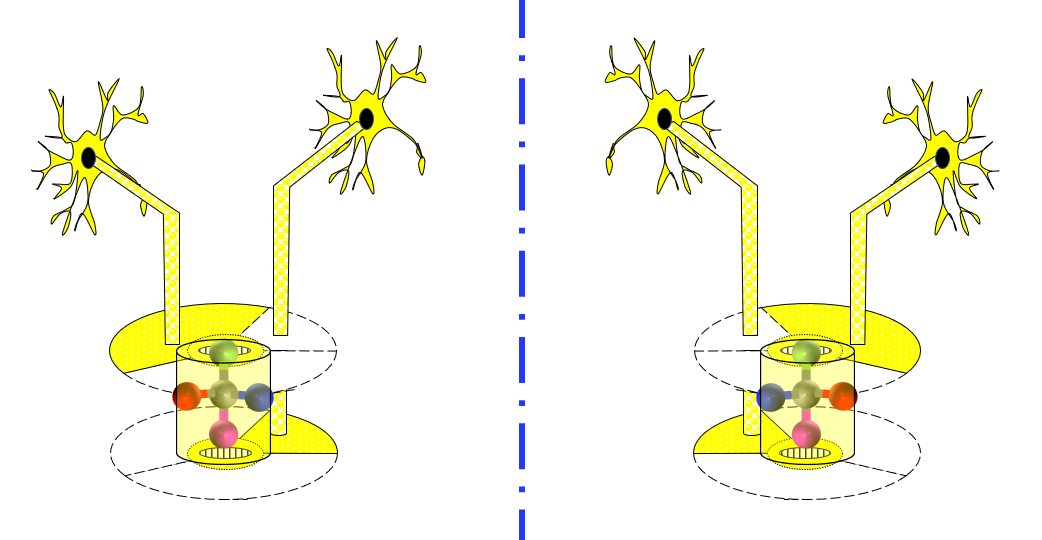,height=3.0in}
\caption{Example of form chiral circuit with chrial molecular clusters with the process described in ~\cite{Zhang2021A,Zhang2018H}}
\label{fig:chiral}
\end{center}
\end{figure}

\begin{figure}[H]
\begin{center}
\epsfig{file=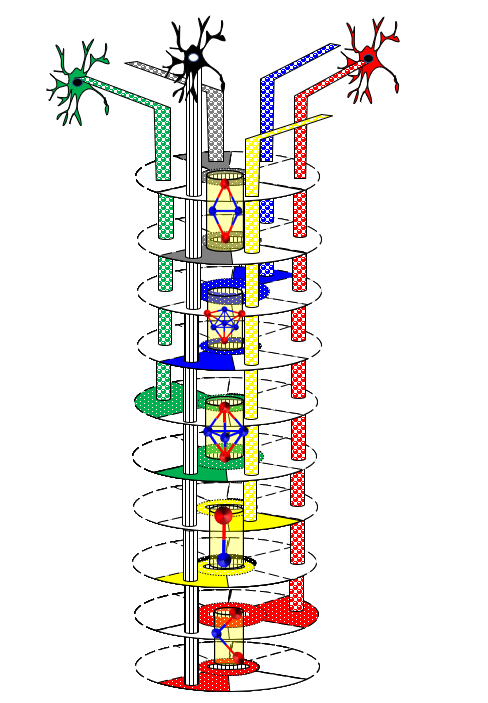,height=3.0in}
\caption{Example of form QCS integrated thermal, optical, electrical, magnetic and mechanical diode devices using various type molecular clusters}
\label{fig:t-diode}
\end{center}
\end{figure}

As technology node continues to shrink, molecular device with different type switch become very important. In our previous work, we proposed a novel future concept device with quantum confinement switch (QCS) ~\cite{Zhang2019B,Zhang2019C,Zhang2019A}. Here we further develop this kind of devices from topological point of view.  Figure ~\ref{fig:i-di-tri} shows the individual thermal, optical, electrical, magnetic and mechanical diode with QCS. Both 2D disc and 1D connection are the topologically simplified drawings. In the practice, the 2D disc can be constructed based on the process we described in ~\cite{Zhang2020N,Zhang2018C}; the 1D connection can be the carbon nano tubes that can be used to conduct heat, optical, electrical, magnetic and acoustic wave flow between two biological cells based on the process we described in ~\cite{Zhang2017D,Zhang2018A}. The molecular clusters that used for the quantum confinement switch can be cation, anion and neutral molecular clusters described in ~\cite{Zhang2019C,Zhang2021A} and can be deposited on the wafers based on the method described in ~\cite{Zhang2021A, Zhang2018H}. This kinds of concepts can also be easily  used to construct the QCS individual thermal, optical, electrical, magnetic and mechanical gate controlled transistor using various type molecular clusters as shown in the bottom of the picture. In the future, this kind of device can be used to form both thermal, optical, electrical, magnetic and mechanical classical  bits as we described in~\cite{Zhang2019C} and quantum bits (Qbits) as we described here for classical and quantum computation as described in section~\ref{subsec:Topological Quantum Computation} depending on the cluster type and size. Also if the chiral molecular clusters are deposit in the center of the vertical transitors with process descirbed in~\cite{Zhang2021A, Zhang2018H} or in the center of the vaccum transitors with process descirbed in~\cite{Zhang2017G,Zhang2020K}, the chiral symmetry integrated circuits can also be formed as shown in Figure~\ref{fig:chiral}. In the same way, we can form the anti-particle integrated circuits, the Majorana integrated circuits and various singularity integrated circuits if the anti-particle pairs or Majorana particles etc are filled in. As discussed in section ~\ref{subsubsec:Singular}, these kinds of devices with inner singularities will play key roles in the realization of technology singularities and  future communication built up among the billions
of galaxies and planets.   Figure ~\ref{fig:t-diode} shows integrated thermal, optical, electrical, magnetic and mechanical diode devices using various type molecular clusters. With this kind of device, thermal, photonic, electric, magnetic and mechanical energy can be redistributed among different biological cells, potentially with better therapeutic effects~\cite{Zhang2017D,Zhang2018A}. Figure ~\ref{fig:t-diode} device can also be used to do nanoscale thermal, optical, electrical, magnetic, mechanical and other various energy transport and conversion ~\cite{Chen,Chen2004}.

\begin{figure}[H]
\begin{center}
\epsfig{file=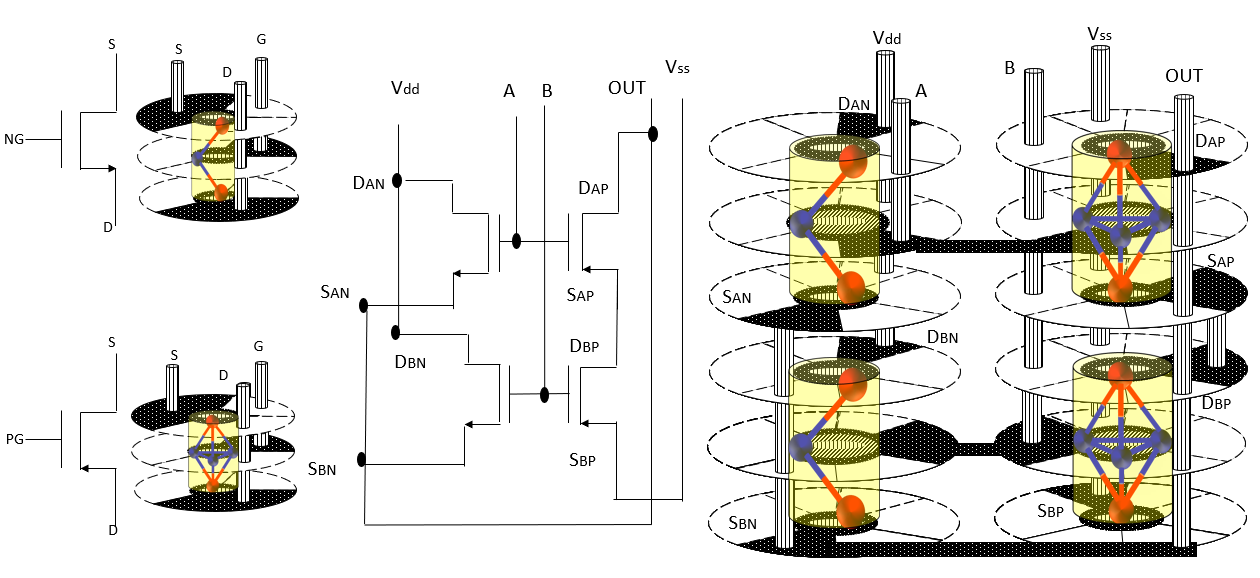,height=2.6in}
\caption{Example of form QCS NG device using anions molecular cluster and PG devices using cation molecular cluster; form NOR gate device with NG and PG. }
\label{fig:b-diode}
\end{center}
\end{figure}

Figure ~\ref{fig:b-diode} shows example of form QCS N gate device using anions molecular cluster and P gate devices using cation molecular cluster; form NOR gate device with N gate and P gate devices. Since NOR gate is an universal gate. All Boolean function can be implemented using NOR gate~\cite{Zhang2020N,Zhang2018C}. This can be used to form the quantum computer CPU described in  section~\ref{subsec:Topological Quantum Computation} and quantun sensing~\cite{Yuhua,Degen}.  With this kind of device,  various type of energy such as heat, photons and charges can also be redistributed among different biological cells with logical gate control as described in~\cite{Zhang2017D,Zhang2018A}, this will further enhance the potentiality of therapeutic effects.

\section{Conclusion}
Time and tide come to a critical step where the technology can take the lead ahead of science and human society activity. How did a single genesis qubit create billions of galaxies, black holes, stars, and planets? How the communications among the billions of galaxies and planets can be built up? How are atoms assembled on the earth and other possible planets to form complex organisms that attract us to explore their roots~\cite{Rees,Barrow}? Technology development will drive us to find the solutions of these questions and the deep connections among these questions. The technological singularity will dominate the world future and can be controlled through the topological duality connections between singularities and their control codes. Duality and trinity are viewed as intrinsically topological objects and are recognized as common knowledge shared among human society activity, mathematics, physics,chemistry,biology and many other kinds of nature science and technology. Topology is precisely the mathematical discipline that allows the passage from local to global. 

Long-range entanglement can give rise to both gauge interactions and Fermi statistics. In contrast, the geometric point of view can only lead to gauge interactions. We should use entanglement pictures to understand our world. This way, we can get both gauge interactions and fermions from a single origin- qubits. We live in a truly quantum world and the universes communication among the billions of galaxies and planets can be built up through the long-range entanglement.

Technologically the long-range entanglement can be realized through the devices that have inner singularities. Various future integrated circuits devices have been reviewed from the point views of topology.

\end{document}

%% file: econfmacros.tex
\def\beq{\begin{equation}}
\def\eeq#1{\label{#1}\end{equation}}
\def\eeqn{\end{equation}}

\def\beqa{\begin{eqnarray}}
\def\eeqa#1{\label{#1}\end{eqnarray}}
\def\eeqan{\end{eqnarray}}

\let\bar=\overbar

\def\Dslash{\not{\hbox{\kern-4pt $D$}}}
\def\dslash{\not{\hbox{\kern-2pt $\del$}}}

\def\msb{{\bar{\ssstyle M \kern -1pt S}}}

